\documentclass[aps,twocolumn,floatfix,longbibliography,superscriptaddress]{revtex4-2}
\usepackage{graphics,graphicx,epsfig}
\usepackage{amssymb,color}
\usepackage{epsf,epstopdf}
\usepackage {amsmath}
\usepackage[final,hidelinks]{hyperref}
\usepackage{ragged2e}
\usepackage{siunitx}

\newcommand{\avg}[1]{\langle{#1}\rangle{}}
\newcommand{\beq}{\begin{equation}}
\newcommand{\eeq}{\end{equation}}
\newcommand{\beqn}{\begin{eqnarray}}
\newcommand{\eeqn}{\end{eqnarray}}

\begin{document}

\title{Gene activity fully predicts transcriptional bursting dynamics} 

\author{Po-Ta Chen}
\thanks{These authors contributed equally and are ordered alphabetically.}
\affiliation{Joseph Henry Laboratories of Physics \& Lewis-Sigler Institute for Integrative Genomics, Princeton University, Princeton, NJ 08544, USA}
\author{Michal Levo}
\thanks{These authors contributed equally and are ordered alphabetically.}
\affiliation{Joseph Henry Laboratories of Physics \& Lewis-Sigler Institute for Integrative Genomics, Princeton University, Princeton, NJ 08544, USA}
\affiliation{Department of Biochemistry and Molecular Biophysics, Columbia University, New York, NY, USA}
\author{Benjamin Zoller}
\thanks{These authors contributed equally and are ordered alphabetically.}
\affiliation{Joseph Henry Laboratories of Physics \& Lewis-Sigler Institute for Integrative Genomics, Princeton University, Princeton, NJ 08544, USA}
\affiliation{Department of Stem Cell and Developmental Biology, CNRS UMR3738 Paris Cité, Institut Pasteur, 25 rue du Docteur Roux, 75015 Paris, France}
\author{Thomas Gregor}
\affiliation{Joseph Henry Laboratories of Physics \& Lewis-Sigler Institute for Integrative Genomics, Princeton University, Princeton, NJ 08544, USA}
\affiliation{Department of Stem Cell and Developmental Biology, CNRS UMR3738 Paris Cité, Institut Pasteur, 25 rue du Docteur Roux, 75015 Paris, France}
 
\date{\today}
 
\begin{abstract}
Transcription commonly occurs in bursts, with alternating productive (ON) and quiescent (OFF) periods, governing mRNA production rates. Yet, how transcription is regulated through bursting dynamics remains unresolved. Here, we conduct real-time measurements of endogenous transcriptional bursting with single-mRNA sensitivity. Leveraging the diverse transcriptional activities in early fly embryos, we uncover stringent relationships between bursting parameters. Specifically, we find that the durations of ON and OFF periods are linked. Regardless of the developmental stage or body-axis position, gene activity levels predict individual alleles' average ON and OFF periods. Lowly transcribing alleles predominantly modulate OFF periods (burst frequency), while highly transcribing alleles primarily tune ON periods (burst size). These relationships persist even under perturbations of cis-regulatory elements or trans-factors and account for bursting dynamics measured in other species. Our results suggest a novel mechanistic constraint governing bursting dynamics rather than a modular control of distinct parameters by distinct regulatory processes. 
\end{abstract}

\maketitle
 
 
Eukaryotic transcriptional regulation is an inherently dynamic and stochastic process, orchestrated by a series of molecular events governing productive transcription initiation by individual RNA polymerases (Pol II complexes) \cite{Lelli2012,Cramer2019}. This process culminates in nascent RNA synthesis, which in turn shapes protein production, and thus dictates cellular identity and behavior in both space and time. Consequently, revealing the fundamental principles underpinning transcriptional dynamics is paramount for understanding and predicting cellular phenotypes.

Research on various biological systems, from yeast to mammalian cells, revealed that transcription occurs in bursts. These bursts involve the release of multiple Pol II complexes during an active phase, known as the ``ON'' period, followed by a quiescent ``OFF'' period \cite{Raj2006,Chubb:2006kz,Zenklusen2008,Suter2011,Bothma2014,Tantale2016,Wan2021}. However, several critical questions remain unanswered: How does the regulation of bursting kinetics determine mRNA production and transcriptional dynamics across developmental time and cell types? Is the transcription rate primarily regulated by adjusting the durations of the ON or OFF periods, the initiation rate (i.e., the rate of Pol II release during active phases), or a combination of these parameters? What is the spectrum of parameters utilized by tightly regulated genes, such as developmental genes? Do different genes employ distinct bursting strategies? Do these strategies vary in temporal and spatial (tissue-specific) transcriptional control, and how do they depend on the regulatory factors at play?  

One hypothesis that has emerged from previous work suggests that different regulatory factors, including transcription factor (TF) binding, \emph{cis}-regulatory elements, nucleosome occupancy, histone modification, Pol II pausing, and enhancer--promoter interactions, may influence distinct aspects of bursting dynamics \cite{Senecal:2014dz,Bartman2016,Li:2018ha,Nicolas:2018if,Donovan2019,FaloSanjuan:2019bp,Hoppe2020,Tantale:2021,Bass:2021,Brouwer2023}. For instance, it has been proposed that enhancers primarily impact burst frequency, while promoters primarily affect burst size \cite{Fukaya2016, Pimmett2021,Larsson2019}. However, integrating diverse observations into a unified and quantitatively predictive understanding of transcriptional control through bursting dynamics has proven to be challenging.

Our previous study \cite{Zoller:2018gj}, which relied on inference from a static snapshot of mRNA abundance, raised the possibility of a simple and unified control of bursting kinetics, even in complex developmental systems. Specifically, mRNA abundance of key developmental genes in fixed \textit{Drosophila} embryos within a narrow developmental window ($\sim 10$ min) could be accounted for by a straightforward two-state kinetic model of transcription, with a single free parameter. This motivated our current efforts to directly measure bursting dynamics rather than rely on specific kinetic assumptions to infer dynamics from a fixed sample. Moreover, by conducting measurements throughout cell cycles and under conditions where key regulatory determinants are perturbed, we aim to assess the impact of developmental time and regulation on bursting behavior.

Here we present real-time transcription measurements with single-mRNA sensitivity, characterizing the bursting parameters of cell fate-determining genes in early \textit{Drosophila} embryos. Our findings unveil unexpected relationships between bursting parameters, revealing tight coupling between ON and OFF periods (or, alternatively, burst size and frequency). This coupling suggests that these distinct parameters are not solely governed by independent regulatory processes. Instead, any gene activity level is achieved through a specific combination of ON and OFF periods, irrespective of gene identity or spatiotemporal coordinates within the developing embryo. This stringent coupling thus holds across diverse regulatory landscapes. Indeed, we demonstrate through cis- and trans-perturbations that predominant modulation of burst size or frequency correlates with activity level rather than specific regulatory determinants. Low-activity alleles primarily modulate OFF periods, altering burst frequency, while mid-to-high-activity alleles primarily modulate ON periods, altering burst size. While previous measurements in various organisms, ranging from yeast to mammalian cells, sample a narrow range of activities, we find that they largely adhere to the elucidated dependencies. However, as these dependencies do not trivially arise from commonly implicated molecular events in transcription initiation, our work prompts a re-examination, guiding future investigations into the mechanisms of mRNA production.

\begin{figure*}
\centering
\includegraphics[scale=0.95]{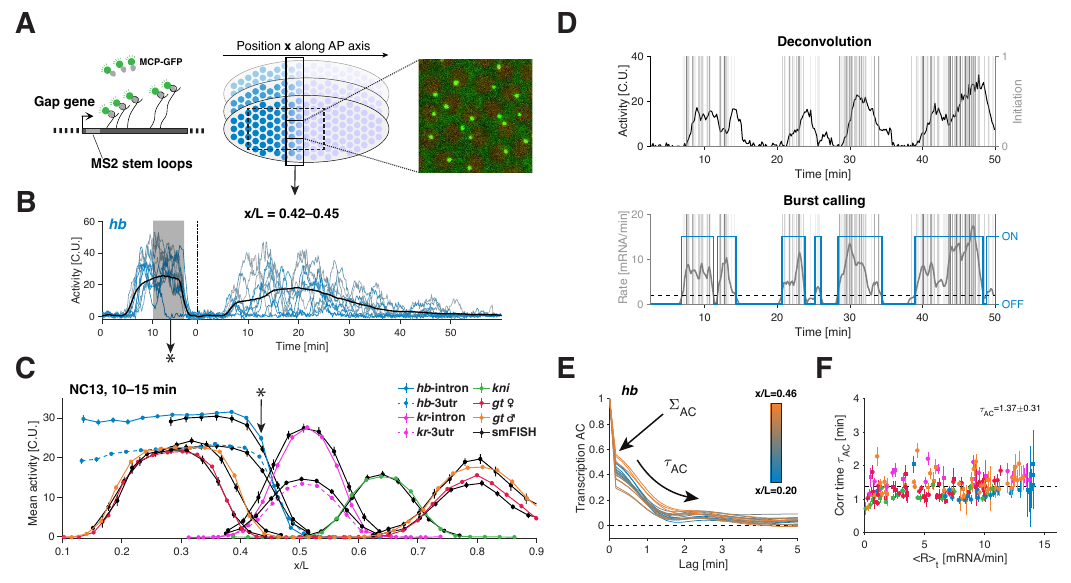}
\caption{{\bf Live single-cell transcription rate measurements of endogenous gap genes.}
(A) Live fluorescence imaging of nascent transcripts using MS2 stem-loops measures single allele transcriptional activity (green hotspots) along the anterior-posterior (AP) axis of the fly embryo (also see Fig. \ref{figS1}A and methods).
(B) Transcription time series for 10 alleles (blue) of the gap gene \emph{hunchback} (\emph{hb}) at position $x/L=0.435\pm0.010$ sampled every $10\,$s. Low embryo-to-embryo variability (Fig. \ref{figS1}F) enables pooling alleles from multiple spatially and temporally aligned embryos ($n=$10--20) to average over 200--350 alleles at a given position (black).
(C) Calibration of transcriptional activity in absolute units performed by matching mean spatial activity profiles from previously calibrated fixed smFISH measurements \cite{Zoller:2018gj} (black) with 5-min-interval averages (gray shade in B) of live time series at each AP position (color) for all examined gap genes. A single global conversion factor matches live and fixed profiles to within $5\%$ error (Fig. \ref{figS1}B), defining a unit for transcriptional activity (i.e., the cytoplasmic unit, C.U. \cite{Little:2013dr}) equivalent to a fully tagged transcript.
(D) Reconstruction of transcription initiation events (gray bars) and underlying bursts from single allele transcription time series (black). (Top) Bayesian deconvolution enables sampling possible configuration of initiation events (see Methods and Fig. \ref{figS2}A-B). (Bottom) Clustering of sampled initiation events (using a moving average of width $\sim\!1\,$min (gray curve) and a threshold at two mRNA/min (dashed line)) identifies individual bursts (blue).
(E) Auto-correlation (AC) functions of single allele \emph{hb} transcription rates averaged over time for different positions along the AP axis (color). AC functions are normalized by the variance; uncorrelated ($\Sigma_{\rm AC}$) and time-correlated ($\tau_{\rm AC}$) components of the rate fluctuations are highlighted (see Fig. \ref{figS5}). 
(F) Correlation time $\tau_{\rm AC}$ (from fitted exponentials, see Methods) of the single allele transcription rate as a function of mean transcription rate $R$ (color code as in C). Dashed line corresponds to overall mean correlation time $\bar{\tau}_{\rm AC}=1.37\pm0.31$ min. Error bars are $68\%$ confidence intervals.}
\label{fig1}
\end{figure*}

\section*{Results}
\noindent \textbf{Instantaneous single allele transcription rate measurements.} 
We developed a quantitative approach to measure endogenous bursting dynamics at a single allele level in living \emph{Drosophila} embryos. To achieve this we utilized a versatile CRISPR-based scheme \cite{Levo2022} to incorporate MS2 cassettes into intronic or 3' untranslated regions (3'UTRs) of the gap genes. These cassettes form stem-loops in the transcribed nascent RNA, which are subsequently bound by fluorescent coat-proteins (Fig. \ref{fig1}A, \ref{figS1}A, and Methods) \cite{Bertrand1998,Larson2011,Garcia2013,Lucas2013}. We employed a custom-built two-photon microscope to generate fluorescence images, allowing us to capture RNA synthesis from one tagged allele per nucleus with nearly single-mRNA sensitivity (Fig. \ref{figS1}D-E). Our optimized field-of-view provided 10-second interval time-lapses (Fig. \ref{fig1}B) for hundreds of nuclei per embryo during a critical 1.5-hour period of embryonic development, specifically nuclear cycles 13 (NC13) and 14 (NC14), essential for robust statistical analysis (Fig. \ref{fig1}A-B; Videos V1-V4). 

We calibrated our fluorescence signal using smFISH data to express our dynamic transcription measurements in terms of absolute mRNA counts (Fig. \ref{fig1}C and \ref{figS1}B-C, and details in Methods). This calibration, combined with the nearly single-transcript sensitivity of our measurements enabled us to reconstruct the underlying Pol II transcription initiation events for each allele using Bayesian deconvolution (see Methods). The convolution kernel we employed describes the fluorescent signal resulting from the release of Pol II complexes onto the gene, which subsequently engages in the elongation process \cite{Tantale:2021,Pimmett2021} (assuming constant and deterministic elongation, Fig.~\ref{figS2}A). For each time trace, our Bayesian approach generates multiple configurations of transcription initiation events (Fig.~\ref{fig1}D). By averaging these configurations, we obtained a time-dependent instantaneous single allele transcription rate, denoted as $r(t)$ (Fig.~\ref{figS2}B). Importantly this approach also provides corresponding error estimates, which we propagated in all subsequent analyses. 

Our kernel-based deconvolution approach was validated by control measurements involving dual-color tagging of the gene body, both at 5' and 3' regions (Fig.~\ref{figS2}C). These measurements support our key assumptions regarding the elongation process and the absence of co-transcriptional splicing (see Methods). Furthermore, they allow us to extract a Pol II elongation rate, denoted as $K_{\rm elo}$, which we determined to be $1.8\pm0.1\,$kb/min. This value aligns with previous measurements reported in the literature \cite{Garcia2013,Liu:2021} (Fig. \ref{figS2}D-G). 

With our approach, the extracted single allele transcription rates are no longer masked by the Pol II elongation dwell time, unlike the directly measured intensities. Instead, they capture initiation events (i.e., Pol II release for productive elongation). Consequently, these rates are independent of gene length, allowing for direct comparisons across different genes. This facilitated the intriguing observation that the gap genes average transcription rate, computed over nuclei per spatial position and time, denoted as $R=\avg{r}$, reach a similar maximum $R_{\rm max}=14.8\pm0.9\,$mRNA/min (Fig. \ref{figS3}A, Video V5). Moreover, these average transcription rates closely mirror the well-documented average protein dynamics \cite{Dubuis2013}. Simple assumptions related to diffusion and lifetime, without the need for explicit post-transcriptional regulation, are sufficient to quantitatively predict protein patterns from the mean transcription rates $R$ (Fig. \ref{figS4} and Video V6). Thus, in this system, the functional output, namely protein synthesis, predominantly relies on transcription regulation. Our quantitative imaging and deconvolution approach paves the way for uncovering how this regulation emerges from the single-allele transcription dynamics.

\begin{figure*}
\centering
\includegraphics[scale=1.0]{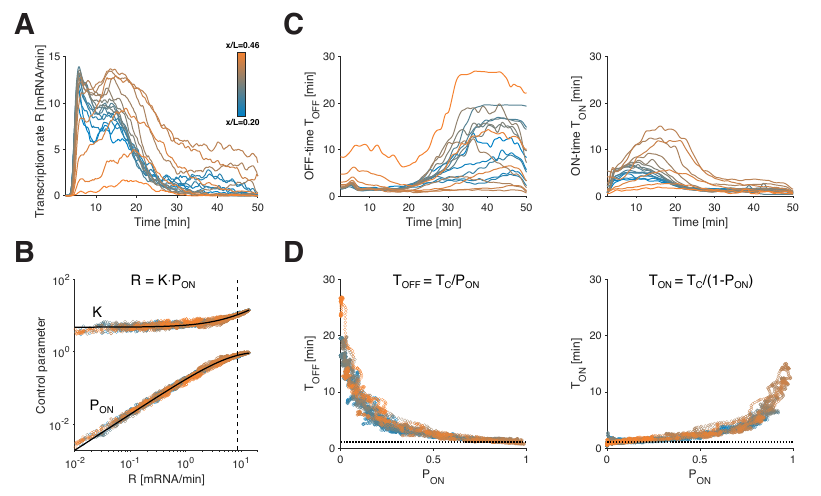}
\caption{{\bf Direct estimation of instantaneous mean transcription parameters.}
(A) \emph{Hb} mean transcription rate $R$ as a function of time in NC14 (color encodes AP position). 
(B) Control parameters $P_{\rm ON}$ and $K$ as a function of mean transcription rate $R$ in log space. Since $\log{(R)}=\log{(K)}+\log{(P_{\rm ON})}$ by construction, changes in $P_{\rm ON}$ determine changes in $R$ below the dashed line ($R\sim\!8.5\,$mRNA/min, corresponding to $P_{\rm ON}=0.75$).
(C) \emph{hb} mean OFF-time $T_{\rm OFF}$ and mean ON-time $T_{\rm ON}$ as a function of time in NC14 for all AP positions (color code as in (A)).
(D) Mean OFF-time $T_{\rm OFF}$ and mean ON-time $T_{\rm ON}$ as a function of $P_{\rm ON}$, for all positions and time points beyond the $7.5\,$min mark in (C). Dotted line corresponds to mean $T_{\rm C}=1.1\pm0.2$ (see Fig. \ref{figS8}E), which sets a lower bound on possible $T_{\rm OFF}$ and $T_{\rm ON}$ values.}
\label{fig2}
\end{figure*}

\vspace{.3 cm}
\noindent \textbf{Single allele transcription rates hint at a universal bursting regime.} 
The gap genes differ in their transcriptional activities both spatially and temporally. However, when we examine the distributions $P(r|R)$ of single-allele transcription rates $r$ that yield a similar mean transcription rate $R$, an intriguing pattern emerges. These distributions converge consistently across different genes (Fig. \ref{figS3}C, see Methods), suggesting the existence of a common transcriptional regime. For transcription rates in the low- to mid-range of $R$, we observe an abundance of non-transcribing or barely transcribing alleles. These distributions starkly contrast with a regime characterized by constitutive transcription. Conversely, for transcription rates in the high range of $R$, the distributions converge towards the constitutive, or Poissonian, regime (Fig. \ref{figS3}D), indicating a higher proportion of active ON alleles. These observations are consistent with the concept of transcriptional bursting, where an allele dynamically transitions between productive ON and quiescent OFF periods \cite{Peccoud1995, Raj2006}.

We obtain additional support for a common bursting regime when we analyze the temporal dynamics of single-allele time traces. Bursting is expected to introduce temporal correlations in transcriptional activity, reflecting the persistence of the ON and OFF periods (Fig. \ref{figS5}A-B). To characterize such correlations, we compute auto-correlation functions for the deconvolved single-allele transcription rates. By using the deconvolved rates we effectively remove the correlated component arising from Pol II elongation along the gene and isolate only the correlations stemming from the initiation and the ON--OFF switching process. When we calculate these auto-correlation functions for different anterior-posterior (AP) bins, nuclear cycles, and various genes (Fig. \ref{fig1}E), we find striking similarities. An initial sharp drop at our sampling time scale ($\sim\!10\,\si{s}$) indicates the presence of uncorrelated noise, consistent with independent Pol II initiation events (Fig. \ref{figS5}C). This drop is followed by a longer decay of correlated noise at a time scale denoted as $\tau_{\rm AC}$, which we find to be confined within 1- to 2-minute range (Fig. \ref{fig1}F). The remarkable consistency of $\tau_{\rm AC}$ across different spatial locations, genes, and transcriptional activity levels (spanning $R$, Fig. \ref{fig1}F) implies the preservation of this fundamental time scale of transcription dynamics. To delve deeper into potential regularities in bursting dynamics, our next step involves directly extracting individual bursts from single-allele time traces.


\vspace{.3 cm}
\noindent \textbf{Allele ON-probability is the primary transcription control parameter.}
From the deconvolved initiation events along individual time traces (Fig. \ref{fig1}D top) we identify distinct ON and OFF periods of active and inactive transcription, respectively. The ON periods are characterized by consecutive initiation events (i.e., multiple Pol IIs released for productive elongation), while the OFF periods are transcriptionally inactive (Fig. \ref{fig1}D bottom). To delineate the transition of an allele from an OFF to an ON state, we employ a simple threshold on the moving average of the single allele transcription rate, set to $2\,$mRNA/min over a one-minute window. This criterion is selected based on our detection sensitivity, allowing us to reliably detect $1-2\,$mRNA molecules, and the window is consistent with the time scale derived from the auto-correlation analysis (see Methods). 

We conducted extensive testing to evaluate the impact of these detection parameters on our analysis, and our results confirm that our intuitive choice minimizes errors in burst characterization (see below, Fig. \ref{figS9}E). Importantly, the primary strength of this burst-calling routine lies in its exclusive reliance on a minimal clustering model. Consequently, it is inherently devoid of assumptions about the distributions of ON and OFF periods. As a result, our burst detection process remains agnostic to the underlying transcription models, as long as transcription can be described by at least one ON and OFF state (which is the case for common N-state models \cite{Zoller:2015im,Corrigan:2016kq,Li:2018ha,Lammers:2020,Wan2021,Tantale:2021,Pimmett2021}.

The next goal of our analysis is to elucidate how the consecutive switches between ON and OFF periods quantitatively govern the transcription rate $R$. Specifically, the mean transcription rate at time $t$, denoted as $R(t)$, can be decomposed into two distinct parameters: the instantaneous probability of an allele being in the ON state ($P_{\rm ON}(t)$), representing the fraction of ON alleles at time $t$, and the mean initiation rate ($K(t)$) for ON alleles. Given the above decomposition, most of the variation in $R$ could arise from changes in either $K$, $P_{\rm ON}$, or both.

Starting with the gene \emph{hb}, we estimate the time-dependent parameters $R(t)$ and $P_{\rm ON}(t)$ for each AP bin. To obtain $R(t)$, we calculate the average of $\sim\! 250$ single-allele instantaneous transcription rates (Fig. \ref{figS6}A). Concurrently, we determine $P_{\rm ON}(t)$ by quantifying the fraction of alleles in the ON state at each time point (Fig. \ref{figS6}B). To compute $K(t)$, we average initiation events restricted to the ON state. By repeating this procedure for all AP positions, we reveal the spatiotemporal variations in $P_{\rm ON}$ and $K$ (Fig. \ref{fig2}A, Fig. \ref{figS6}C-D). We validate our approach for burst calling and the recovery of bursting parameters from transcription time traces across a wide range of simulated data. We achieve an overall median error of $10\%$, gaining insights into the robustness of our analysis across the potential parameter space (see Methods, Fig. \ref{figS9} and Fig. \ref{figS10}).

We find that all three parameters vary significantly across space and time (Fig. \ref{fig2}A and \ref{figS6}C-D). While we observe the expected interdependence $R=K \cdot P_{\rm ON}$ (Fig. \ref{figS6}F), our analysis indicates that changes in $R$ are primarily governed by changes in $P_{\rm ON}$, while the influence of $K$ is more moderate and less predictive of $R$ (Fig. \ref{fig2}B and Fig. \ref{figS7}A). $K$ only covers a two-fold change in dynamic range, which is marginal compared to $R$ spanning from 0 to 15 (mRNA/min) (Fig. \ref{figS6}H-I). This two-fold change is largely due to the existence of two optically unresolved sister chromatids and a modest time dependence of $K$ throughout the nuclear cycle (Fig. \ref{figS7}B-G, see Methods). With these considerations, we estimate for a single active chromatid the mean initiation rate at $6.3\pm1.4\,$mRNA/min and the mean Pol II spacing at $287\pm68\,$bp, consistent with the classic Miller spreads with average Pol II spacing of 330±180 bp \cite{McKnight1979}. These results for \emph{hb} suggest that transcriptional activity is predominantly controlled by the probability of an allele being in the ON state, and once in the ON state, transcription initiates at a quasi-constant rate.

\begin{figure*}
\centering
\includegraphics[scale=0.93]{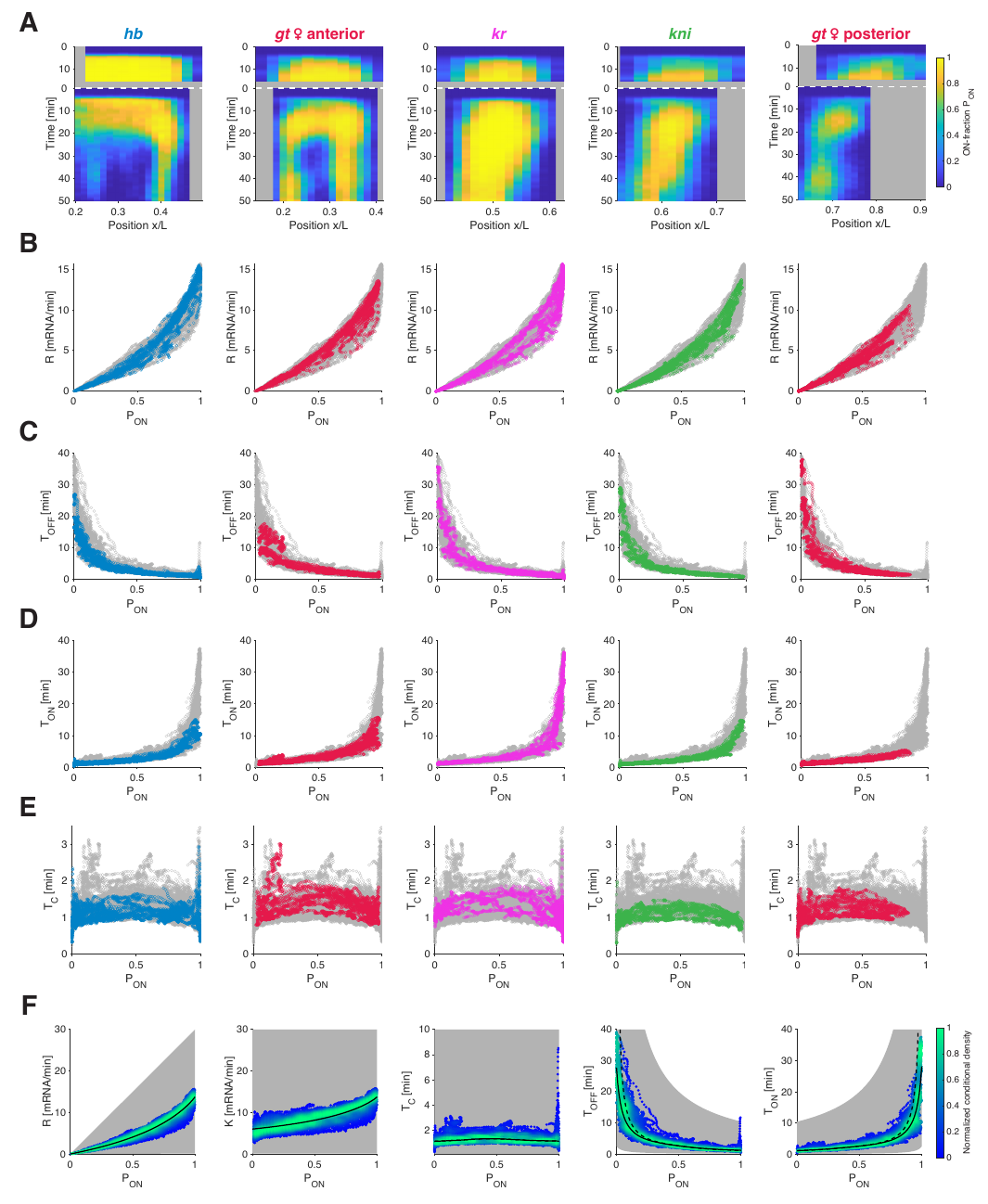}
\caption{{\bf Transcription parameters collapse for all gap genes.}
(A) Kymographs of ON-probability $P_{\rm ON}$ for all gap genes as a function of position and time for NC13 and NC14. The spatiotemporal transcriptional pattern of the gap genes arises from a complex regulation of $P_{\rm ON}$ (color map).
(B-E) Transcription parameters collapse for all gap genes across time and position. Transcription rate $R$ (B), Mean OFF-time $T_{\rm OFF}$ (C), ON-time $T_{\rm ON}$ (D), and switching correlation time $T_{\rm C}$ (E) as a function of $P_{\rm ON}$. Colored data points represent individual gap genes (same color code as in (A), see Fig. \ref{figS11}A-B for \emph{gt} male data). Each panel shows all the remaining genes in gray.
(F) Density (color) of all data points across space and time of the transcription parameters for all gap genes, normalized by the maximum density. Potentially accessible space (gray shade) for plausible ranges of $K$ (0.1--30 mRNA/min) and $T_{\rm C}$ (0.1--10 min). $P_{\rm ON}$ almost fully determines $R$ and sets the combinations of $T_{\rm OFF}$ and $T_{\rm ON}$. For $T_{\rm OFF}$ and $T_{\rm ON}$, the dashed lines are the 2-state model predictions based on $T_{\rm C}$, and the solid lines take the finite recording length into account (see Fig. \ref{figS11}D).}
\label{fig3}
\end{figure*}

Any given ON-probability can result from various combinations of mean ON and OFF periods, denoted as $T_{\rm ON}$ and $T_{\rm OFF}$. Indeed, since $P_{\rm ON}$ = $T_{\rm ON}/(T_{\rm ON}+T_{\rm OFF})$ near steady state, re-scaling both $T_{\rm ON}$ and $T_{\rm OFF}$ would achieve the same $P_{\rm ON}$. This raises the possibility that alleles at different spatial positions and times employ distinct combinations, or there could be underlying regularities governing these periods. We computed the mean ON and OFF times $T_{\rm ON}$ and $T_{\rm OFF}$ (Fig. \ref{figS8}A-C; \ref{figS8}F for full distributions; see Methods), and found that these also vary substantially across space and time (Fig. \ref{fig2}C). However, when we plot $T_{\rm ON}$ and $T_{\rm OFF}$ against $P_{\rm ON}$, we find that all data points collapse onto two tight anti-symmetric relationships (Fig. \ref{fig2}D). Despite the potential for multiple combinations of $T_{\rm ON}$ and $T_{\rm OFF}$ for any given $P_{\rm ON}$, these relationships consistently associate a given $P_{\rm ON}$ value with a unique pair of $T_{\rm ON}$ and $T_{\rm OFF}$ values, irrespective of spatial position or time.

The dynamic switching between ON and OFF states is associated with a correlation time $T_{\rm C}$, which determines the time separation required for the transcription rate of a single allele to become uncorrelated. $T_{\rm C}$ can be computed directly from the mean ON and OFF times using the equation $1/T_{\rm C} = 1/T_{\rm ON} + 1/T_{\rm OFF}$ (Fig. \ref{figS8}E). For the \textit{hb} gene, we find that  $T_{\rm C}$ is confined around $1.1\pm0.2$ min across all positions and time points, and seems independent of $P_{\rm ON}$. Since $T_{\rm ON}$ and $T_{\rm OFF}$ can be expressed as functions of $P_{\rm ON}$ and $T_{\rm C}$ (via $T_{\rm ON} = T_{\rm C}/(1-P_{\rm ON})$ and $T_{\rm OFF} = T_{\rm C}/P_{\rm ON}$, Fig. \ref{fig2}D and Fig. \ref{figS8}B), the value of $T_{\rm C}$ sets the lower limit of $T_{\rm ON}$ and $T_{\rm OFF}$ (when $P_{\rm ON}$ approaches one, or zero, accordingly). We thus find that mean ON and OFF periods share a similar minimal value on the order of 1-1.5min, which is larger than the minimal ON period we can reliably detect ($\sim30$ s, see Methods). Furthermore, the constancy of $T_{\rm C}$ effectively links the mean ON and OFF times, and provides a mathematical explanation for the tight anti-symmetric relationships between $T_{\rm ON}$, $T_{\rm OFF}$, and $P_{\rm ON}$ (Fig. \ref{fig2}D). Thus, not only does $P_{\rm ON}$ govern the mean transcription rate $R$, but also the entire transcriptional bursting dynamics, with a characteristic combination of ON and OFF periods associated with any $P_{\rm ON}$ value.


\vspace{.3 cm}
\noindent \textbf{Common bursting relationships underlie the regulation of all gap genes.} 
The gap genes differ in the composition, number, and arrangements of their \emph{cis}-regulatory elements (Fig. \ref{figS1}A), resulting in distinct regulatory binding events (e.g., by transcription factors, transcription machinery components, and chromatin modifiers) \cite{Lagha:2012fz}. Consequently, each gene displays unique spatiotemporal transcriptional activities (Fig. \ref{fig1}C, Video V5). Despite these differences, we find that the relationships governing bursting parameters for \emph{hb} appear to generalize to other gap genes.

When we applied our burst calling procedure (Fig. \ref{fig1}D) to the transcription time traces of other gap genes (\emph{gt}, \emph{Kr}, and \emph{kni}), we obtained distinct spatiotemporal $P_{\rm ON}$ profiles (Fig. \ref{fig3}A) that closely mirrored the gene-specific transcription rates $R$. Indeed, all genes exhibit nearly identical relationships between $R$ and $P_{\rm ON}$ (Fig. \ref{fig3}B) and between $K$ and $P_{\rm ON}$ (Fig. \ref{figS11}C), affirming that $P_{\rm ON}$ is the predominant factor governing transcriptional activity across time, space, and genes. 

Furthermore, the genes display similar relationships between $T_{\rm OFF}$ and $P_{\rm ON}$ (Fig. \ref{fig3}C) and between $T_{\rm ON}$ and $P_{\rm ON}$ (Fig. \ref{fig3}D). Thus, when different genes exhibit a specific $P_{\rm ON}$ value, potentially at different spatiotemporal coordinates, the underlying $T_{\rm ON}$ and $T_{\rm OFF}$ periods are nonetheless largely identical. This finding can be related to the conservation of the switching correlation time $T_{\rm C}$ across all positions, times, and genes (Fig. \ref{fig3}E), with the average $T_{\rm C}$ value ($1.25\pm0.37$ min) aligning quantitatively with the time scale predicted by the auto-correlation analysis above (Fig. \ref{fig1}F). Notably, the common bursting relationships apply not only across genes but also across distinct spatiotemporal domains of activity of a single gene, known to be driven by distinct enhancers \cite{Schroeder:2004cs,Perry:2012dr}. This is particularly evident in the large and yet distinct anterior and posterior domain of the gene \emph{gt} (Fig \ref{fig3}A and Fig. \ref{figS11}). The use of similar bursting parameters in distinct spatial patterns of a single gene was indeed proposed in a study of a reporter construct of the gene \emph{eve} \cite{Berrocal2020,Berrocal_2023}. 

Pooling the parameters derived from all genes, times, locations, and embryos (comprising over $10^6$ data points) accentuates the limited subset of the parameter space utilized and underscores the stringent quantitative relationships emerging from our dataset (Fig. \ref{fig3}F). Overall, our analysis shows that, across all data points, the mean transcription rate $R$ is primarily governed by $P_{\rm ON}$, with only limited modulation of $K$. When we segregate the data into three developmental time windows, these relationships tighten even further, hinting at a modest developmental time dependence (Fig. \ref{figS12}A). Notably, the relationships obtained within the first developmental window generalize with greater precision (2-fold improvement) our sparse estimates using fixed measurements in a previous study, restricted to this window \cite{Zoller:2018gj} (Fig. \ref{figS13}). Furthermore, once we convert our single allele parameters into their single gene copy equivalent, accounting for the combined contribution of two sister chromatids, the resulting relationships display even greater simplicity (Fig. \ref{figS12}B), with a near constant $K$ and an almost linear $R$ to $P_{\rm ON}$ relationship. 

The near-constant switching correlation time $T_{\rm C}$ observed across the pooled data set, is associated with an apparent inverse proportionality between $T_{\rm ON}$ and $T_{\rm OFF}$, with a predominant modulation of one of these two parameters when $P_{\rm ON}$ changes. While lowly-transcribing alleles (as characterized by $P_{\rm ON}$) tend to achieve higher expression levels mainly by reducing $T_{\rm OFF}$, medium-to-high-transcribing alleles are predominantly tuned by extending $T_{\rm ON}$ (Fig. \ref{fig3}F). This observation means that changes in burst frequency ($F=1/(T_{\rm ON}+T_{\rm OFF})$) govern tuning of the transcriptional activity of low-transcribing alleles, while changes in burst size ($B=K \cdot T_{\rm ON}$) exert greater influence on the tuning of medium-to-high-transcribing alleles (Fig. \ref{figS11}E,F). 

The association between gene activity level and bursting parameters is discernible due to the wide range of transcriptional activities exhibited by the measured genes. This association holds consistently across genes, spatial locations, and time points (see Fig. \ref{fig3}F, Fig. \ref{figS11}, Fig. \ref{figS12}), indicating its relevance across diverse regulatory landscapes, which may involve varied TF concentrations, chromatin accessibility, and distinct activating enhancers. However, predicting the direct impact of individual regulatory determinants on bursting kinetics is challenging, given the multitude of factors collectively shaping transcriptional outcomes. Thus, targeted perturbation experiments are necessary to elucidate the specific effects of each determinant on bursting behavior.


\vspace{.3 cm}
\noindent \textbf{Common bursting relationships predict effects of \emph{cis}- and \emph{trans}-perturbations.}
Diverse regulatory determinants, including \emph{cis}-regulatory elements like enhancers and \emph{trans}-factors such as TF repressors, contribute to controlling transcriptional activity. It is commonly assumed that distinct regulatory mechanisms directly influence specific bursting parameters. Thus, we sought to perturb various regulatory determinants to assess whether they produce distinct effects on bursting dynamics or if the relationships identified from wild-type measurements can account for the modified transcriptional activity upon perturbations.

\begin{figure*}[t!]
\centering
\includegraphics[scale=1.0]{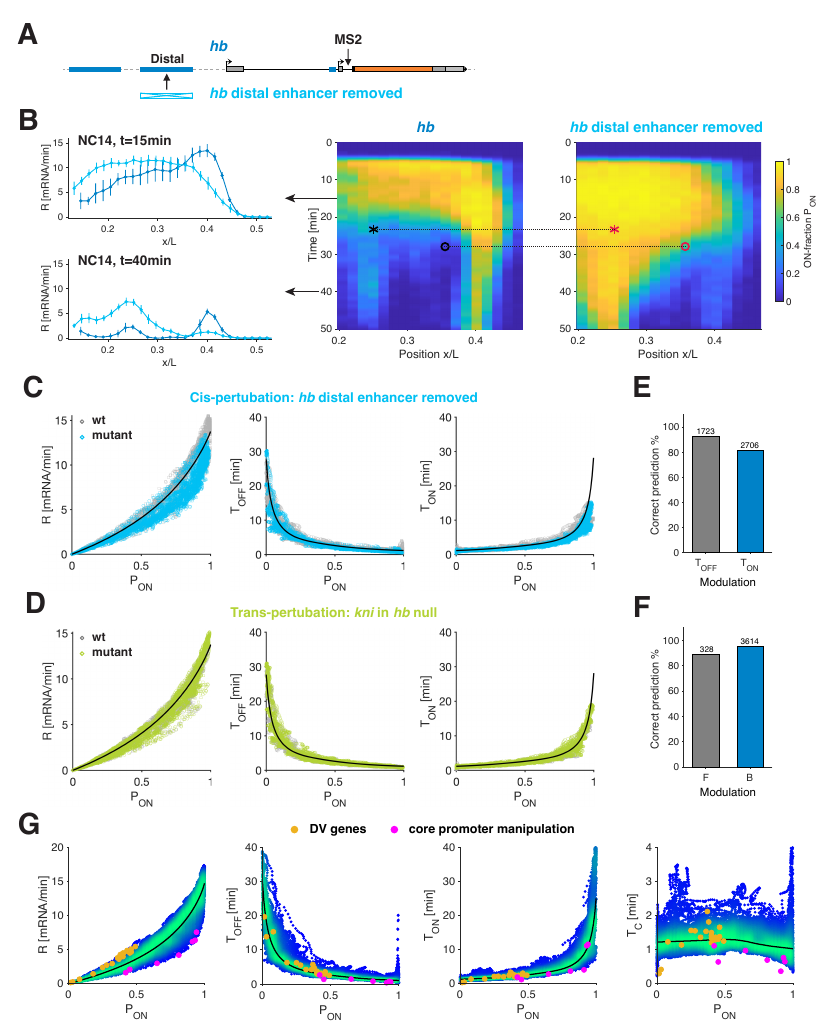}
\caption{{\bf Effect of \emph{cis}- and \emph{trans}-perturbations on ON and OFF times.}
(A) Distal \emph{hb} enhancer removal from the fly line carrying MS2 stem-loops in the endogenous \emph{hb} locus.
(B) Quantification of \emph{hb} wild-type and mutant phenotypes. Both transcription rate $R$ (left) and $P_{\rm ON}$ (right) levels display significantly different expression patterns for the enhancer deletion mutant. Black arrows indicate the time points (15 and 40 min into NC14) in the kymograph at which rate profiles are depicted. ``o'' and ``$\star$'' mark two bins with predominant $T_{\rm OFF}$ modulation and predominant $T_{\rm ON}$ modulation, respectively.
(C) Transcription parameters for \emph{hb} enhancer deletion (cyan) collapse on corresponding wild-type parameters (gray). 
(D) Transcription parameters for \emph{kni} (green) in a \emph{hb} null background (i.e. a \emph{trans}-mutation, Fig. \ref{figS14}D-E) collapse on corresponding wild-type parameters (gray). Solid black lines in (C) and (D) correspond to the endogenous bursting relationships from Fig. \ref{fig3}F. 
(E-F) Verification of predicted changes in $T_{\rm OFF}$ and $T_{\rm ON}$ (E) (or changes in burst size $B$ and frequency $F$ (F)) for all wild-type and mutant $P_{\rm ON}$ pairs (two example pairs shown in (A) for \emph{hb}).
For most pairs ($>85\%$) the prediction is correct (see also Fig. \ref{figS15}A-C.
(G) Transcription parameters computed from two other \emph{Drosophila} studies (yellow circles \cite{Hoppe2020} and pink circles \cite{Pimmett2021}, respectively) are consistent with the gap genes relationships (black lines, data density color coded as in Fig. \ref{fig3}F).}
\label{fig4}
\end{figure*}

Upon endogenous deletion of the distal enhancer of \emph{hb}, we observe significant alterations in transcriptional activity (Fig. \ref{fig4}A-B, Video V7), including increased or decreased activity, at different times and locations along the AP axis, consistent with previous findings \cite{Fukaya2021}. However, we find that bursting dynamics in this mutant still adhere to the relationships identified in the wild-type context. Specifically, transcription rates across different spatial and temporal coordinates are again governed by $P_{\rm ON}$, the stringent relationships between $T_{\rm ON}$/$T_{\rm OFF}$ and $P_{\rm ON}$ hold, and the switching correlation time $T_{\rm C}$ remains broadly conserved around $0.9\pm0.2$ min (Fig. \ref{fig4}C and \ref{figS15}A-D).

Two additional perturbations further confirmed these findings. Deleting the distal enhancer of \emph{kni} results in a significant reduction in \emph{kni} activity (Fig. \ref{figS14}A-B, Video V8). Although the mutant exhibits a narrower dynamic range of activity, we observe a similar data collapse within this reduced range (Fig. \ref{figS14}C and \ref{figS15}A-D). Next, we explored the effect of a \emph{trans}-perturbation by measuring \emph{kni} activity in embryos with a \emph{hb} null background (Fig. \ref{figS14}D-E, Video V9). This \emph{trans}-perturbation significantly alters \emph{kni} activity, consistent with earlier studies \cite{Hulskamp:1994}). However, the underlying bursting dynamics again collapse onto the same busting relationships (Fig. \ref{fig4}D and \ref{figS15}A-D).
  
The consistency of $P_{\rm ON}$ to $T_{\rm ON}$ and $T_{\rm OFF}$ relationships in wild-type and mutants suggests that we can predict how ON and OFF periods change upon a perturbation. Such a prediction relies solely on how the activity level, captured by $P_{\rm ON}$, changed, and thus should remain valid at any given spatiotemporal coordinate. Remarkably, for each type of perturbation, we observe both predominant $T_{\rm ON}$ and predominant $T_{\rm OFF}$ modulation, at different spatiotemporal coordinates (Fig. \ref{fig4}E). Comparing predictions based only on the wild-type-derived relationships with the directly measured $T_{\rm ON}$ and $T_{\rm OFF}$ from the mutant, we find agreement as to which parameter was primarily altered in more than $\sim\!86$\% of cases for all spatiotemporal coordinates (Fig. \ref{fig4}E and \ref{figS15}E,G and I). Additionally, similar successful predictions are achieved when assessing the change in transcriptional activity in terms of altered burst size versus burst frequency (Fig. \ref{fig4}F and \ref{figS15}F,H and J). These findings challenge previous intuitions linking perturbations of specific regulatory elements or mechanisms to changes in a particular bursting parameter. Instead, these findings suggest the predictive power of $P_{\rm ON}$, a proxy of the transcription activity, across different perturbations. 
 
To further explore the generality of these observations, we examined data from two previous studies in the early fly embryo. One focused on the transcriptional effect of BMP signaling, a dorsoventral (DV) morphogen \cite{Hoppe2020}. Transcription of a BMP target gene, u-shaped (ush), was measured across different DV positions and under ectopic signaling.  A second study employed synthetic reporter constructs to examine the transcriptional effect of two core promoter motifs (TATA box and Initiator) \cite{Pimmett2021}. These studies pointed to the modulation of distinct bursting parameters, and while the analyzed genes and perturbed regulatory determinants differ from those we measured, we found the datasets collapse onto our identified bursting relationships (Fig. \ref{fig4}G). 

As suggested in these studies, the first dataset shows predominantly $T_{\rm OFF}$ modulation, while the latter study has primarily $T_{\rm ON}$ changes. Intriguingly, when plotted in the context of the full spectrum of $P_{\rm ON}$ values captured by our measurements, the two independent datasets cluster in disjoint halves (Fig. \ref{fig4}G). Our analysis raises the possibility that the predominantly changed parameter ($T_{\rm OFF}$ versus $T_{\rm ON}$) might not be inherent to the examined regulatory manipulation (e.g., input TF concentrations or core promoter elements), but rather a consequence of the expression range (the $P_{\rm ON}$ regime) of these genes.

\vspace{.3 cm}
\noindent \textbf{Data across different organisms is consistent with the identified bursting relationships}

The conserved nature of the transcription machinery and regulatory mechanisms across eukaryotes, suggests that fundamental properties, likely reflecting molecular constraints, apply to numerous systems \cite{Sanchez:2013gg}. However, technical and biological factors currently hinder the ability to perform direct measurements of bursting dynamics and quantitatively compare parameter values or dependencies across diverse settings. For example, our analysis above highlights the necessity of a large dynamic range of gene activity to reveal underlying relationships. Yet, a large dynamic range is not readily observed in many setups. Using absolute units (e.g., calibrating arbitrary fluorescent units to mRNA counts) is crucial for estimating measurement sensitivity and facilitating comparisons across genes, perturbations, and systems. Additionally, an analysis that decouples the contribution of different biological steps to the measured signal (e.g., transcription initiation, elongation and mRNA half lives) is essential. These considerations guided our reexamination of data in yeast and mammalian cells.

\begin{figure*}
\centering
\includegraphics[scale=1.0]{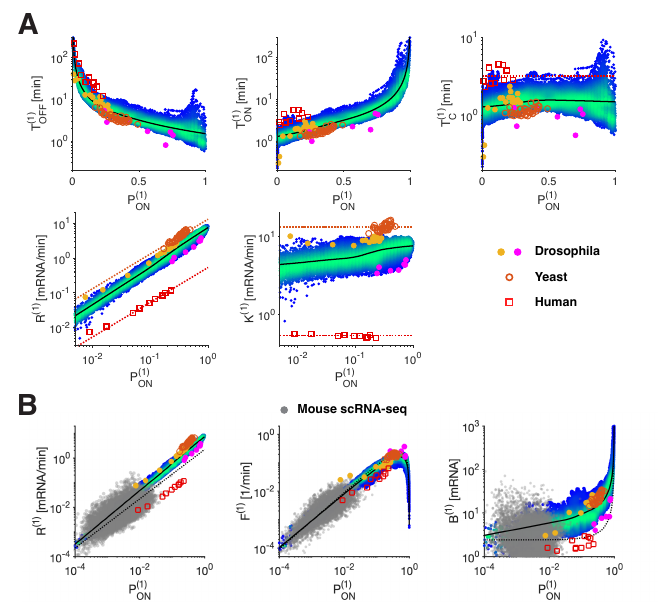}
\caption{{\bf Generality of bursting relationships.}
(A) Scatter plots of the transcription parameters for a single gene copy (see Method) versus $P_{\rm ON}$ (color code as in Fig. \ref{fig3}F). As in Fig. \ref{fig4}G, transcription parameters computed from two other \emph{Drosophila} studies (yellow circles \cite{Hoppe2020} and pink circles \cite{Pimmett2021}.
Transcription parameters resulting from multiple perturbations performed on the yeast \emph{GAL10} gene (orange circles \cite{Brouwer2023}) also closely follow our relationships (black lines), suggesting that these may apply beyond \emph{Drosophila}. Transcription parameters estimated using our deconvolution approach on 11 human genes (red squares \cite{Rodriguez2019,Wan2021}) further highlight the constant nature of $T_{\rm C}$ and $K$ (see Methods).
(B) Transcription rate $R$, burst frequency $F$ and burst size $B$ for a single gene copy versus $P_{\rm ON}$. Transcription parameters estimated from single-cell RNA-seq in mouse cells (gray circles \cite{Larsson2019}) are mostly consistent with our relationships (black lines), though parameters are likely underestimated due to low mRNA recovery rate ($\sim10$--$30\%$) as demonstrated in original study. Black dotted lines correspond to expected relationships in scRNA-seq data assuming constant $K$ and $T_{\rm C}$.}
\label{fig5}
\end{figure*}

To facilitate comparison, we computed the equivalent single gene copy parameters for all the fly data (Fig. \ref{figS12}B and Fig. \ref{figS16}, see Methods), assuming the independence of the two unresolved sister chromatids \cite{Little:2013dr}. We examined bursting parameters derived from extensive perturbations of a yeast gene \cite{Brouwer2023}, and found strong agreement with our observations from the early fly embryo (Fig. \ref{fig5}A and methods). While the yeast data points span a relatively small range of bursting parameters, compared to the developmentally regulated \textit{Drosophila} genes, they are consistent with the $T_{\rm ON}$-to-$P_{\rm ON}$ and $T_{\rm OFF}$-to-$P_{\rm ON}$ relationships and show a highly constrained $T_{\rm C}$ value of $1.1\pm0.1\,$min, within our observed range. 

Additionally, we re-analyzed transcription time traces from 11 endogenous human genes imaged with the MS2 system in cell lines \cite{Rodriguez2019,Wan2021}. We aimed to extract promoter ON-time, capturing initiation, rather than previously reported ON-times encompassing initiation and mRNA dwell-time. We calibrated the signal in absolute units using smFISH measurements of nascent mRNA from one of these genes (\emph{TFF1}) \cite{Rodriguez2019}, and we performed a fluctuation analysis to assess measurement noise and mean mRNA dwell time (see Methods). We deconvolved initiation events and estimated bursting parameters for these genes, acknowledging longer mRNA dwell-time (6–16 min vs 1–2 min in flies), lower temporal resolution (100 s), and measurement noise as contributors to further uncertainties in these estimates (see Methods). The resulting parameters closely match those obtained from the fluctuation analysis performed in the original study on the gene \emph{TFF1} \cite{Rodriguez2019}. 

Across genes, we find that both the initiation rate $K$ and the switching correlation time $T_{\rm C}$ are mainly constant, around $K=0.54\pm0.02\,$mRNA/min and $3.2\pm0.8\,$min, respectively (Fig. \ref{fig5}A). While absolute values for $K$, and to a lesser extent $T_{\rm C}$, appear to differ from fly genes (12- and 2-fold difference, respectively), the predictive power of $P_{\rm ON}$ is preserved, and the human genes follow very similar relationships.

Thus, while the initiation rate $K$ appears mostly conserved across genes and conditions in a given organism, we observe a clear difference between the three species that we probed (yeast $K=13.2\pm1.3$, fly $K=6.2\pm1.8$ and human $K=0.54\pm0.02$; Fig. \ref{fig5}A). Such differences may highlight species-specific metabolic variations \cite{Rayon:2020,Diaz:2023}. However, the $T_{\rm C}$ values seem very close across species. Importantly, we find that the extracted relationships between bursting parameters hold across these datasets. They are therefore general, likely reflecting conserved underlying mechanisms. 

Two high-throughput studies conducted in mammalian systems utilized vastly different approaches (library of reporters and single-cell RNAseq), but also indicate trends that align with our established bursting relationships \cite{Dar2012,Larsson2019}. Yet, long mRNA dwell times (on the order of hours) complicate the extraction of initiation dynamics in these studies, and the quantitative comparison of bursting parameters. Nevertheless, we were able to re-analyze the scRNA-seq data in mouse cells \cite{Larsson2019}, by fitting the steady-state mRNA distributions (as in the original study), with an additional weak prior on $T_{\rm C}$ (see Methods). The resulting fits are as good as the original ones (marginal loss in likelihood), and provide parameter estimates with plausible physical scales. Overall, scRNA-seq parameters are consistent with our bursting relationships (Fig. \ref{fig5}B), highlighting the predictive power of $P_{\rm ON}$. Furthermore, our analysis highlights the potential of using parameters derived from live imaging to interpret scRNA-seq data in terms of physical kinetic rates, which can be linked to the underlying molecular events.

\section*{Discussion}

\begin{figure*}
\centering
\includegraphics[scale=1.0]{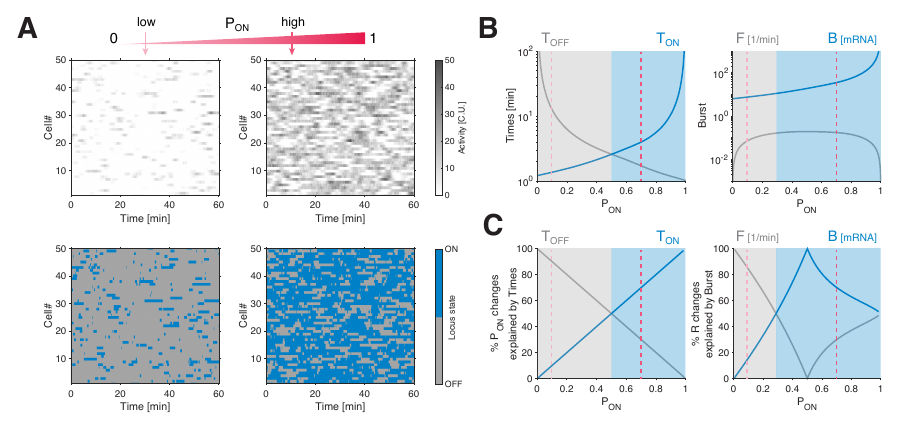}
\caption{{\bf Gene activity as a predictor of bursting dynamics}
(A) Heatmaps of simulated single allele activity (top) and of corresponding ON-OFF gene state (bottom) for a low and high $P_{\rm ON}$ regime, as predicted from our bursting relationships (see B).
(B) The bursting relationships extracted from our data characterize single-allele bursting dynamics, as shown in A. 
(C) Fractional changes in activity explained by $T_{\rm OFF}$ and $T_{\rm ON}$ (left), and explained by burst frequency $F$ and size $B$ (right). For low $P_{\rm ON}<0.5$, changes in activity are predominantly dictated by changes in $T_{\rm OFF}$, while for high $P_{\rm ON}>0.5$ by changes in $T_{\rm ON}$. Similarly, for low $P_{\rm ON}<1/3$, changes in activity are mainly dictated by changes in $F$, while for high $P_{\rm ON}>1/3$ by changes in $B$.}
\label{fig6}
\end{figure*}

In this study, we developed a method to quantify real-time single allele transcriptional bursting in the developing early \emph{Drosophila} embryo. The system's wide range of transcriptional activities allowed us to uncover fundamental relationships governing bursting dynamics. We found a highly restrictive regime with strong dependencies between bursting parameters, characterized by tight asymmetric relationships between $T_{\rm OFF}$, $T_{\rm ON}$, and $P_{\rm ON}$, underpinned by a relatively low and largely constant time scale ($T_{\rm C}$). Importantly, these relationships are consistent with data from yeast and mammalian cells\cite{Brouwer2023, Rodriguez2019, Larsson2019}, indicating that gene activity, predominantly governed by $P_{\rm ON}$, predicts the characteristic ON and OFF times utilized across systems.

Low activity alleles exhibit longer OFF times and shorter ON times, while high activity alleles ($P_{\rm ON}>0.5$) display longer ON times and shorter OFF times (Fig. \ref{fig6}A-B). Changes in activity level are associated predominantly with shortening OFF periods at low activity and primarily with prolonging ON periods at mid-to-high activity (Fig. \ref{fig6}C, left). Correspondingly, transcriptional tuning of either very low or very high activity alleles is associated with larger changes in burst frequency ($F=1/(T_{\rm ON}+T_{\rm OFF})$). However, while changes in burst frequency predominate over changes in burst size at low activity ($B=K\cdot T_{\rm ON}$), at mid-to-high activity, changes in burst size take over (Fig. \ref{fig6}C, right).

Our findings do not rely on parameters inferred by fitting specific mechanistic models of transcription to the data. Instead, we directly identify ON and OFF periods from single-allele initiation events. The elucidated bursting dynamics further align with single-allele time-trace auto-correlation analysis, which refrains from any ON-OFF calling. Thus, our bursting relationships are effective and broadly applicable across various models. However, while not originating from model-specific kinetic assumptions, these relationships impose constraints on the kinetic parameters of any underlying transcription model. Any detailed N-state model must adhere to these relationships once coarsely adapted to produce ON and OFF periods akin to those observed in the data \cite{Zoller:2015im, Li:2018ha, Wan2021, Pimmett2021}.

Since, to the best of our knowledge, the observed relationships do not trivially map to commonly invoked molecular processes shaping transcriptional dynamics, our findings suggest a novel mechanistic constraint. Our work raises the need to elucidate molecular mechanisms that integrate various regulatory processes into the control parameter $P_{\rm ON}$. Furthermore, the largely constant and relatively small value of $T_{\rm C}$ raises intriguing questions about its molecular implementation. The measured value of ~1 min is reliably above our detection capability (see Methods), yet small given the theoretically accessible parameter space (Fig. \ref{fig4}E). This value further corresponds to the shortest ON and OFF durations that we observe (at $P_{\rm ON}$ close to 0 or 1, correspondingly, and given $T_{\rm ON} = T_{\rm C}/(1-P_{\rm ON})$ and $T_{\rm OFF} = T_{\rm C}/P_{\rm ON}$). It is non-trivial that these periods display a similar minimal duration, as potentially distinct molecular events might govern the ON and OFF states. Our analysis across genes, perturbations, and organisms suggests a mechanism that operates independently of the specific gene locus and might be conserved across eukaryotic systems. It is intriguing to consider that the tight linkage between ON and OFF periods could stem from more recently appreciated aspects of the transcriptional environment such as nuclear architecture or the assembly and disassembly of transcription machinery components \cite{Tsai2017, Cho2018, Li:2020, Henninger:2021, Nguyen:2021, Bruckner2023}.

Intriguingly, the identified bursting relationships align with a naive ON-OFF bursting regime where periods have a lower limit, and any $P_{\rm ON}$ is encoded with the shortest ON-OFF combination, i.e., minimizing $T_{\rm ON}+T_{\rm OFF}$ (or maximizing burst frequency, Fig \ref{figS17}). These two simple assumptions give rise to a narrow parameter space utilized, asymmetric relationships for $P_{\rm ON}$ versus $T_{\rm OFF}$ and $T_{\rm ON}$, and a consistently low $T_{\rm C}$ throughout the full $P_{\rm ON}$ range. Such a regime could benefit developmental regulation by permitting fine-tuning gene expression with minimal resource expenditure. 

More generally, any bursting regime characterized by a preserved low value of $T_{\rm C}$ offers noise minimization as bursts are easily buffered by longer mRNA lifetimes. It also offers swift responsiveness as transcription outcomes rapidly adjust to input TF changes (Fig. \ref{figS8}C-D). Thus, similar to other fields where organizing principles are emerging, such as optimized information flow and others \cite{Tkacik:2008ir, Jones:2014pe, Hausser:2018hd, Petkova:2019hx, Balakrishnan:2022pb}, the elucidated relationships offer insights into the functionality encoded by complex processes and guide future investigations into conserved mechanisms at their core.

\begin{acknowledgments}
We thank members of the Gregor laboratory for discussions and comments on the manuscript; and Eric Wieschaus for suggestions at various stages of the project. We also thank Tineke Lenstra, Mounia Lagha, Edouard Bertrand, Ovidiu Radulescu, and Daniel Larson for discussions and sharing data. This work was supported in part by the U.S. National Science Foundation, through the Center for the Physics of Biological Function (PHY-1734030), and by National Institutes of Health Grants R01GM097275, U01DA047730, and U01DK127429. M.L. received a Human Frontier Science Program fellowship (LT000852/2016-L), EMBO long-term postdoctoral fellowship (ALTF 1401-2015), and Rothschild postdoctoral fellowship.
\end{acknowledgments}
 
 
\bibliography{Bursting2024}


\onecolumngrid

\newpage


\section*{Supplemental Figures}

\makeatletter 
\renewcommand{\thefigure}{S\@arabic\c@figure}
\renewcommand{\thetable}{S\@arabic\c@table}
\setcounter{figure}{0}

\begin{figure*}[h!]
\centering
\includegraphics[scale=1.0]{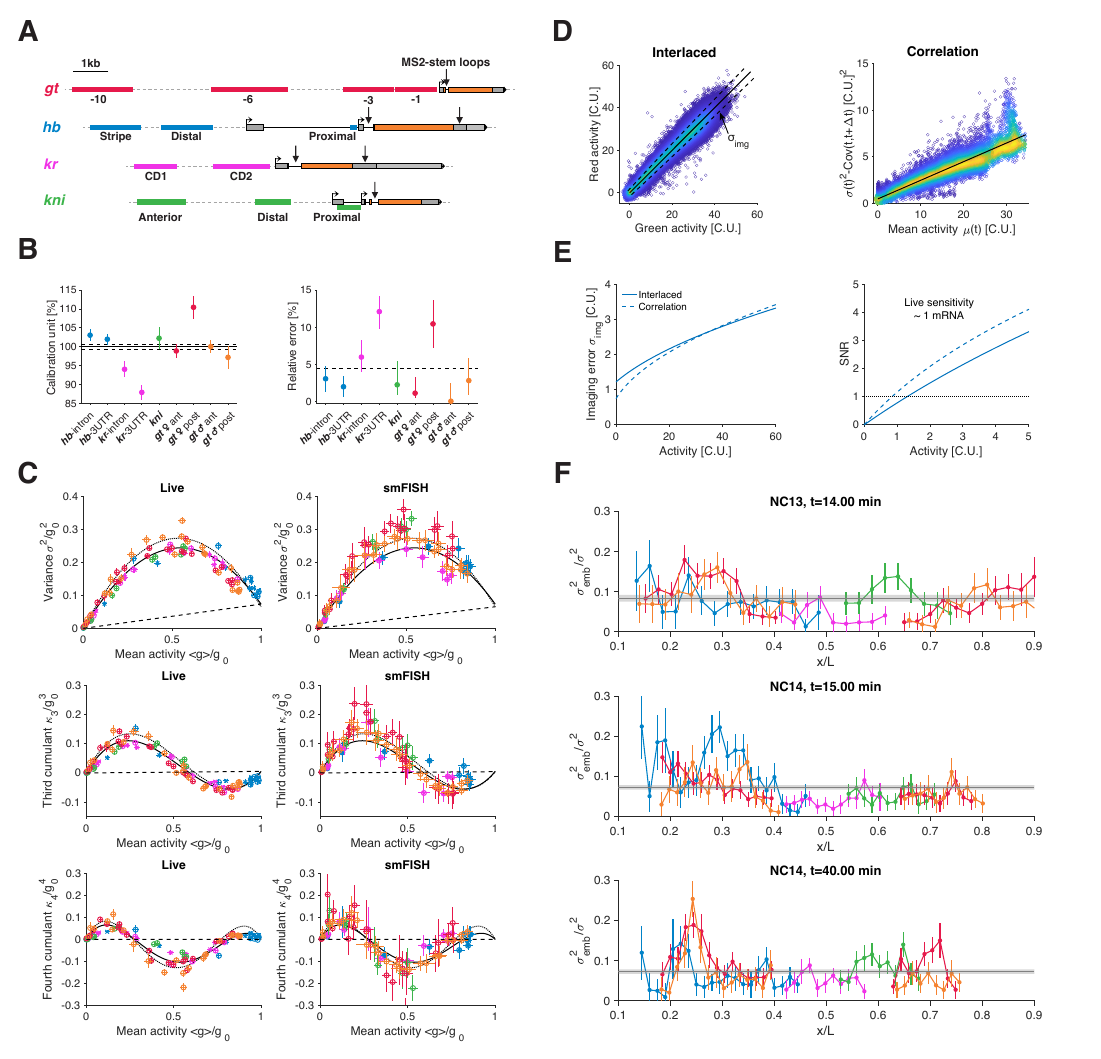}
\caption{{\bf Signal calibration, measurement error and embryo-embryo variability.} [Caption see next page.]}
\label{figS1}
\end{figure*}

\newpage
\begin{figure*}
\centering
\begin{justify}
FIG. S1. {\bf Signal calibration, measurement error and embryo-to-embryo variability.}
(A) The four trunk gap genes, \emph{giant} (\emph{gt}), \emph{hunchback} (\emph{hb}), \emph{Kruppel} (\emph{Kr}) and \emph{knirps} (\emph{kni}) were imaged using the MS2/PP7 stem-loop labeling systems. Stem-loop cassettes (vertical black arrow) were inserted either in the first/second intron or in the 3'UTR of each gene. Gap genes harbor different \emph{cis}-architectures as characterized by the number of promoters, number of enhancers (color boxes), and composition of these \emph{cis}-regulatory elements (TF binding motifs, core promoter elements, etc.). 
(B) Relative calibration unit (left) and relative error (right) for each gap gene construct related to Fig. \ref{fig1}C. The conversion of the live signal to absolute units is performed by comparison to a smFISH-based measurement. Calibration was performed by matching the mean full embryo length transcriptional activity profiles (reconstructed by averaging over all nuclei in 2.5\% AP bins, within a 5 min time window in NC13) measured by live imaging to those previously measured with smFISH (Zoller et al., 2018). (Left) The procedure was performed using all measured gap profiles at once, leading to a final calibration unit (horizontal back line, dashed lines are plus/minus one standard error). We then repeated the procedure for each individual construct separately (color circles), and the derived units are expressed in percent of the global fit. (Right) Relative error for the calibration unit of each individual gene construct with respect to the global unit and mean relative error (dashed line), which is below 5\%. Error bars are 68\% confidence intervals.
(C) Comparison of higher cumulants versus mean activity relationships obtained by live imaging and smFISH measurements; right column panels are reproduced from Figure 3B-D in Zoller et al. (2018), left column panels are from live data analyzed equivalently. Live cumulants of transcriptional activity (mean, variance, 3rd and 4th cumulant) are estimated over all nuclei in 2.5\% AP bins, within a 5 min time window in NC13. Cumulants are converted from equivalent cytoplasmic mRNA units (C.U.) to Pol II counts for a single gene copy of average length (3.3 kb). The cumulants are normalized with respect to $g_0$ defined as the intercept of the Poisson background (dashed line) and the polynomial fit to the data (black solid line for live and doted for smFISH). The number $g_0$ can be interpreted as the mean number of Pol II on a 3.3 kb long gap gene at maximal activity. We get $g_0=13.6$ for live and $g_0=15.2$ with smFISH measurements, a difference of $12\%$. Overall, the higher cumulants versus mean relationships obtained from live (left column) and from smFISH (right column) are extremely close (black solid versus dotted line), confirming the quantitative nature and the proper calibration of our live assay. Two independent methods (one being non-invasive genetically but involving fixation, while the other involves gene editing and stem-loop cassette insertions) leading to the same quantitative conclusions validate each other reciprocally. It strongly suggests that our synthetic modifications of the endogenous gap gene loci have no currently measurable effect on the transcriptional output of the system.
(D) Two independent methods to assess the imaging error. (Left) An interlaced cassette of alternating MS2 and PP7 stem-loops, labeled with two differently colored coat proteins (MCP-GFP and PCP-mCherry), is inserted in the first intron of \emph{Kr}. In absence of imaging error, the transcriptional activity in the green and red channels when calibrated to C.U. should perfectly correlate (on the diagonal). We fitted the spread $\sigma_{\rm img}$ orthogonal to the diagonal (black line, slope one) to characterize the imaging error; assuming $\sigma_{\rm img}^2$ scales as $\sigma_b^2+\alpha I$ with mean intensity $I$, where $\sigma_b^2$ is the background noise and $\alpha I$ a Poisson shot noise term. The resulting fit for $\sigma_{\rm img}$ is highlighted by the dashed lines (plus minus one std around the diagonal). (Right) Imaging error estimation from the single allele transcriptional time series (with the assumption that the measured transcriptional fluctuations result from the sum of uncorrelated imaging noise and correlated noise due to the elongation of tagged nascent transcripts). We computed the time-dependent mean activity $\mu(t)$, variance $\sigma^2 (t)$ and covariance between consecutive time points $\text{Cov}(t,t+\Delta t)$ (where $\Delta t$ is 10 s), over all nuclei within 1.5--2.5\% AP bins for all measured genes. The uncorrelated imaging variability $\sigma_{\rm img}^2$ is then approximated by $\sigma(t)^2-\text{Cov}(t,t+\Delta t)$, which is plotted as a function of $\mu(t)$ for all time points. We characterized $\sigma_{\rm img}^2$ by fitting the data with a line $\sigma_b^2+\alpha \mu$. Fitting results are shown in E. Overall, the fractional imaging variability $\sigma_{\rm img}^2/\sigma^2$ is $\sim5\%$.
(E) Our two estimates for the imaging error (interlaced dual-color construct (solid line) and correlation-based approach (dashed line)) are consistent. The signal-to-noise-ratio (SNR), defined as $\mu/\sigma_{\rm img}$, is close to one (dotted line) when $\mu \approx 1$, indicating that the sensitivity of our live measurements is close to one mRNA molecule.
(F) Fractional embryo variability profiles as a function of AP position and developmental time, for all gap genes. We define embryo variability $\sigma_{\rm emb}^2$ as the variance of the mean activity across embryos, and report the fractional embryo variability as the ratio $\sigma_{\rm emb}^2/\sigma^2$, where $\sigma^2=\sigma_{\rm emb}^2+\sigma_{\rm img}^2+\sigma_{\rm nuc}^2$ is the total variance, and $\sigma_{\rm nuc}^2$ corresponds to the transcriptional allele-to-allele noise across nuclei. Overall, the fractional embryo variability $\sigma_{\rm emb}^2/\sigma^2$ is $\sim10\%$, meaning that most of the variability arises from $\sigma_{\rm nuc}^2$. Thus, together D, E, and F show that $\sigma^2=+\sigma_{\rm emb}^2+\sigma_{\rm img}^2+\sigma_{\rm nuc}^2$ is a good proxy for $\sigma_{\rm nuc}^2$, which is the relevant noise contribution that contains all the bursting phenomenology.
\end{justify}
\end{figure*}

\begin{figure*}
\centering
\includegraphics[scale=1.0]{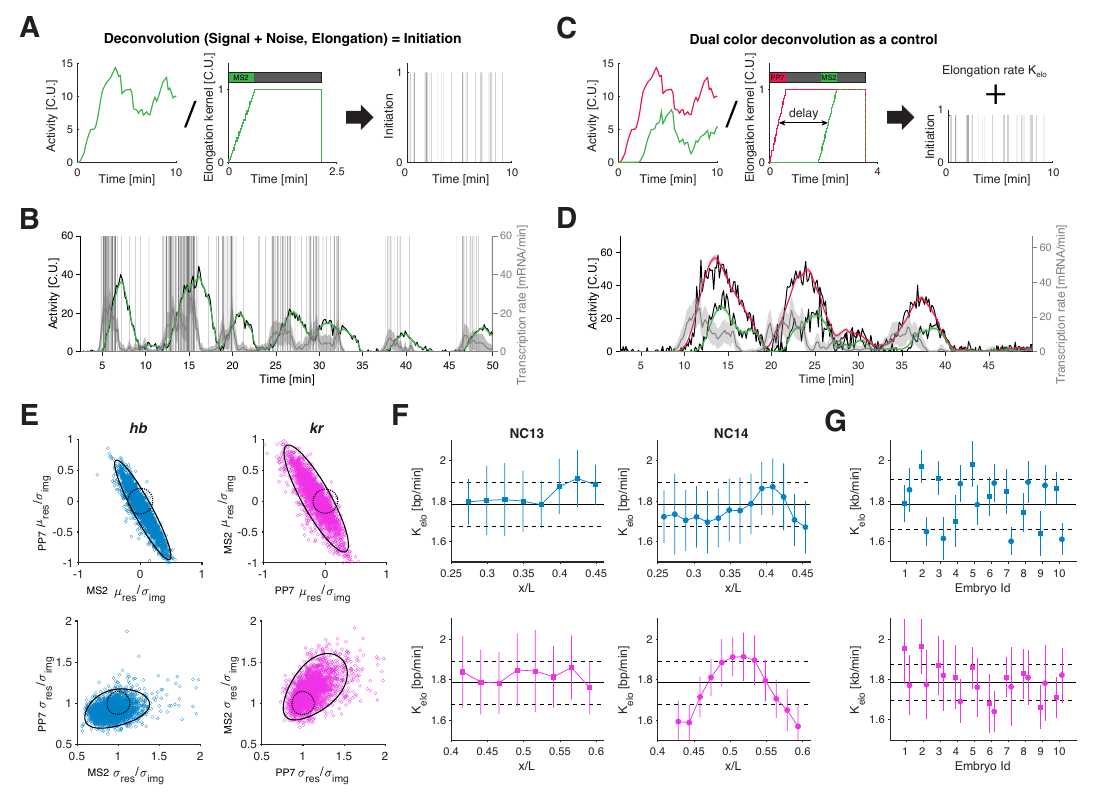}
\caption{{\bf Dual color measurements to validate single-cell deconvolution and measure elongation rate.} [Caption see next page.]}
\label{figS2}
\end{figure*}

\newpage
\begin{figure*}
\centering
\begin{justify}
FIG. S2. {\bf Dual color measurements to validate single-cell deconvolution and measure elongation rate.}
(A) Reconstruction of transcription initiation events from deconvolution of single allele transcription time series. 
The signal is modeled as a convolution between transcription initiation events and a kernel accounting for the elongation of a single Pol II through the MS2 cassette and the gene body (using an elongation rate $K_{\rm elo}=1.8\,$kb/min, see D-E). Bayesian deconvolution is performed by sampling from the posterior distribution of possible configuration of initiation events given the measured activity and measurement noise (Fig. \ref{figS1}F-G).
(B) Example deconvolved initiation configuration (vertical gray bars) and corresponding reconstructed signal (green) from a single allele transcription time series (black). Single allele transcription rate (gray line) is estimated by counting the number of initiation events within 10$\,$s intervals for a given sampled configuration and averaged over 1'000 of such configurations. The displayed solid line and envelope for transcription rate (gray) and reconstructed signal (green) correspond to the mean and one standard deviation of the posterior distribution.
(C) Validation of the kernel assumption for the deconvolution of initiation events from single allele transcription time series using a dual-color (confocal) imaging approach for \emph{hb} and \emph{Kr}. For \emph{hb} (\emph{Kr}), we generated fly lines with dual insertions of an MS2 (PP7) stem-loop cassette in the respective first intron and a PP7 (MS2) stem-loop cassette in the 3'UTR. In both cases, the two cassettes were labeled using two different colors (MCP-GFP green and PCP-mCherry red). Since the two signals are correlated through the elongation process, the simultaneously measured pair of time series has a further constrained set of underlying initiation configurations and represents thus a good test for the approach. To deconvolve single allele dual color time series together (i.e., a single train of polymerases needs to match two signals), using two kernels modeling each loop-cassette location and satisfying our key assumptions (i. constant and deterministic elongation rate; ii. no Pol II pausing/dropping in gene body; iii absence of co-transcriptional splicing; iv. fast termination). In addition, the dual-color strategy allows estimation of the average elongation rate from the overall delay between the two signals (using the known genomic distance between the MS2 and PP7 insertion sites).
(D) Dual-color signal reconstructed from deconvolved single allele transcription time series (black lines for raw measured data). single allele transcription rate (gray line with one std envelope) is deconvolved from the single depicted pair of measured time series (black lines). The signal (red and green lines with one std envelope) is devoid of imaging noise (as it was modeled from Fig. \ref{figS1}D during the deconvolution process) and is reconstructed by convolving back the resulting transcription rate with the kernel of each channel. Qualitatively, the signal (color) matches well (see E) the measured time series (black) in strong support of our kernel assumptions.
(E) Distribution of residuals from the dual-color reconstruction. We quantified the mean and standard deviation of the normalized residuals, i.e., of the difference between the measured signal (black in D) and the reconstructed signal (color in D) divided by the standard deviation of the imaging noise, for each recorded individual allele (for \emph{hb} $N=2666$ (blue) and for \emph{Kr} $N=2594$ (pink)). Overall, the dispersion of the means and standard deviations of normalized residuals (black line, $95\%$ confidence ellipse) is close to the expected dispersion of a perfect model (dotted line, $95\%$ confidence ellipse).
(F-G) Estimated elongation rate $K_{\rm elo}$ from dual-color measurements. (F) Average elongation rate computed over nuclei across 10 embryos as a function AP position (both \emph{hb} (blue) and \emph{Kr} (pink)) in NC13 (square) and NC14 (circle), with error bars representing one standard deviation across the embryo means. (G) Average elongation rate was computed for individual embryos (color code and symbols as F), with error bars representing the standard deviation across the means over positions. The elongation rate is globally conserved across genes and nuclear cycles, with $K_{\rm elo}=1.8\pm0.1$ kb/min (corresponding to the mean across embryos (black line) plus/minus one standard deviation (dashed line).
\end{justify}
\end{figure*}

\begin{figure*}
\centering
\includegraphics[scale=1.0]{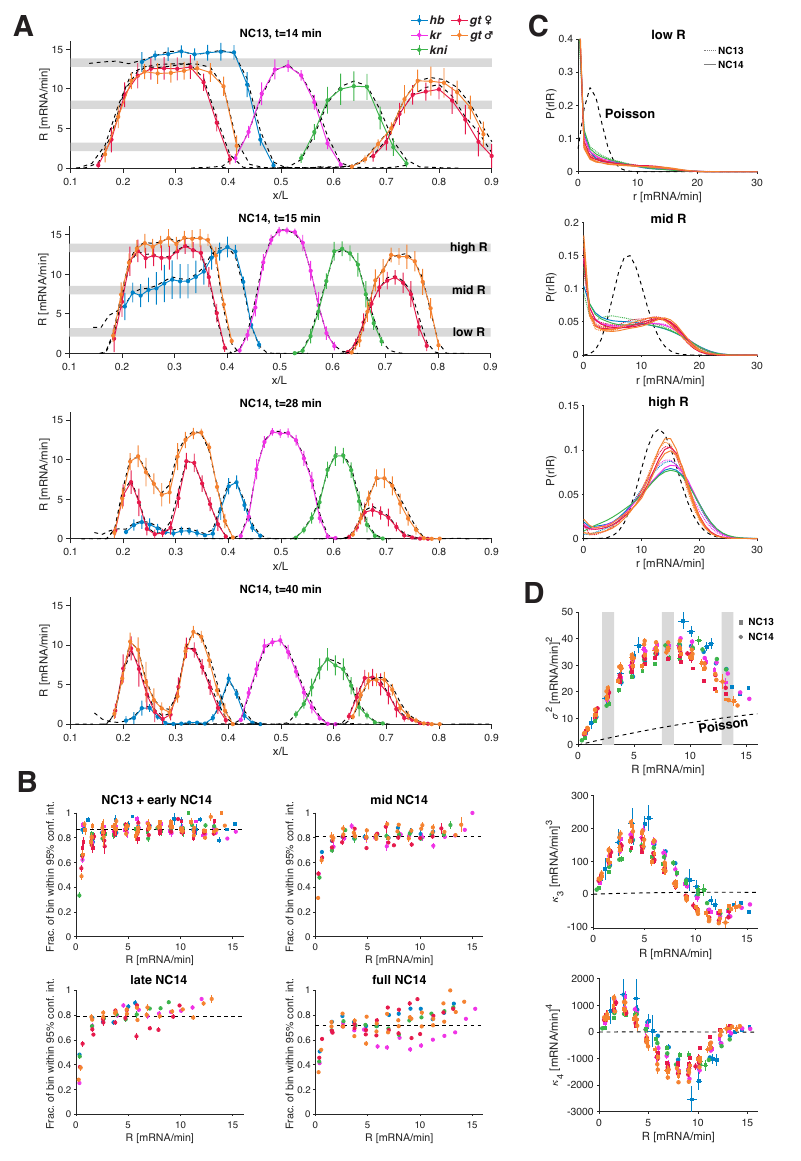}
\caption{{\bf Single-allele transcription rate distributions reveal common bursting characteristics.} [Caption see next page.]}
\label{figS3}
\end{figure*}

\newpage
\begin{figure*}
\centering
\begin{justify}
FIG. S3. {\bf Single-allele transcription rate distributions reveal common bursting characteristics.}
(A) Snapshots of the gap gene mean transcription rate $R$ as a function of AP position in late NC13, as well as early, mid and late NC14 (time $t$ min after mitosis). Gap gene profiles (color) are obtained by averaging the deconvolved single allele transcription rate over all nuclei within each AP bin (width of $2.5\%$ and $1.5\%$ embryo egg length in NC13 and NC14 respectively) and at each time point (10 s temporal resolution). The black dashed lines correspond to the mean activity (Fig. \ref{fig1}C) of each gap gene at the same position and time normalized by the effective elongation time (see Methods, Fig. \ref{figS4}A). Both the colored and dashed profiles agree, justifying our deconvolution approach. Error bars are one standard deviation across embryo means. Overall, we have effectively deconvolved $N_g = 7$ ``genes'' (4+\emph{gt} male and female and anterior and posterior regions), over $N_t=362$ time points (NC13+NC14), across $N_x=$9--18 positions, leading to a total of $33'214$ bins, each averaging $\sim200$ nuclei (with a single allele per nucleus). Strikingly, the gap genes reach a similar maximum average transcription rate $R_{\rm max}=14.8\pm0.9$ mRNA/min.
(B) Fraction of spatial and temporal bins whose single allele transcription rate distribution $P(r)$ is consistent with the conditional transcription rate distribution $P(r|R)$ determined by pooling nuclei over multiple bins at a given mean transcription rate $R$ (see C). We computed the $95\%$ confidence interval on the cumulative distribution of $P(r|R)$ and checked for all the underlying bins at a given $R$ whether their individual cumulative distribution was within the overall confidence interval. We repeated this process for four distinct developmental time windows: NC13 ($6.5\leq t$ min after mitosis) plus early NC14 ($7.5\leq t<20.5$ min), mid NC14 ($20.5\leq t<34.5$ min), late NC14 ($34.5\leq t < 48$ min), and a wider NC14 window ($7.5\leq t < 48$ min). Overall, bins that share similar $R$ within the same time window have very similar $P(r)$ distribution (median given by dashed line over $80\%$), which justifies the pooling of these bins. On the other hand, when pooling bins over the whole NC14 we observe further dissimilarities between bins, suggesting that $P(r|R)$ might moderately change over time.
(C) Distribution $P(r|R)$ of single allele transcription rates estimated within 1-min-intervals in both NC13 (color dotted lines) and early NC14 (color solid lines). These distributions are computed over all the nuclei from time points and AP bins whose mean transcription rate $R$ corresponds either to a low $[2.1,3.2]$, mid $[7.5,8.5]$ or high transcription level $[12.8,13.9]$ (as gray shade in A and D). The various gap gene distributions collapse at all transcription levels indicating an underlying common mode of transcription. Furthermore, these distributions differ from the Poisson distribution (black dashed line), which is the expected distribution for a constitutive regime (in which the gene is continuously active and always ON). The difference is most-pronounced for low- to mid-levels of $R$, where the gap distributions highlight two modes (instead of one for Poisson): a large probability mass around zero suggesting an abundance of non-transcribing or barely transcribing alleles and an enrichment of highly transcribing alleles beyond the Poisson expectation. Such features are highly suggestive of a universal bursting regime.
(D) Variance ($2^{\text nd}$ cumulant), $3^{\text rd}$ cumulant and $4^{\text th}$ cumulant of single allele transcription rate as a function of mean transcription rate $R$ in NC13 (square) and early NC14 ($7.5\leq t<20.5$ min; circle). The single allele transcription rates are estimated within 1-min-intervals, highlighting a strong departure of the cumulants from a constitutive Poisson regime (dashed line, $\sigma^2$, $\kappa_3$ and $\kappa_4=R$). Our data approaches the Poissonian regime only on the extreme ends of the $R$ spectrum. Moreover, the mean-variance relationship has a marked concave parabolic shape. Such a mean-variance relationship is consistent with the prediction of a 2-state model of bursting, where changes in $R$ results from modulation of $P_{\rm ON}$ \cite{Zoller:2018gj}. Together, these results suggest that the gap genes transition all the way from fully OFF ($P_{\rm ON}=0$) to fully ON ($P_{\rm ON}=1$), following a common bursting regime. Vertical gray bars correspond to low, mid, and high $R$, as in A.
\end{justify}
\end{figure*}

\begin{figure*}
\centering
\includegraphics[scale=1.0]{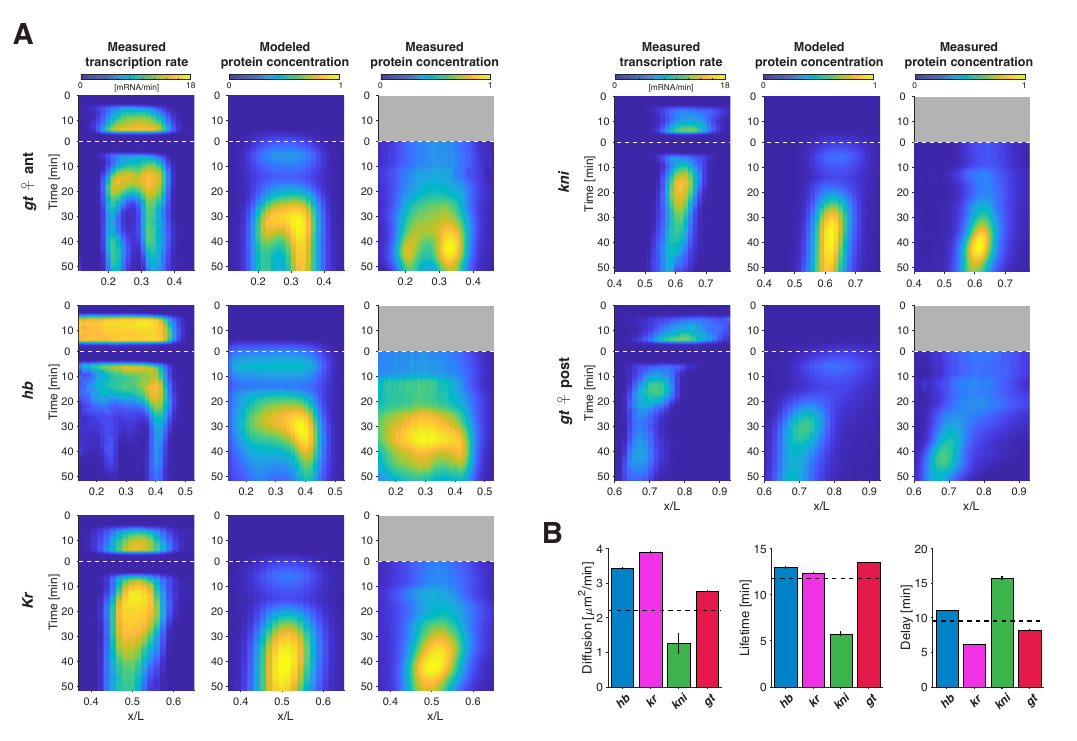}
\caption{{\bf Mean transcription rate explains dynamic pattern establishment.}
(A) A simple modeling attempt for protein accumulation from mean transcription rate measurements. The mean transcription rate (left column) across space and time is estimated by normalizing the measured mean activity by the elongation time and applying a minor correction for the delay ($<1$ min) resulting from the loop insertion location. Horizontal white dashed lines correspond to the transition (mitosis) from NC13 to NC14. Protein accumulation (middle column) is computed from the mean transcription rate as the convolution of the latter with a kernel modeling protein decay, diffusion, and delay due to mRNA export, translation, and nuclear import. This simple model introduces three free parameters, a protein lifetime, a diffusion constant, and a time delay (see B). These three parameters were set by minimizing the mean squared error with previously measured protein patterns from carefully staged gap gene antibody staining (right column; Dubuis et al., 2013). Small residual deviations between the middle and right columns might be due to post-transcriptional regulatory processes that our simple model does not account for. 
(B) Parameters estimated for the modeled accumulation of effective proteins as described in A. The three parameters were either estimated for each gene separately (color bars) or all genes together (dashed lines, used for middle column in A). Overall, the effective parameters are mostly in line with previous estimates \cite{Becker:2013}.}
\label{figS4}
\end{figure*}

\begin{figure*}
\centering
\includegraphics[scale=1.0]{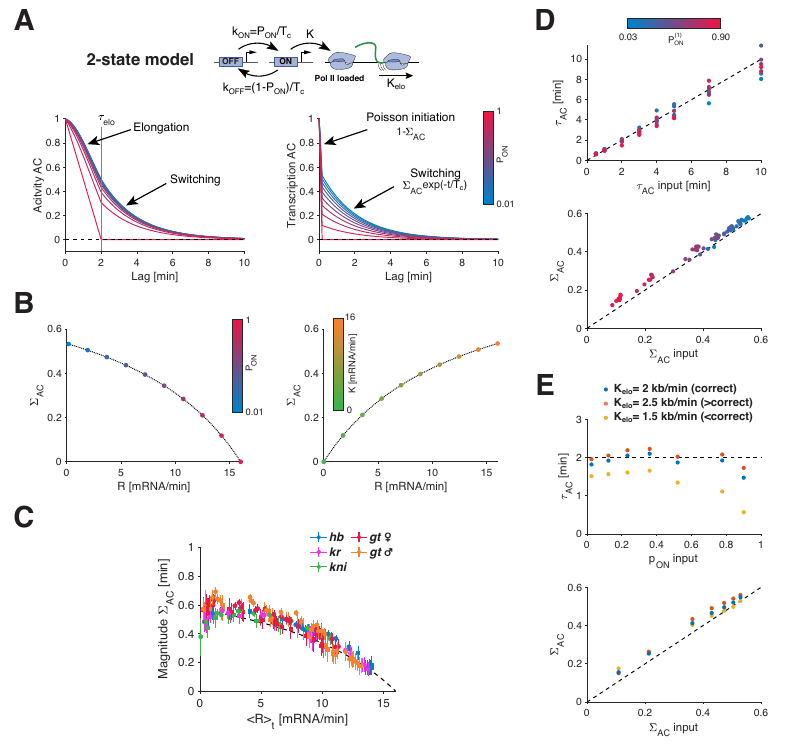}
\caption{{\bf Single-allele transcription rate fluctuations reveal key bursting characteristics.}
(A) Interpreting auto-correlation function using the 2-state model of transcriptional bursting. In this model for a single gene copy (top), the gene promoter switches stochastically between an OFF and ON state with rates $k_{\rm OFF}$ and $k_{\rm ON}$, where in the latter Pol II can be loaded at rate $K^{(1)}$ and elongate at rate $K_{\rm elo}$. The superscript $^{(1)}$ denotes the parameter for a single gene copy. The auto-correlation functions were computed from the model using a switching correlation time $T_{\rm C}^{(1)}=1/(k_{\rm OFF}+k_{\rm ON}) = 2$ min, a Pol II elongation time $\tau_{\rm elo} = L_g/K_{\rm elo} = 2$ min (where $L_g$ is the gene length) and an initiation rate $K^{(1)} = 8$ mRNA/min; the steady state ON-probability $P_{\rm ON}^{(1)} = k_{\rm ON}/(k_{\rm ON}+k_{\rm OFF}$) varies from $0$ to $1$ (i.e., blue to red color code, a fraction of nuclei in ON state or fraction of time a nucleus is in ON state). In principle, promoter switching (generating bursts) leads to temporal correlations in the transcriptional activity time series (Activity AC). However, from the raw live measurements, these correlations are hard to distinguish from the ones introduced by elongation (left), specifically when the switching correlation time $T_{\rm C}^{(1)}$ is close to or smaller than the elongation time $\tau_{\rm elo}$. Instead, performing auto-correlation analysis on deconvolved single allele transcription rates resolves the switching correlations (Transcription AC, right), since correlations due to elongation have been removed. Thus, the switching correlation time $T_{\rm C}^{(1)}$ can be estimated by fitting an exponential to the decay of the Transcription AC.
(B) Expected effect of the ON-probability (left) and Pol II initiation rate (right) on the magnitude of the correlated fluctuations, $\Sigma_{\rm AC}$, for $T_{\rm C}^{(1)}=2$ min. (Left) As $P_{\rm ON}^{(1)}$ increases, the transcription rate $R = 2 K^{(1)} P_{\rm ON}^{(1)}$ increases (here for 2 sister chromatids), and $\Sigma_{\rm AC}$ decreases until it vanishes at $P_{\rm ON}^{(1)}=1$ (Poisson regime). This behavior is consistent with our data shown in C. (Right) At fixed $P_{\rm ON}^{(1)}$ ($P_{\rm ON}^{(1)}=0.5$) and varying initiation rate $K^{(1)}$, the $\Sigma_{\rm AC}$ increases with growing transcription rate $R$. This behavior is the opposite of what we observed in our data. In both cases (Left and Right) the dotted line corresponds to the exact solution for $\Sigma_{\rm AC}$, which is well-approximated by $\Sigma_{\rm AC} = \Delta t K^{(1)} (1-P_{\rm ON}^{(1)}) / (1+\Delta t K^{(1)} (1-P_{\rm ON}^{(1)}) )$, where $\Delta t=10$ s corresponds to the data sampling time.
(C) Magnitude $\Sigma_{AC}$ of the correlated fluctuations in single-allele gap data as a function of mean transcription rate $R$. All gap data (color) collapses showing a universal trend (dashed line, guide to the eye). The fraction of correlated variability decreases as $R$ increases, as expected when approaching a constitutive regime of uncorrelated Poisson initiation (see B).
(D) Correlation time and correlated magnitude are properly retrieved after single allele deconvolution. Using the Gillespie algorithm, we generated simulated data ($N=200$ and $50$ min long cell recordings per condition) according to the 2-state model in A. For each input condition ($P_{\rm ON}^{(1)}$ from $0.03$ to $0.9$ and $T_{\rm C}^{(1)}$ from $0.5$ to $10$ min), we performed single allele deconvolution and computed the auto-correlation on the resulting transcription rates. We estimated the correlation time $T_{\rm C}^{(1)}$ and the magnitude $\Sigma_{\rm AC}$ by fitting exponential. Both parameters are properly retrieved with minimal biases. Color code stand for $P_{\rm ON}^{(1)}$ and dashed line for slope 1.
(E) Estimating deconvolution biases due to elongation rate measurement bias. As in D, we generated simulated data ($P_{\rm ON}^{(1)}$ from $0.03$ to $0.9$ and $T_{\rm C}^{(1)}=2$ min) and aimed to deconvolve the data with elongation rate higher (orange dot) or lower (yellow dot) than the correct value (blue dot, used to generate the data as in F). Overall, the parameters are estimated correctly, with larger biases at large $P_{\rm ON}^{(1)}$ and when underestimating the elongation rate (yellow).}
\label{figS5}
\end{figure*}

\begin{figure*}
\centering
\includegraphics[scale=1.0]{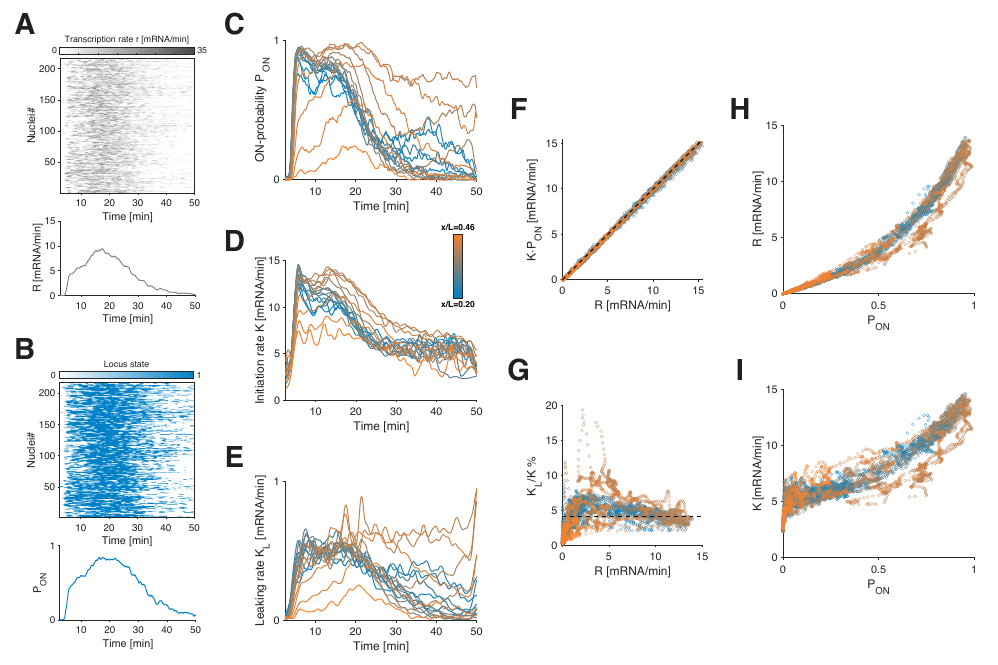}
\caption{{\bf Direct estimation of instantaneous mean transcription parameters for \emph{hunchback} in NC14.}
(A-B) Heatmaps of deconvolved single allele transcription rates (A) (estimated over 10$\,$s intervals) and of corresponding ON-OFF periods (B) (obtained from burst calling) as a function of time during NC14 for $N=217$ nuclei expressing \emph{hb}-MS2 at AP position $x/L=0.43$. Instantaneous mean transcription parameters such as transcription rate $R$ (A, bottom) and ON-probability $P_{\rm ON}$ (B, bottom) are obtained by vertically averaging the heatmaps (top) over all nuclei, respectively.
(C-E) \emph{hb} ON-probability $P_{\rm ON}$ (C), initiation rate $K$ (D), and leaking rate $K_L$ (E), as a function of time in NC14 for all AP positions (color coded). $P_{\rm ON}$ is computed as in B, while $K$ and $K_L$ are obtained by averaging over all nuclei in each AP bin the single allele transcription rate (A) conditioned on the locus being ON or OFF (B), respectively (as opposed to $R$, obtained by averaging regardless of allele state).
(F) Transcription rate $R$ versus the product of the initiation rate $K$ and the ON-probability $P_{\rm ON}$ for \emph{hb} in NC14 at all time points and positions. The color code stands for AP position as in Fig. \ref{fig3} and \ref{fig4}. As it should be by construction, $R$ can be decomposed into the product of $K$ and $P_{\rm ON}$.
(G) Leaking rate $K_L$ over initiation rate $K$ in \% as a function of the transcription rate $R$. The leaking rate $K_L$ never exceed 5\% of $K$ on average, supporting our ability to identify well-demarcated bursts over the whole range of transcription rate.
(H-I) Transcription rate $R$ (I) and initiation rate $K$ (J) as a function of $P_{\rm ON}$, for all time points and positions, demonstrating a massive data collapse, suggesting that $P_{\rm ON}$ is the central regulatory parameter for transcriptional bursting.}
\label{figS6}
\end{figure*}

\begin{figure*}
\centering
\includegraphics[scale=1.0]{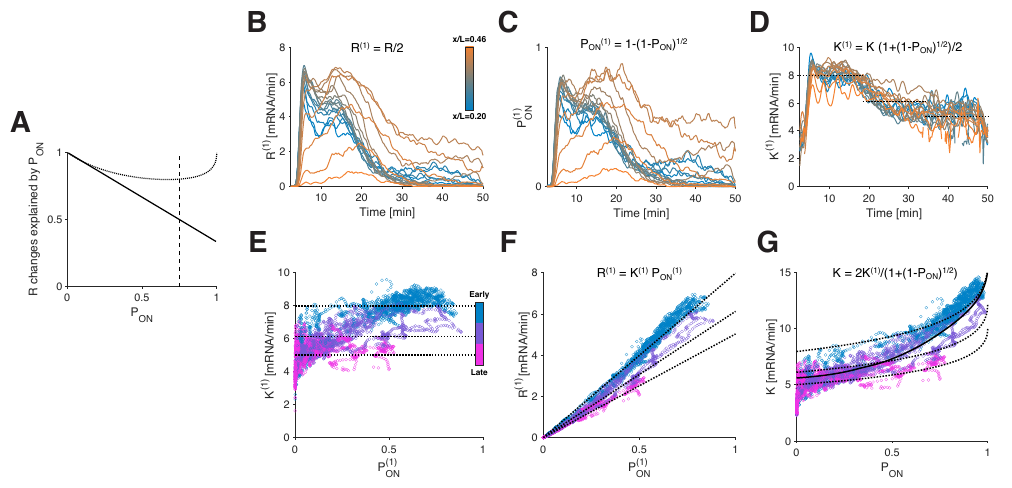}
\caption{{\bf Single gene copy parameter for \emph{hunchback}.}
(A) Fraction of changes in $R$ explained by $P_{\rm ON}$ as a function of $P_{\rm ON}$. Black line shows $d\log{P_{\rm ON}}/dP_{\rm ON}$ normalized by the sum $d\log{K}/dP_{\rm ON} + d\log{P_{\rm ON}}/dP_{\rm ON}$. The dotted black line is the corrected contribution assuming two independent sister chromatids (see B-G). Most of the changes in $R$ are thus dictated by $P_{\rm ON}$.
(B-D) Single gene copy (SGC) parameter computed from the estimated effective parameters for \emph{hb} in NC14, assuming two independent sister chromatids. The color code stands for AP as B. The spatiotemporal regulation of the SGC transcription rate $R^{(1)}$ (B) and the SGC ON-probability $P_{\rm ON}^{(1)}$ (C) are very similar to their corresponding effective parameters, $R$ and $P_{\rm ON}$. On the other hand, the SGC initiation rate $K^{(1)}$ (D) no longer varies across position, but only as a function of time. During NC14, the SGC initiation rate $K^{(1)}$ decreases by $38\%$, from 8.0 mRNA/min (first dotted line) to 5.0 mRNA/min on average (third dotted line), with most of the decrease happening between the $16^\text{th}$ and $34^\text{th}$ minute mark (second dotted line at 6.1 mRNA/min).  Interestingly, the distribution of $K^{(1)}$ leads to a mean Pol II spacing for a single active chromatid at $K_{\rm elo}/K^{(1)}=303\pm73$ bp, which is consistent with average the Pol II spacing of 330$\pm$180 bp in the classic Miller spreads \cite{McKnight1979}.
(E-G) Temporal changes in SGC initiation rate and independent sister chromatid assumption explain the dependence of the effective initiation on the ON-probability. Color code stands for three time windows in NC14: early (cyan, 2.5-16.7 min), mid (purple, 16.7-34.2 min), and late (magenta, 34.2-50 min). (E) Most of the variation in $K^{(1)}$ is explained by time, rather than $P_{\rm ON}^{(1)}$. The dotted lines are drawn at the same $K^{(1)}$ values as in D. (F) The SGC transcription rate $R^{(1)}$ appears almost linearly dependent on $P_{\rm ON}^{(1)}$. The nonlinearity is mostly explained by temporal changes in $K^{(1)}$, as highlighted by dotted lines whose slopes are the $K^{(1)}$ values from D and E. (G) Under the two independent sister chromatids assumption, the effective initiation rate $K$ depends on $P_{\rm ON}$ and on $K^{(1)}$, which varies as a function of time (see D). As $P_{\rm ON}$ increases, the propensity to observe two gene copies initiating transcription at the same time increases, which explains up to a factor of two in the dependence of $K$ on $P_{\rm ON}$. Indeed, the dotted lines correspond to the predicted behavior using the same three constant values of $K^{(1)}$ as in D. In addition, $K^{(1)}$ varies by up to $38\%$ along time during NC14, as can be seen in D. Together, it explains close to a factor of $3.2$ in $K$ variation with $P_{\rm ON}$. This reasoning explains the observed relationships very well (black line) and argues for a very weak dependence of $K$ on $P_{\rm ON}$.}
\label{figS7}
\end{figure*}

\begin{figure*}
\centering
\includegraphics[scale=1.0]{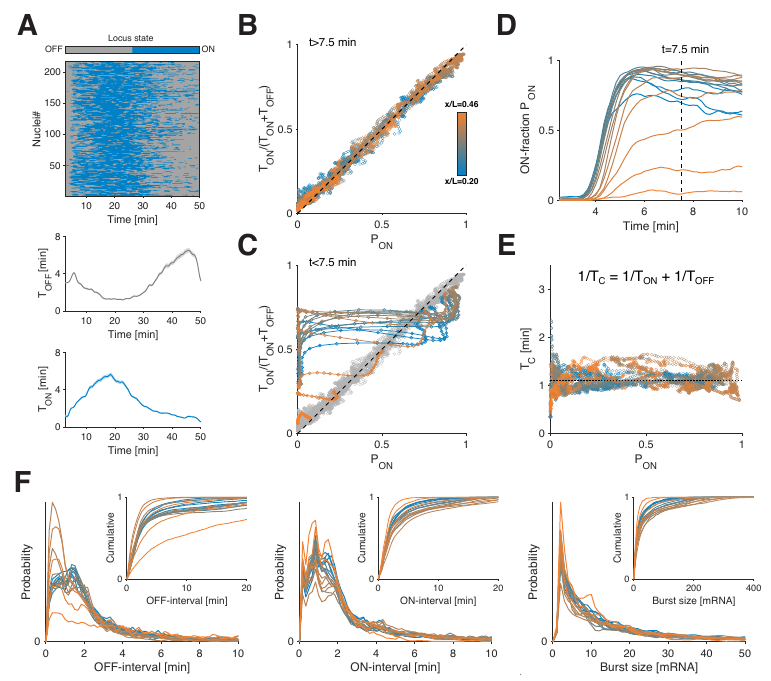}
\caption{{\bf Direct estimation of instantaneous mean ON and OFF-time for \emph{hunchback} in NC14.}
(A) Binarized heatmap from Fig. \ref{figS6}B. Instantaneous mean OFF-time $T_{\rm OFF}$ (bottom, gray) and mean ON-time $T_{\rm ON}$ (bottom, blue) are obtained by the weighted average of the ON and OFF times over all nuclei (see methods). The weights are given by the inverse of the number of time points within each period.
(B) The ratio of $T_{\rm ON}$ over the sum of $T_{\rm ON}$ and $T_{\rm OFF}$ versus ON-probability $P_{\rm ON}$ for all positions and time points beyond the 7.5 min mark in Fig. \ref{fig2}B-C. The 2-state model predicts that near-steady state (after an initial transient) both quantities match (C and D). The agreement between these quantities is a strong indication of the self-consistency of our general approach for extracting these bursting parameters and that the system operates near steady state. Thus, temporal changes in transcription parameters (due to developmental regulation) beyond the 7.5 min mark must be slow enough to allow relaxation.
(C) Evidence for initial out-of-steady state transient for \emph{hb} in early NC14. Near the steady-state regime, $P_{\rm ON}$ should be well-approximated by $T_{\rm ON}/(T_{\rm ON}+T_{\rm OFF})$, as it is the case beyond the 7.5 min mark in NC14 (gray circles tracing the diagonal; see B). However, at the post-mitotic onset of transcription ($\sim$3--7.5 min into NC14) we observe strong deviations from the expected near-steady state relationship at all positions (color curves). The system is undergoing a fast transient relaxation that drives it near-steady state within the first 8 min after mitotic exit, see D.
(D) Close-up of the first 10 min of Fig. \ref{fig3}D shows a rapid transient in the ON-probability for \emph{hb} at the onset of NC14. The vertical dashed line at 7.5 min marks the transition between the transient and the near-steady-state regime as observed in C.
(E) Effective switching correlation time $T_{\rm C}$ (defined as: $1/T_{\rm C} = 1/T_{\rm ON}+1/T_{\rm OFF}$) as a function of $P_{\rm ON}$, computed using data points in Fig. \ref{fig2}B-C. $T_{\rm C}$ is mostly conserved across time points and position and is $P_{\rm ON}$ independent. Dotted line corresponds to mean $T_{\rm C}=1.1\pm0.2$.
(F) Distributions and cumulative distributions of OFF-intervals, ON-intervals and burst size for all positions (color coded). These distributions are computed using all alleles and time points at a given position (see heatmpas in Fig. \ref{figS6}A  and \ref{figS8}A). As such, they represents non-stationary transcriptional dynamics (see Fig. \ref{fig2}A,C and \ref{figS6}C-D),  and are consequently less amendable to direct interpretation (e.g. of the non-exponential nature OFF- and ON-interval distribution) as deviations from the 2-state model stationary expectation.}
\label{figS8}
\end{figure*}

\begin{figure*}
\centering
\includegraphics[scale=1.0]{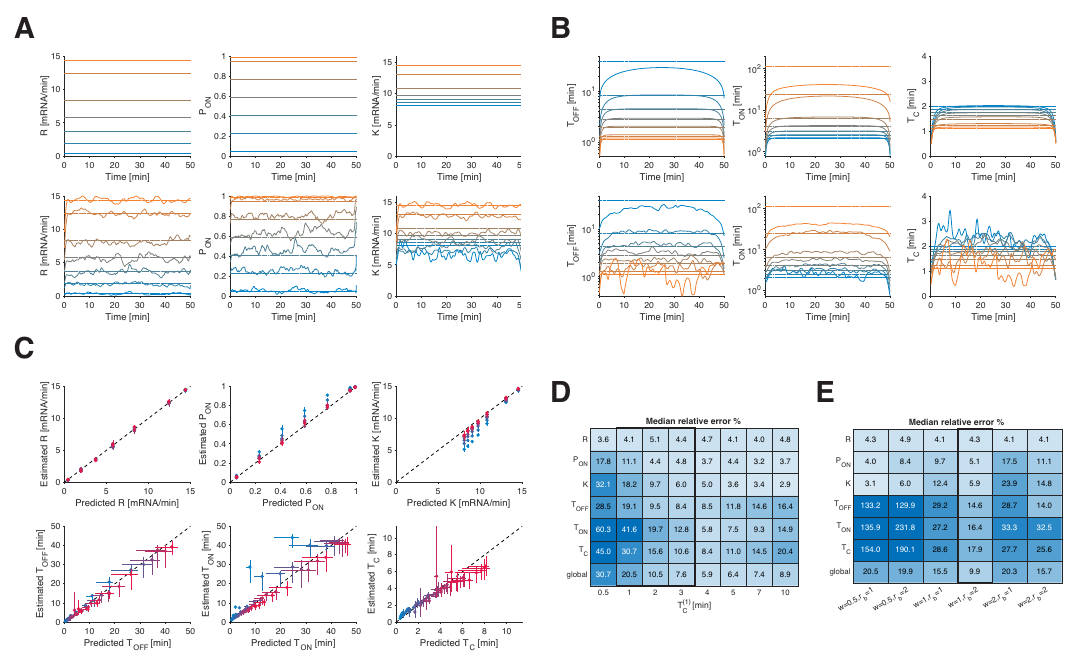}
\caption{{\bf Validation of transcription parameter estimation using stationary simulated data.}
Burst calling permits effective estimation of transcription parameters. For each set of input parameters ($K^{(1)}$, $T_C^{(1)}$, $P_{\rm ON}^{(1)}$), we generated simulated data (200 alleles and 50 min long time series at 10 s intervals and with the same imaging noise as measured in real data). Using the 2-state model (Fig. \ref{figS3}D) and the Gillespie algorithm, we generated time series for each individual sister chromatid. We summed the initiation events of both chromatids assuming independence. The resulting initiation events were convolved with the elongation kernel to generate synthetic single allele signal data over which measurement noise was added. We performed single allele deconvolution and burst calling as described (Fig. \ref{fig3}A) to estimate the effective mean transcription parameters ($R$, $K$, $P_{\rm ON}$, $T_{\rm ON}$, $T_{\rm OFF}$ and $T_{\rm C}$) for each simulated set of alleles.
(A-B) Comparison between the theoretically expected (top row) and the estimated from burst calling effective parameters (bottom row) as a function of time, using stationary (constant in time) input parameters ($T_C^{(1)}=2$ min, $K^{(1)}=8$ mRNA/min and $P_{\rm ON}^{(1)}$ varies from $0.03$ to $0.9$ that is color coded from blue to orange). Overall, the estimated effective parameters are recovered very well. (A) Despite sampling fluctuations, the constancy of the parameters in time is properly preserved (bottom row), as it should be based on input (top row). (B) Despite the stationary nature of the input parameters $T_C^{(1)}$ and $P_{\rm ON}^{(1)}$, biases for $T_{\rm ON}$, $T_{\rm OFF}$ and $T_{\rm C}$ (top row) is expected due to the finite length of the simulated time series (censoring), especially noticeable near the beginning and the end, where the estimations are “bent” (the resulting mean time estimate cannot exceed the width of the time window used to perform the estimate). However, our parameter estimation (bottom row) is very much in line with the expected biases (top row).
(C) Global comparison between expected and estimated parameters for the stationary case (input parameters are constant in time). The parameter estimation was performed on a large simulated data set that includes data in A and B ($T_C^{(1)}=2$ min, $P_{\rm ON}^{(1)}$ from $0.03$ to $0.9$) and data for other values of $T_C^{(1)}$ comprised between $T_C^{(1)}=0.5$ (blue dots) to $T_C^{(1)}=10$ (red dots). Each dot results from one combination of input parameters ($K^{(1)}$, $T_C^{(1)}$, $P_{\rm ON}^{(1)}$) and corresponds to the median effective parameter and the error bars to the $68\%$ confidence interval estimated over 50 min. Our deconvolution and burst calling approaches lead to an excellent estimation of the effective parameters over a large range of $T_C^{(1)}$ and $P_{\rm ON}^{(1)}$ values, albeit with noticeable biases in $P_{\rm ON}$, $K$ and $T_{\rm ON}$ when $T_C^{(1)}$ approaches 0.5 min (blue dots). Importantly, biases for the effective switching correlation time $T_C$ are small, supporting our ability to detect its constancy in real data.
(D) Summary of median relative error for each effective parameter estimated from the data in C as a function of input $T_C^{(1)}$. Parameter estimated from real data (Fig. \ref{figS12}B) suggests that $T_C^{(1)}$ lies within 1 and 3 min (black border).
(E) Summary of median relative error for each effective parameter estimated from the whole data set in C as a function of the burst calling parameters. Burst calling depends on two free parameters: the time window $w$ over which the rate is estimated, and the rate threshold $r_b$ applied to call the burst (see Fig. \ref{fig3}A). Our default parameter values are $w=1u$ and $r_b=2/u$ (with $u=5\Delta t=5/6$ min), which should be close to optimal given the estimated correlation time of $\tau_{\rm AC}\sim 1$ min (Fig. \ref{fig2}F) and our measurement sensitivity of 1--2 mRNA. When testing the effect of different $w$ and $r_b$ values on the median relative error, our default choice leads to the lowest global relative error.}
\label{figS9}
\end{figure*}

\begin{figure*}
\centering
\includegraphics[scale=1.0]{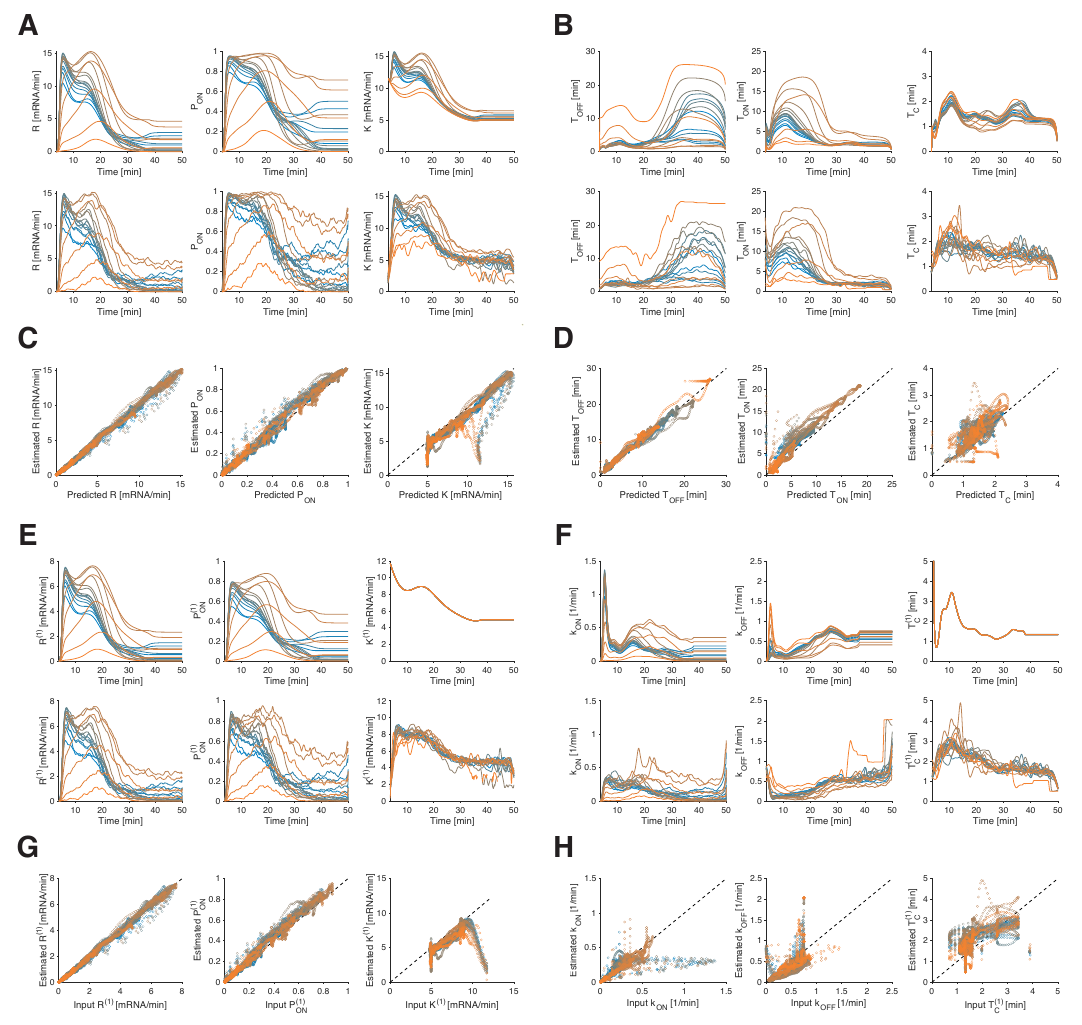}
\caption{{\bf Validation of transcription parameter estimation using \emph{hb}-like non-stationary simulated data.}
Simulated data is generated as in Fig. \ref{figS9}, except that the input parameters are non-stationary (time-dependent).
(A-B) Comparison between expected (A, B, top row) and estimated effective parameters (A, B, bottom row) for time-dependent input parameters (A, B, top row), mimicking the transcriptional output of \emph{hb} in NC14. Color code (blue to orange) stands for virtual AP position. Estimated effective parameters tightly recapitulate the time-dependence of the expected parameters, allowing us to also capture the temporal regulation of the gap genes. (C, D) We show with this realistic test case that our estimated parameters (A, B, bottom row) using burst calling very closely match the expected effective ones (A, B, top row). Indeed, most of the time points (circles) lie on the line of slope one (dashed line), which strongly supports our ability to precisely characterize the transcription parameters from real data. 
(E–F) Comparison between input (E, F, top row) and burst calling estimated single-gene copy parameters (E, F, bottom row) as a function of time. The time-dependent input parameters (E, F, top row) were used to generate the synthetic \emph{hb} data in A and B. Color code (blue to orange) stands for virtual AP position as in A and B. The estimated single-gene copy parameters were computed from the effective ones (A, B, bottom row) assuming the latter originate from two independent sister chromatids. Namely, we get $R^{(1)}=R/2$, $P_{\rm ON}^{(1)}=1-(1-P_{\rm ON} )^{1/2}$, $K^{(1)}=K(1+(1-P_{\rm ON})^{1/2})/2$, which are exact, and assuming steady state $T_{\rm C}^{(1)}=2 T_{\rm C}/(1+(1-P_{\rm ON})^{1/2})$, $k_{\rm ON}=P_{\rm ON}^{(1)}/T_{\rm C}^{(1)}$ and $k_{\rm OFF}=(1-P_{\rm ON}^{(1)})/T_{\rm C}^{(1)}$. (G-H) Even though the single gene copy parameters are deeply buried in the data, our simple burst calling procedure still manages to recover them correctly. Some discrepancies are observed for $k_{\rm ON}$, $k_{\rm OFF}$ and $T_{\rm C}^{(1)}$, mostly near the beginning when the transient after mitosis violates the near-steady-state assumption. But these are expected as the relationships $k_{\rm ON}=P_{\rm ON}^{(1)}/T_{\rm C}^{(1)}$ and $k_{\rm OFF}=(1-P_{\rm ON}^{(1)})/T_{\rm C}^{(1)}$ are only valid for near-steady state.}
\label{figS10}
\end{figure*}

\begin{figure*}
\centering
\includegraphics[scale=1.0]{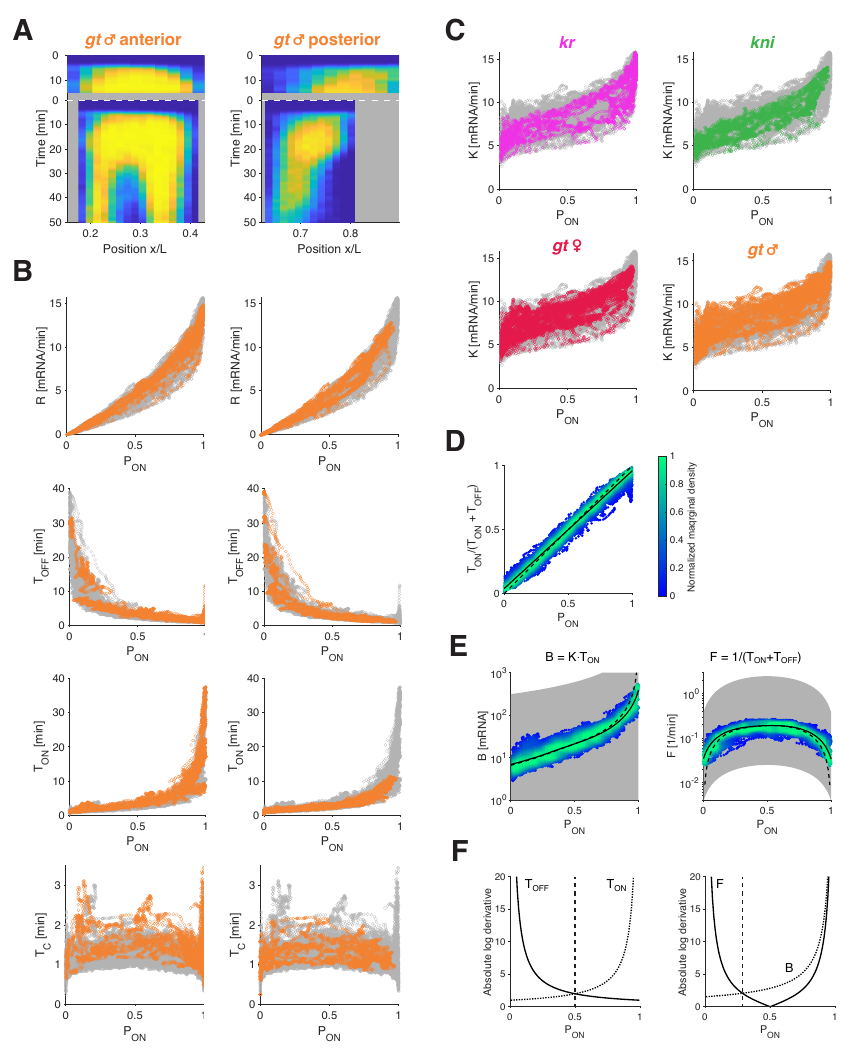}
\caption{{\bf Further transcription parameters collapse.}
(A) Kymograph of ON-probability for \emph{gt} transcription in male embryos, as a function of position and time. As for the other gap genes, the spatiotemporal transcription patterns arise from complex regulation of the ON-probability.
(B) transcription parameters for \emph{gt} in male embryos for NC13 and NC14 as a function of PON (orange data points) and all the other gap gene data sets (gray).
(C) Initiation rate $K$ as a function of $P_{\rm ON}$. $K$ collapses for all gap genes across time and position. 
Colored data points represent individual gap genes (same color code as in Fig. \ref{fig5}A-E and \ref{figS11}A-B); underlying is the remaining data of all other genes (gray). The $K$--$P_{\rm ON}$ relationship for \emph{hb} is shown in Fig. \ref{fig3}H.
(D) Near-steady state relationship between ON-probability $P_{\rm ON}$ and $T_{\rm ON}/(T_{\rm ON}+T_{\rm OFF})$ for all gap genes in NC13 ($t\geq6.5$min) and NC14 ($t\geq7.5$min). Although the data is distributed near the expected relationship (dash line), we observe a slight but clear bias at the extreme ends of the $P_{\rm ON}$ spectrum (solid line), namely $T_{\rm ON}/(T_{\rm ON}+T_{\rm OFF})$ is slightly above zero at $P_{\rm ON}=0$ and slightly below one at $P_{\rm ON}=1$. This is a consequence of the finite nature of our recording (50 min in NC14 and 18.4 min in NC13). Thus, our recording time sets an upper limit on the length of measurable ON and OFF intervals. That limit leads to the observed bias.
(E) Global scatter of the burst size $B$ and burst frequency $F$ (color) as a function of $P_{\rm ON}$ for all gap genes in both NC13 and NC14 and putative accessible space (gray region). See definitions of $B$ and $F$ at top of respective panels. Color code and accessible space are defined as in Fig. \ref{fig5}F. Solid line stands for the bursting relationships derived with bias (solid line in C) and dashed line without.
(F) Absolute derivatives with respect to $P_{\rm ON}$ of $\log{T_{\rm ON}}$ and $\log{T_{\rm OFF}}$ (left), and $\log{F}$ and $\log{B}$ (right), computed from bursting relationships in Fig. \ref{fig3}F and E above. Decomposition in terms of log derivative is convenient since the sources of changes in $P_{\rm ON}$ and $R$ become additive, i.e., $\log{(1-P_{\rm ON})/P_{\rm ON}} = \log{T_{\rm OFF}} - \log{T_{\rm ON}}$ and $\log{R} = \log{F} + \log{B}$. Thus, transition in the predominant type of expression modulation occurs at the crossing of the derivatives: at $P_{\rm ON}=0.5$ for ON and OFF-times and $P_{\rm ON}=1/3$ for burst size and frequency.}
\label{figS11}
\end{figure*}

\begin{figure*}
\centering
\includegraphics[scale=1.0]{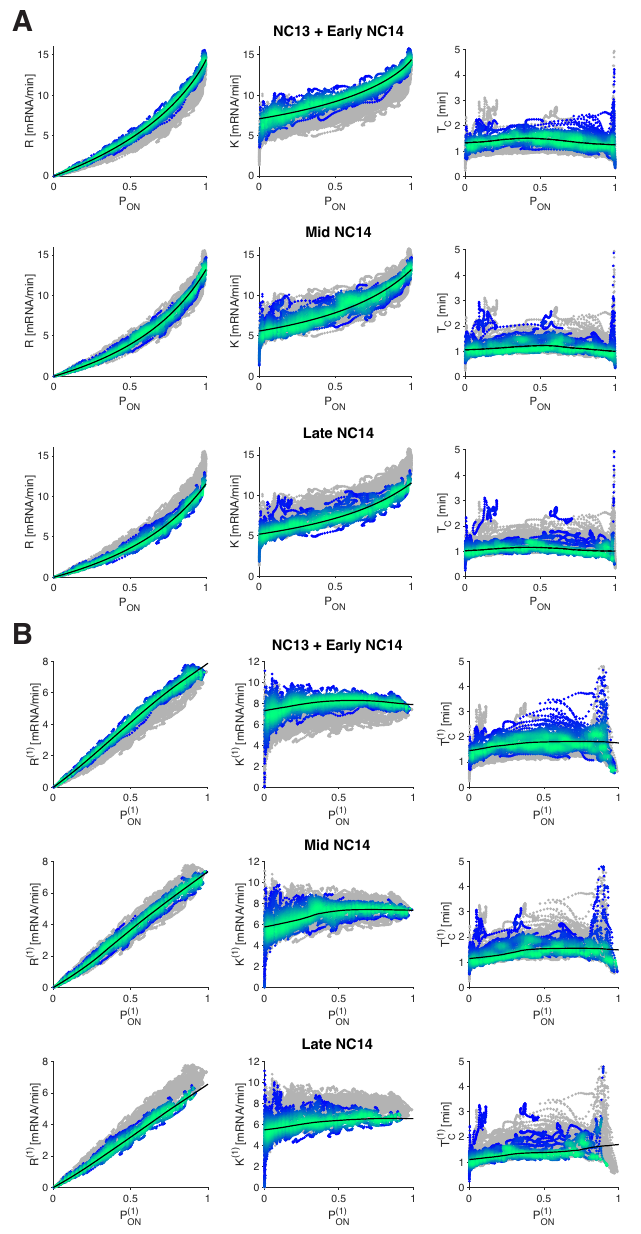}
\caption{{\bf Common bursting relationships across developmental time and for single gene copy.}
(A) Global scatter of effective transcription parameters as a function of $P_{\rm ON}$ for all gap genes estimated within three time windows, corresponding to NC13 ($6.5\leq t$ min) plus early NC14 ($7.5\leq t<20.5$ min), mid NC14 ($20.5\leq t<34.5$ min), and late NC14 ($34.5\leq t<48$ min). The observed bursting relationships are further refined when accounting for possible temporal changes (color scatter) compared to all time-pooling (gray scatter, Fig. \ref{fig5}F). Indeed, small changes in $K$ and $T_{\rm C}$ over developmental time ($\sim40\%$ decrease) explain part of the observed spread in Fig. \ref{fig5}F.
(B) As in A, but for single gene copy parameters computed from the effective ones assuming independent sister chromatids. Interestingly, the relationship between SGC transcription rate $R^{(1)}$ and SGC ON-probability $P_{\rm ON}^{(1)}$ is almost linear, confirming that the SGC initiation rate $K^{(1)}$ does not depend strongly on $P_{\rm ON}^{(1)}$. Thus, the apparent dependence of $K^{(1)}$ on $P_{\rm ON}^{(1)}$ is only effective and results from measuring two sister chromatids (two gene copies) together, instead of an isolated single gene copy.}
\label{figS12}
\end{figure*}

\begin{figure*}
\centering
\includegraphics[scale=1.0]{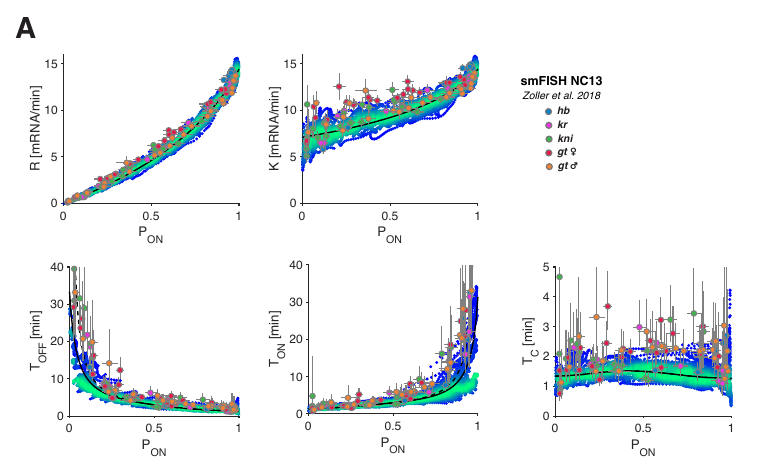}
\caption{{\bf Validating bursting relationships with smFISH data.}
(A) Bursting relationships in NC13 and early NC14 are consistent with parameters inferred from previous fixed measurements using smFISH. We converted the single gene copy parameters of the gap genes in mid-late NC13 (Zoller et al.) into effective parameters for two sister chromatids (color dots), with error bars corresponding to 68\% confidence intervals. The resulting smFISH effective mean correlation time $T_{\rm C}=2.06\pm0.66$ min is slightly smaller than in the original study ($T_{\rm C}=3.0\pm1.2$ min). Two reasons explain this difference: 1) we corrected for our updated elongation rate of 1.8kb/min compared to 1.5kb/min used previously, leading to a $T_{\rm C}$ reduction of 17\%; 2) the effective correlation time is up to two times smaller for large $P_{\rm ON}$ compared to the SGC correlation time $T_{\rm C}^{(1)}$ used in the original study ($T_{\rm C}^{(1)}=2 T_{\rm C}/(1+(1-P_{\rm ON})^{1/2})$, see Methods), leading to a further 17\% reduction on average. Overall, the smFISH effective parameters closely verify our relationships derived from live measurements ($T_{\rm C}=2.06\pm0.66$ min fixed versus $T_{\rm C}=1.25\pm0.37$ min live), albeit with small deviations likely stemming from a technical origin (differences in experimental protocol and microscopy, as well as limitations imposed by fixed measurements on parameter estimation). Thus, most likely, the width of the data clustering around our relationships reflects estimation errors with limited biological variation in parameters, as can also be verified from simulated data (Fig. \ref{figS5} and Fig. \ref{figS6}).}
\label{figS13}
\end{figure*}

\begin{figure*}
\centering
\includegraphics[scale=1.0]{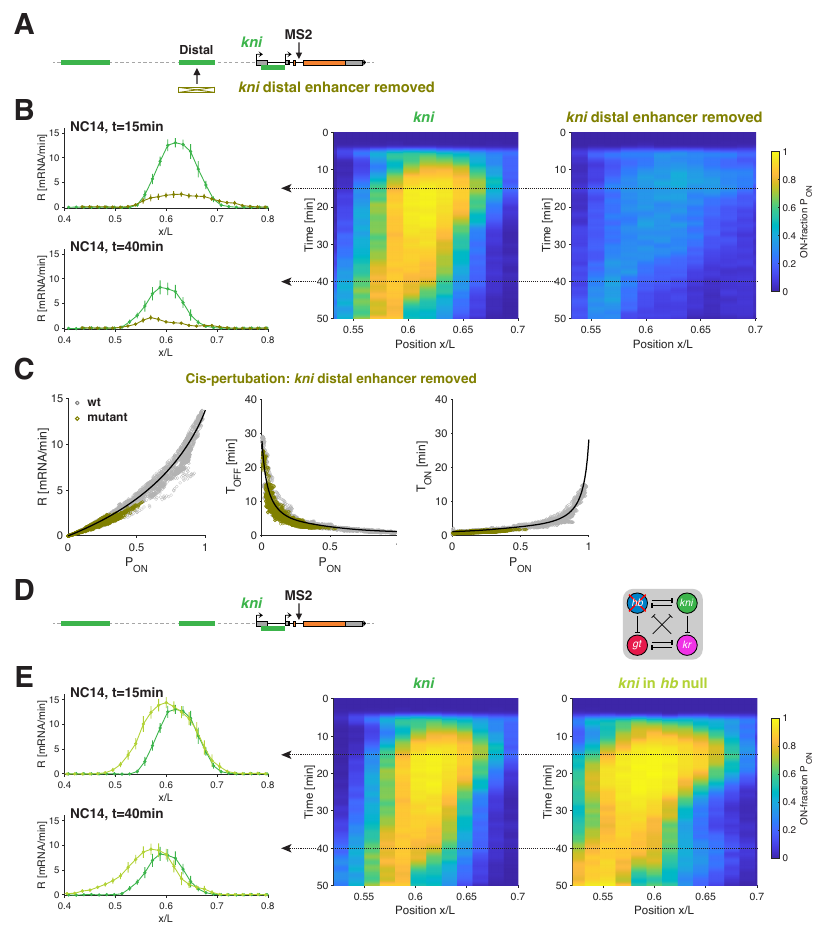}
\caption{{\bf Bursting relationships verified by \emph{cis}- and \emph{trans}-perturbations.}
(A) Distal \emph{kni} enhancer removal. The MS2-stem loops are inserted at the same location in the mutant (enhancer deletion) and wild-type fly lines.
(B) Quantification of \emph{kni} wild-type and mutant (A) phenotypes. Both transcription rate $R$ as a function of $x/L$ (left) and the kymograph for $P_{\rm ON}$ (right) display a significant level decrease underlying the expression patterns of the mutant. Dotted arrow indicates time point in kymograph at which rate profiles (left) are depicted.
(C) Transcription parameters for \emph{kni} \emph{cis}-mutant (olive) collapse on corresponding wild-type parameters (gray), as for \emph{hb} \emph{cis}-mutant (see Fig. \ref{fig4}C). Solid black lines correspond to the endogenous bursting relationships from Fig. \ref{fig5}F.
(D) \emph{kni} measurements in a \emph{hb} null background. The absence of \emph{hb} expression alters the network, namely the concentration of input transcription factors sensed by \emph{kni} in the mutant fly line.
(E) Quantification of \emph{kni} wild-type and mutant (D) phenotypes. Both transcription rate $R$ (left) and $P_{ON}$ kymograph (right) display a significant shift of the anterior boundary in the mutant expression patterns. Dotted arrow indicates time point in kymograph at which rate profiles are depicted.}
\label{figS14}
\end{figure*}

\begin{figure*}
\centering
\includegraphics[scale=1.0]{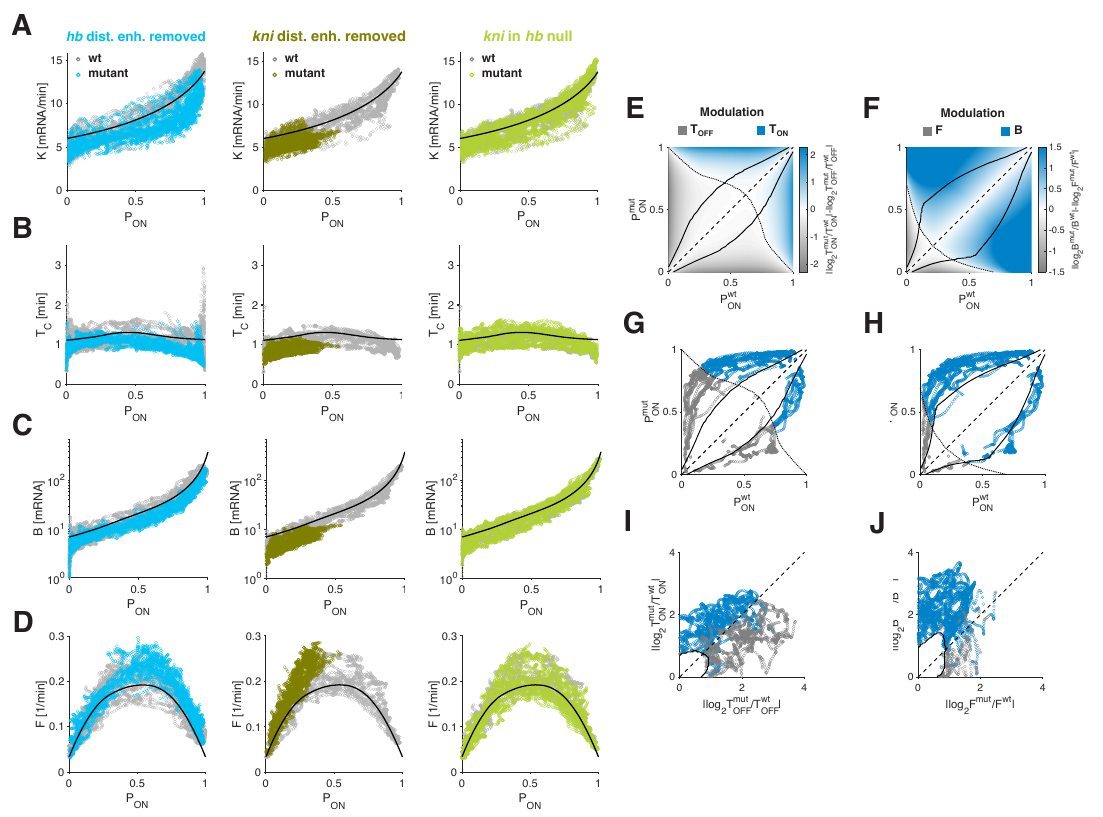}
\caption{{\bf Bursting relationships predict ON and OFF modulation by \emph{cis}- and \emph{trans}-perturbations.}
(A-D) Initiation rate $K$ (A), correlation time $T_{\rm C}$ (B), burst size $B$ (C), and burst frequency $F$ (D) for \emph{hb} in \emph{cis}-mutation (cyan), \emph{kni} in \emph{cis}-mutation (olive) and for \emph{kni} in \emph{trans}-mutation (light green) collapse on corresponding wild-type parameters (gray). Solid black lines correspond to the endogenous bursting relationships from Fig. \ref{fig5}F and \ref{figS11}E.
(E-J) Predicted $T_{\rm OFF}$ versus $T_{\rm ON}$, and $F$ versus $B$ bursting modulation for mutant (\emph{hb} in \emph{cis}-mutation and \emph{kni} in \emph{trans}-mutation) based on wild-type-derived relationships (Fig. \ref{fig5}F and \ref{figS11}E, black lines). Color code stands for type of modulation.
(E) The type of modulation is predicted by first approximating $T_{\rm OFF}$ and $T_{\rm ON}$ as a function of $P_{\rm ON}$ using the wild-type relationships. The predicted fold change in $T_{\rm OFF}$ ($T_{\rm OFF}^{\rm mut}/T_{\rm OFF}^{\rm wt}$) and $T_{\rm ON}$ ($T_{\rm ON}^{\rm mut}/T_{\rm ON}^{\rm wt}$) are then computed for all possible pairs of $P_{\rm ON}$ (i.e. $P_{\rm ON}^{\rm wt}$ and $P_{\rm ON}^{\rm mut}$). The dotted line delimits the regions where changes in transcription rate are either dominated by changes in $T_{\rm OFF}$ (gray region, $|\log{(T_{\rm OFF}^{\rm mut}/T_{\rm OFF}^{\rm wt})}| > |\log{(T_{\rm ON}^{\rm mut}/T_{\rm ON}^{\rm wt})}|)$ or $T_{\rm ON}$ (blue region, $|\log{(T_{\rm OFF}^{\rm mut}/T_{\rm OFF}^{\rm wt})}| < |\log{(T_{\rm ON}^{\rm mut}/T_{\rm ON}^{\rm wt})}|)$). The solid black lines delimit the region, where changes in $T_{\rm OFF}$ and $T_{\rm ON}$ are not significant given the “thickness” of our relationships ($95\%$ confidence intervals, see Methods). Thus, this procedure defined a look-up table enabling prediction of the type of modulation using pairs of $P_{\rm ON}$. (F) Same as (E), except using $F$ and $B$ instead of $T_{\rm OFF}$ and $T_{\rm ON}$.
(G) Scatter plot of all the $P_{\rm ON}$ pairs from \emph{hb} wt and \emph{cis}-mutant (at same spatiotemporal location). Colors correspond to the predicted modulation ($T_{\rm OFF}$ dominated in gray and $T_{\rm ON}$ dominated in blue) using the look-up table in A. (H) Same as (G), except using $F$ and $B$ instead of $T_{\rm OFF}$ and $T_{\rm ON}$.
(I) Verification of predicted modulation in G. For each $P_{\rm ON}$ pair, we computed the $T_{\rm OFF}$ ($T_{\rm OFF}^{\rm mut}/T_{\rm OFF}^{\rm wt}$) and $T_{\rm ON}$ ($T_{\rm ON}^{\rm mut}/T_{\rm ON}^{\rm wt}$) fold change using the estimated $T_{\rm OFF}$ and $T_{\rm ON}$ from data (Fig. \ref{fig4}C). Supporting our ability to predict the modulation, almost all the blue data points (predicted as $T_{\rm ON}$ modulation) are located above the slope 1 diagonal (dashed line), whereas most of the gray ones (predicted as $T_{\rm OFF}$ modulation) are below. Thus, for most data points ($>85\%$) the prediction is correct (Fig. \ref{fig4}E). (J) Same as (I), except using $F$ and $B$ instead of $T_{\rm OFF}$ and $T_{\rm ON}$. Supporting our ability to predict the modulation, almost all the blue data points (predicted as $B$ modulation) are located above the slope 1 diagonal (dashed line), whereas most of the gray ones (predicted as $F$ modulation) are below. Thus, for most data points ($>95\%$) the prediction is correct (Figure \ref{fig4}G).}
\label{figS15}
\end{figure*}

\begin{figure*}
\centering
\includegraphics[scale=1.0]{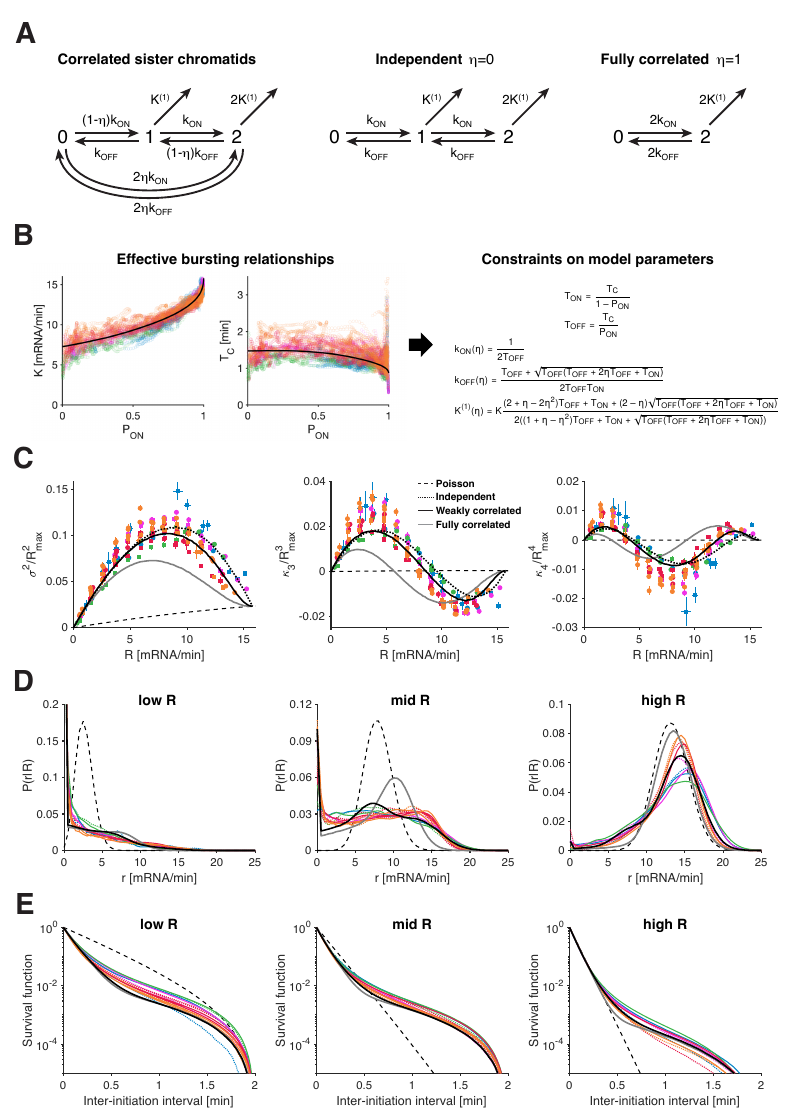}
\caption{{\bf Bursting relationships provide strong empirical constraints on transcription models.} [Caption see next page.]}
\label{figS16}
\end{figure*}

\newpage
\begin{figure*}
\centering
\begin{justify}
FIG. S16. {\bf Bursting relationships provide strong empirical constraints on transcription models.} This figure demonstrates how the derived bursting relationships help to discriminate between transcription models. Also shown how models that specifically account for two indistinguishable sister chromatids provide a better explanation for the data, suggesting that transcription loci in the embryo are formed by two weakly correlated sister chromatids.
(A) Minimal class of models describing two identical and possibly correlated sister chromatids. The gene copy on each sister chromatid behaves as a 2-state model whose transitions between the active and inactive state can be coupled ($\eta>0$). Assuming two identical gene copies, the model reduces to three effective states corresponding to 0, 1 and 2 active copies. Transition rates between the effective states are function of the single gene copy switching rates $k_{\rm ON}$ and $k_{\rm OFF}$. Production of mRNAs in the active states (`1' or `2') is determined by the single gene copy initiation rate $K^{(1)}$. The coupling parameter $\eta\in[0,1]$, together with $k_{\rm ON}$ and $k_{\rm OFF}$, determine the correlation coefficient $\rho\in[0,1]$ between the two bursting gene copies, which is given by $\rho=\frac{P(2)-(P(2)+P(1)/2)^2}{(P(2)+P(1)/2)(1-P(2)-P(1)/2)}$ with $P(1)$ and $P(2)$ the occupancy probability of states `1' and `2'. In the limiting case $\eta=0$ the two gene copies are independent and $\rho=0$, whereas in the case $\eta=1$ the two gene copies are fully correlated and $\rho=1$.
(B) The identified bursting relationships (derived without fitting) impose strong constraints on possible models. Any model of transcription (characterized by a set of active and inactive states) must satisfy the relationships between the effective parameters $K$, $T_{\rm C}$ and $P_{\rm ON}$ (left, data from NC13 + early NC14, see Fig. \ref{figS12}A). Using the class of models in A, we can map the effective parameters onto the single copy parameters by satisfying the following equations: $T_{\rm OFF} = T_0$, $T_{\rm ON} = T_{1|2}$ and $K P_{\rm ON} = K^{(1)}P(1)+2K^{(1)}P(2)$. Doing so, we find expressions for $k_{\rm ON}$, $k_{\rm OFF}$ and $K^{(1)}$, as functions of the sole free parameter $\eta$. We have thus reduced a four-parameters model into a single-parameter one, which can easily be tested against data (see C,D and E).
(C) Variance ($2^{\text nd}$ cumulant), $3^{\text rd}$ cumulant and $4^{\text th}$ cumulant of single allele transcription rate $r$ as a function of mean transcription rate $R$ in NC13 (square) and early NC14 ($7.5\leq t<20.5$ min; circle). The single allele transcription rates are estimated within 2-min-intervals and the cumulants are normalized by $\max R = R_{\rm max}=15.77$ mRNA/min. Data color code as in Fig. \ref{figS3}. The dashed line corresponds to the Poisson limit (i.e., a single gene copy that is constitutively transcribed), whereas the dotted and solid curves are predictions made by the models in A that satisfy the derived bursting relationships in B. The dotted curve corresponds to independent sister chromatids ($\rho=0$), the solid black line to slightly correlated sister chromatids ($\rho=0.2$) and the solid gray line to fully correlated sister chromatids ($\rho=1$). We immediately see that not accounting for pair of sister chromatids, or equivalently only considering highly correlated pair $\rho\sim1$, provides a poor explanation for the data. On the contrary, models that include small correlations between chromatids $\rho\sim0.2$ provide a good match to the data. The predicted cumulants are computed by sampling from each model to account for the Pol II footprint estimated around 60bp.
(D) Distribution $P(r|R)$ of single allele transcription rates estimated within 2-min-intervals in both NC13 (color dotted lines) and early NC14 (color solid lines). These distributions are computed over all the nuclei from time points and AP bins whose mean transcription rate $R$ corresponds either to a low $[2.1,3.2]$, mid $[7.5,8.5]$ or high transcription level $[12.8,13.9]$ (as in Fig. \ref{figS3}). Black dashed line corresponds to the Poisson distribution for a single constitutive gene. Solid black and gray lines are predicted distributions by the models as in C (slightly correlated and fully correlated sister chromatids, respectively). While the model with fully correlated sister chromatids fails to account for the empirical distributions at mid and high $R$, a small amount of correlation $\rho$ leads to a good match.
(E) Survival function of elapsed time between successive Pol II initiation events (inter-initiation intervals) estimated within 2-min-intervals in both NC13 (color dotted lines) and early NC14 (color solid lines). These survival functions (1-``the cumulative distribution'') are computed over the same nuclei, time points and AP bins as in D. The black dashed line corresponds to the Poisson for a single constitutive gene. Solid black and gray lines are predicted survival functions by the models as in C and D. A model describing a pair of slightly correlated bursting sister chromatids generate survival functions that mimic closely the data on a 2-min-interval (over which the system can be considered roughly stationary, i.e. $T_{\rm C}\sim1.5$ min)
\end{justify}
\end{figure*}

\begin{figure*}
\centering
\includegraphics[scale=1.0]{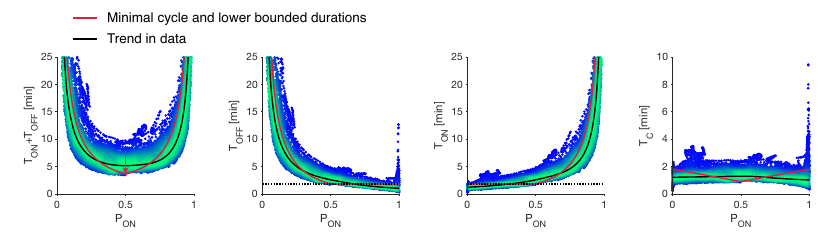}
\caption{{\bf A bursting regime with a minimal ON-OFF cycle recapitulated the data.}
An ON-OFF transcriptional regime, with a lower limit on these period durations and a minimal mean cycle duration $T_{\rm ON}+T_{\rm OFF}=1/F$ shown in red, closely recapitulates the observed bursting relationships in black. The horizontal dotted line corresponds to a lower bound of 1.8 min providing the best-fit to the data. An allele at mid-activity level ($P_{\rm ON}=0.5$) minimizes $T_{\rm ON}+T_{\rm OFF}$. Various ON-OFF combinations can yield $P_{\rm ON}>0.5$, but increasing OFF periods prolong encoding $P_{\rm ON}$ compared to solely increasing ON durations.
}
\label{figS17}
\end{figure*}

\clearpage
\section*{Supplemental Videos}

\begin{figure*}[h!]
\centering
\begin{justify}
Video V1-V4. Representative videos of 4 gap gene transcription rate measurements: for \emph{hb}, \emph{Kr}, \emph{kni}, and \emph{gt} (female), respectively. Anterior is on the left. We measured the dorsal side of the embryos. Red channel shows nuclei marked by Histone-RFP label; green channel shows the MCP-GFP signal, in particular highlighting one site of nascent transcription in each nucleus when it binds to MS2 stem-loops (see Methods). Each video was mean projected in the axial (z) direction. Projected image size is about 210x120 $\mu\text{m}^2$. Timestamp is in units of minutes. The video display frame rate is 15 Hz, which is equivalent to 2.5 min of imaging time per display second. Video contrast was adjusted for better visualization.
\end{justify}
\end{figure*}

\begin{figure*}[h!]
\centering
\begin{justify}
Video V5. The temporal progression of transcriptional activity along developmental time. X-axis: normalized embryo length (0-1). (Top) The mean activity directly from the MS2 signal. (Bottom) Calculated mean transcription rate (Methods Section 5.2) that is no longer gene length dependent.
\end{justify}
\end{figure*}

\begin{figure*}[h!]
\centering
\begin{justify}
Video V6. The temporal progression of relative protein abundance along developmental time. (Top) Protein accumulation predicted from measured mean transcription rate based on a simple model. (Bottom) Previously measured protein patterns from carefully staged gap gene antibody staining \cite{Dubuis2013}. Major differences observed for \emph{hb} are likely related to maternal mRNAs, whose contribution is not observed in our live measurements reflecting only zygotic gene expression.
\end{justify}
\end{figure*}

\begin{figure*}[h!]
\centering
\begin{justify}
Video V7. The mean transcription rate for wildtype \emph{hb}-MS2 (dark blue) and its distal enhancer deletion mutant (light blue).
\end{justify}
\end{figure*}

\begin{figure*}[h!]
\centering
\begin{justify}
Video V8. The mean transcription rate for wildtype \emph{kni}-MS2 (green) and its distal enhancer deletion mutant (dark yellow-green).
\end{justify}
\end{figure*}

\begin{figure*}[h!]
\centering
\begin{justify}
Video V9. The mean transcription rate for wildtype \emph{kni}-MS2 (green) and its expression in the background of \emph{hb} null mutant (light yellow-green).
\end{justify}
\end{figure*}

\clearpage
\section*{Methods}

\section{Fly strains and genetics}

\subsection{Plasmid construction}
MS2 and PP7 stem-loops cassettes were produced by a series of cloning steps, duplicating the annealed oligos below. The final cassette consists of 24 stem-loops (12 repetitions of the initial annealed oligos) 
\begin{itemize}
    \item MS1 oligo 1: CTAGTTACGGTACTTATTGCCAAGAAAGCACGAGCATCAGCCGTGCCTCC\\AGGTCGAATCTTCAAACGACGACGATCACGCGTCGCTCCAGTATTCCAGGGTTCATCC
    \item MS2 oligo 2: CTAGGGATGAACCCTGGAATACTGGAGCGACGCGTGATCGTCGTCGTTTG\\AAGATTCGACCTGGAGGCACGGCTGATGCTCGTGCTTTCTTGGCAATAAGTACCGTAA
    \item PP7 oligo 1: CTAGTTACGGTACTTATTGCCAAGAAAGCACGAGACGATATGGCGTCCGT\\GCCTCCAGGTCGAATCTTCAAACGACGAGAGGATATGGCCTCCGTCGCTCCAGTATTC\\CAGGGTTCATCC 
    \item PP7 oligo 2: CTAGGGATGAACCCTGGAATACTGGAGCGACGGAGGCCATATCCTCTCGT\\CGTTTGAAGATTCGACCTGGAGGCACGGACGCCATATCGTCTCGTGCTTTCTTGGCAA\\TAAGTACCGTAA
\end{itemize}
All 2attP-dsRed plasmids were made by cloning homology arms into a previously used 2attp-dsRed plasmid \cite{Rogers:2017}. All 2attB-insert plasmid were made by cloning the inserts into a previously used 2attB-insert plasmid \cite{Chen:2018bk}. Plasmid maps and cloning details are available upon request.

\subsection{Transgenic fly generation}
For the endogenous tagging of the gap genes (\textit{hb}, \textit{kni}, \textit{Kr}, and \textit{gt}) a two-step transgenic strategy was used. First, a CRISPR-mediated replacement of each locus was performed. Specifically, the upstream regulatory regions (including annotated enhancers) and coding regions were replaced by a 2attp-dsRed cassette. This CRISPR step was performed with the following guides:
\begin{center}
    \begin{tabular}{ |c|l|l|l| } 
    \hline
    Gene & Guide1 & Guide2 & Stem loops insertion coordinates (in dm6) \\ 
    \hline
    \textit{kni} & CTTGAAGCTCAT & GGGAGGGCTTGA & Intronic (3L:20694142) \\ 
                      & TAATTCCACGG & TTCGGGAAAGG & \\
    \hline
    \textit{hb}  & ATGAACACTCAT & GTCACGGCTAAG & Intronic (3R:8694188), 3utr (3R:8691669*)\\ 
                      & ACATATCCTGG & ACGCCTTAAGG & \\
    \hline
    \textit{gt}  & TCTTACGTGTAA & CGGCCGGCGAGG & Intronic (X:2428789)\\
                      & GAATTCATGGG & AAGTGAACGGG  & \\
    \hline
    \textit{Kr}  & GTAAATCCCAGA & AAGACTTGAACC & Intronic (2R:25227036), 3utr (2R:25228873)\\
                      & TGTATAATTGG & AAATACACAGG & \\
    \hline
    \end{tabular}
\end{center}
* An additional $\sim\!1.5\,$kb fragment of the gene yellow was placed downstream of the stem-loop cassette to insure sufficient length for signal detection.

Homology arms were amplified from genomic DNA of the \textit{nos-Cas9/CyO} injection line (BDSC \#78781). For \textit{kni} and \textit{gt}, loss of gap gene proteins was verified by antibody staining as previously described \cite{Dubuis2013}. For all four genes segmentation defects previously ascribed to the loss of protein were observed. PCR verification was performed (i.e., from the dsRed to the flanking genomic regions). These lines are referred to as a gap gene null line (e.g., \textit{hb}-null).

In a second step, the deleted region for each gap gene was PCR amplified from the \textit{nos-Cas9/CyO} line and cloned into a 2attB plasmid. MS2 stem-loops (see description above) were cloned into the gene (see insertion position in the above table and fig. S1A). This 2attB-insert was subsequently delivered into the 2attp site of the corresponding gap gene line from step one by co-injection with phiC31 integrase (RMCE injection with $\sim\!0.25\mu g/\mu l$ [DNA] and hsp-PhiC31 DNA $\sim\!0.1\mu g/\mu l$). Flies were screened for loss of dsRed and PCR verified for the presence of the insert in the correct orientation, with primers from inside the insert to the flanking genomic regions. 

In addition to MS2 lines for each gap gene, two lines with dual (orthogonal) stem-loop systems (MS2 and PP7) on the same gene were produced. They were primarily used for elongation measurements (see Section \ref{sec:singler_estimate} and fig. S2C-G). We generated 1) a \textit{hb} line with intronic insertion of MS2 and a 3'utr insertion of PP7, 2) a \textit{Kr} line with intronic insertion of PP7 and 3'utr insertion of MS2 (insertion positions are as in the single stem-loop lines, see table above). An additional control line with a cassette of alternating MS2 and PP7 stem-loops \cite{Chen:2018bk} in the intronic \textit{Kr} gene was used to assess imaging error (fig. S1D-E).

A \textit{hb}-MS2 fly line was generated without the distal \textit{hb} enhancer. To this end, the 2attB-insert included the deleted region from the \textit{hb} locus, the MS2 (intronic) insertion, and a replacement of the distal enhancer (3R:8698553-8700369, dm6) by a fragment of the same length with the \textit{lacZ} gene. The modified 2attB-insert was delivered into the \textit{hb}-2attp site as described above. Plasmid maps and cloning details are available upon request.

\subsection{Genetic crosses for live imaging}
A fly line with the fluorophore (\textit{yw; His2Av-mRFP; nanos$>$MCP-eGFP}) \cite{Garcia2013} was crossed with wild-type (Ore-R) flies to reduce the background fluorescence. Female offspring (\textit{yw/+; His2Av-mRFP/+; nanos$>$MCP-eGFP/+}) was subsequently crossed with our CRISPR-MS2 transgenic fly lines. 

\paragraph{Male \textit{gt}-MS2 crossing scheme:}
Since \textit{gt} is on the X-chromosome, the above crossing scheme only works for female \textit{gt-MS2} embryos. For male embryos, we followed a different crossing strategy, where Female \textit{gt-MS2} flies are crossed with a 3-color fluorophore line (\textit{yw; His2Av-mRFP; nanos$>$MCP-eGFP,nanos$>$PCP-mCherry}). The female offspring of that cross (\textit{gt-MS2/+; His2Av-mRFP/+;\\ nanos$>$MCP-eGFP,nanos$>$PCP-mCherry/+}) was then crossed with (\textit{gt-MS2PP7-interlaced}) males. To ensure we measure male embryos, imaging was performed only with embryos expressing \textit{gt-MS2} signal but not PP7-PCP-mCherry signal.

\paragraph{\textit{kni-MS2} in \textit{hb-null} background crossing scheme.}
Since \textit{kni} and \textit{hb} are both on Chromosome-III, we adopted a two-generation crossing scheme for this line. The first generation had two sets of crosses. A first fly line (\textit{hb-null/Tm3sb}) was crossed with \textit{hb-3’UTR-MS2}. A second fly line (\textit{hb-null,kni-MS2/Tm3sb}) was crossed with a dual fluorophore line (\textit{yw; His2Av-mRFP; nanos$>$MCP-eGFP}). Male offspring from the first cross (\textit{hb-null}/\textit{hb-3’UTR-MS2}, selected against Tm3sb) was then crossed with female offspring from the second (\textit{His2Av-mRFP/+; nanos$>$MCP-eGFP/hb-null,kni-MS2}, selected against Tm3sb). Imaging was performed with embryos that had \textit{kni-MS2} expression but without the \textit{hb-3UTR-MS2} signal (spatially distinguishable within embryos). This ensured the measured embryos had a genotype of \textit{hb-null,kni-MS2/hb-null} on Chromosome-III.
    
\section{Live imaging}

\subsection{Sample preparation}

Sample preparation for live imaging is adapted from previous work \cite{Garcia2013,Bothma:2018io}. We prepare an air-permeable membrane (roughly 2 cm by 2 cm) on a sample mounting slide. Heptane glue is evenly distributed on the membrane. Since we record embryos starting from late nuclear cycle 12, the flies are caged on agar plates for 2 to 2.5 hours. Embryos laid on these plates in that time window are transferred to a piece of double-sided tape by a dissection needle with which the embryos are also hand-dechorionated and placed on the glued membrane. After mounting, we immerse embryos in Halocarbon 27 oil (Sigma) and compress the embryos with a cover glass (Corning \#1 1/2, 18x18 mm). Excess oil is removed by a tissue if needed. All embryos are mounted dorsal up (dorsal side facing objective lens).

\subsection{Imaging settings}

Live embryo imaging is performed from late NC12 to the end of NC14 (onset of gastrulation) using a custom-built inverted two-photon laser scanning microscope (similar in design to previous studies \cite{Gregor:2007ce,Liu:2013gv}). The laser sources for two-photon excitation are a Chameleon Ultra (wavelength at 920 nm, for channel 1) and a HighQ-2 from Spectra Physics (wavelength at 1045 nm, for channel 2). Excitation and emission photons are focused and collected by a 40x oil-immersion objective lens (1.3NA, Nikon Plan Fluor). The average laser power measured at the objective back aperture are 20 mW and 4 mW, for Ultra and HighQ-2 lasers, respectively. In dual color Imaging Error Experiments (Section 4.4) 20 mW was used for both channels. Laser scanning and image acquisition are controlled by ScanImage 5.6-1 (Vidrio Technologies, LLC). Fluorescence signal from both channels is simultaneously detected by separated GaAsP-PMTs (both Hamamatsu H10770P-40). The pixel size is 220 nm, with an image size of 960x540 pixels. Each image stack contains 12 frames, separated by 1 $\mu$m in the axial z-direction. Pixel dwell time is 1.4 $\mu$s. The overall temporal resolution for one image stack is around 10 s.

\section{Image processing}


\subsection{Nuclei tracking}

Nuclei tracking was performed on a specifically dedicated image acquisition channel with red fluorophores, i.e. red-fluorescently labeled histone proteins (His2Av-mRFP) that are provided maternally in all imaged embryos. Typical imaging windows span from late NC12 to late NC14 during embryonic development, when the embryos undergo two rounds of nuclear division. As during this period, the size of the nuclei changes continuously (from $\sim\!12\,\mu$m to $\sim\!7\,\mu$m in diameter), we use adaptive parameters for nuclei filtering and identification. First, as a pre-filter, a simple 2D Gaussian filter was applied to the Z-projections of the image stack of the red nuclei channel. This procedure roughly captures the average nuclear size of the population at each time point. Second, this time-dependent mean nuclear size is used to construct dynamic parameters for a 3D difference of Gaussian (DOG) filter to smooth the nuclei images, whose spherical shape is enhanced with a disk filter. For each 3D embryo image stack (per time point), filtered images are normalized to the maximum pixel intensity and subsequently binarized and segmented using a watershed algorithm. From each of these 3D segmented image stacks we constructed a 2D Voronoi diagram. For consecutive time points, the identity for all nuclei is propagated based on the shortest pair-wise distance of centroids and the largest overlapping fraction of Voronoi surfaces.

\subsection{Transcription spot tracking and signal integration}

To detect low-intensity signal-to-noise ratio (SNR) spots, our detection algorithm is tuned to high sensitivity at the cost of increasing the false positive rate when no obvious bright object throughout the detection region is present. Thus, we could find multiple spot candidates per nucleus at a given time point. However, in all single-allele-labeling experiments, we expect at most one spot per nucleus at all times. With this knowledge, a dedicated custom detection and tracking (modified Viterbi) algorithm identifies and tracks the real low SNR transcription spots based on the spot candidates.

\paragraph{Detecting spot candidates and transcription trajectories.} Identification of transcription spots is simplified by dividing the original images into 3D image stacks containing individual nuclei. The transcription channel for each such stack, tracked across time, is filtered for salt-and-pepper noise with a 3D median filter and subsequently, a 3D difference of Gaussian (DOG) filter is applied to detect round, concave objects. These objects are initial candidates for identifying transcription active sites. If consecutive time points present multiple spot candidates, we determined the most likely association using the pair-wise distance between potential pairs and a diffusion-based potential. The latter is constructed iteratively from the random walk of all transcription spots across all time points. This potential is penalizing for less likely displacements. This procedure identifies the most likely trajectory for each nucleus.

\paragraph{Signal integration and transcription time-series.} We calculated the mean pixel intensity of a 3D ellipsoid volume centered at the centroid coordinates of each transcription spot as identified by the spot-tracking procedure. The ellipsoid volume has diameters of (X, Y, Z)=(7, 7, 5) pixels, determined empirically based on the objective point spread function and the image pixel size (see image settings). Similarly, the mean local fluorescence background was estimated by averaging the intensities of all pixels in a larger 3D ellipsoid (i.e. (X, Y, Z)=(19, 19, 7) pixels, also centered at the spot centroid), except for the pixels of the inner spot ellipsoid. The fluorescence intensity of transcription spots is then quantified by subtracting this mean local background value from the mean spot intensity. This procedure is performed to avoid intensity contributions of freely floating MCP-GFP molecules within the nuclei from the transcription spot intensities. Our detection and tracking pipeline results in background-subtracted intensity time series, each corresponding to the real-time transcriptional activity of a single labeled allele within a nucleus.

\paragraph{AP position assignment.} \label{sec:ap_determination} The anterior-posterior axis (AP) was determined based on a mid-sagittal image of the full embryo in the histone-RFP channel. The tips of the long axis of the full embryo set the positions of both the anterior and posterior poles. Both poles are routinely checked manually. The XY coordinates of the nuclei identified in the surface images are registered onto the mid-sagittal image to determine their AP coordinates in that reference frame, relative to the poles.

\section{Data processing, calibration, and measurement errors}

\subsection{Spatiotemporal alignment of embryos}
\label{sec:alignment}

All data sets (consisting of multiple embryos) are realigned in time (with respect to mitotic events) and space to minimize inter-embryo variabilities. These are stemming from differences in the speed of developmental progression (mainly due to temperature fluctuations at the sample) and from misalignment in space resulting from the AP-axis determination. 

\paragraph{Developmental time normalization and temporal alignment.} To determine the speed of developmental progression, we compute for each nucleus in each embryo the time between mitosis 12 and 13, i.e., the duration of nuclear cycle 13 (NC13) $t_{13}$. $t_{13}$ is thus the time for individual nuclei between two mitotic events (that way accounting for the well-documented mitotic waves that propagate along the AP-axis \cite{Hayden:2022}. It is identified when our nucleus-segmentation algorithm detects two separate entities (calling error $\sim\!10\,$s). The embryo-intrinsic developmental time is determined by estimating the average duration of NC13 over all nuclei of a given embryo, $\avg{t_{13}}=T_{13}$. The global mean measured over $n=123$ embryos is $\avg{T_{13}}=18.4\pm1.2\,$min, corresponding to $6.5\%$ variability at $24\pm1$$^{\circ}$C. 

While most of our transcription measurements are performed during NC14, the mean duration of NC14 $T_{14}$ is difficult to measure as the end of the cycle occurs beyond gastrulation and the completion of the cycle is asynchronous among cells. Even the onset of gastrulation (cells no longer form a monolayer) is difficult to time precisely due to the way our imaging is performed: specimen labeling is optimized to measure transcription, which does not allow for optimal determination of the onset of gastrulation. However strong correlations of the durations of earlier cycles ($T_{11}$, $T_{12}$ and $T_{13}$, Pearson correlation $\rho_{11-12}=0.75$ and $\rho_{12-13}=0.81$) suggest that $T_{13}$ serves as a good proxy for the speed of developmental progression. Thus each embryo is aligned in time to match this unified speed. Temporal alignment is achieved in each embryo by normalizing the sampling time $\Delta t$ for both NC13 and NC14 by $T_{13}/\avg{T_{13}}$, leading to $\Delta t'=\Delta t \avg{T_{13}}/T_{13}$. Consequently, all embryos are aligned with respect to the onset and the end of NC13 (and thus the onset of NC14), making further translational temporal shifts between embryos unnecessary. The resulting temporal alignment typically provides a reduction in embryo-to-embryo variability (defined as the variance across mean embryo transcriptional activity) by a factor of two at 15 min into NC13 (near the end of NC13).

\paragraph{Spatial alignment.} All nuclei are projected on each embryo's internally determined AP-axis. However, axis determination is error-prone due to image analysis constraints, small changes in the azimuthal angle of the embryo with respect to the optical axis, and small tissue deformation due to the compression from the coverslip. We thus adjusted the spatial alignment for each embryo by correcting for possible errors stemming from these technical constraints. Indeed, the embryo elliptic mask fit (see Section \ref{sec:ap_determination}; AP position assignment) together with small deformation of the embryo surface and natural variability typically leads to a positional error of $\sim\!1\%$ egg length (corresponding to approximately one cell diameter, which is statistically significant to our subsequent analyses, see also \cite{Zoller:2018gj}). Spatial alignment is achieved directly on the image stack that measures gene activity. The mean and variance of gene activity are measured across nuclei for each embryo as a function of time and space within AP bins (bin sizes of 2.5\% and 1.5\% egg length in NC13 and NC14, respectively). For each embryo $i$ and each nuclear cycle, we thus obtain an activity surface (kymograph) for the mean $\mu_i (x,t)$ and variance $\sigma_i^2 (x,t)$ as a function of AP position $x$ and time $t$ with respect to mitosis. We then introduce new coordinates for each embryo that account for spatial shift $\Delta x$, spatial dilatation $\alpha$, and temporal dilatation $\beta$, i.e., $x'=\Delta x+\alpha(x-\avg{x})+\avg{x}$ and $t'=\beta t$, such that $\tilde{\mu}_i(x',t')=\mu_i (x,t)$ and $\tilde{\sigma}_i^2 (x',t' )=\sigma_i^2 (x,t)$. Next, we define least squares sums based on the shifted and dilated surfaces $\tilde{\mu}_i(x,t)$ and $\tilde{\sigma}_i^2 (x,t )$:
\begin{align*}
\chi_1(x,t) &= \sum_{i=1}^{N_e} (\tilde{\mu}_i(x,t)-\avg{\tilde{\mu}_i(x,t)})^2, \\
\chi_2(x,t) &= \sum_{i=1}^{N_e} (\tilde{\sigma}^2_i(x,t)-\avg{\tilde{\sigma}^2_i(x,t)})^2, \\
\chi(\{ \Delta x_i, \alpha_i, \beta_i \}) &= \int \int (\chi_1(x,t) + \sqrt{\chi_2(x,t)}) dx dt,
\end{align*}
where $N_e$ is the total number of embryos for a given gene.

We then minimized $\chi(\{ \Delta x_i, \alpha_i, \beta_i \})$ to learn $\Delta x_i$, $\alpha_i$ and $\beta_i$ for each embryo, under the following constraints $\sum_i \Delta x_i=0$, $\sum_i \log{\alpha_i}=0$ and $\sum_i \log{\beta_i}=0$. These constraints ensure that on average (over embryos), position and time remain the same, i.e., $\avg{x_i}=0$, $\avg{\alpha_i}=1$ and $\avg{\beta_i}=1$. When learning the new alignment in NC13, no time dilation was needed, and $\beta_i=0$, $\forall i$. The $\beta_i$ were only learned for NC14 since the absence of well-defined events for the exact onset of gastrulation made the temporal alignment harder in that case. In the end, the standard deviation (std) for $\{\Delta x_i \}$ is 1.2\% (NC13) and 0.8\% (NC14) egg length, the std for ${\alpha_i}$ is 8\% (NC13) and 6\% (NC14), and the std for ${\beta_i}$ is 5\% (NC14).

\subsection{Embryo-to-embryo variability and pooling}
\label{sec:pooling}

To increase our allele sample size in each condition, we use a pooling strategy, effectively pooling alleles from multiple spatially and temporally aligned embryos (see Section \ref{sec:alignment}). This is facilitated by the high reproducibility of the \emph{Drosophila} embryos, which intrinsically exhibit low embryo-to-embryo variability.

\paragraph{Embryo-to-embryo variability.} We characterize embryo-to-embryo variability on the measured transcriptional activity across embryos. As we have done for aligning embryos, we compute the mean and variance of gene activity across nuclei for each embryo as a function of time and space within AP bins (bin sizes of 2.5\% and 1.5\% egg length in NC13 and NC14, respectively). We thus obtain for each embryo $i$ and each nuclear cycle the mean $\mu_i(x,t)$ and variance $\sigma_i^2(x,t)$ as a function of AP position $x$ and time $t$ with respect to mitosis. We define embryo-to-embryo variability as the variance of the mean $\sigma_{\rm emb}^2$ across $N_e$ embryos, i.e.,
\begin{equation}
\sigma_{\rm emb}^2(x,t) = \sum_{i=1}^{N_e} w_i(x,t) \left( \mu_i(x,t) - \mu(x,t) \right)^2,
\end{equation}
where the weight are given by $w_i(x,t)=1/N_i(x,t)$ with $N_i(x,t)$ the number of alleles at time $t$ in the AP-bin centered at position $x$ for embryo $i$, and the global mean $\mu(x,t)=\sum_{i=1}^{N_e} w_i(x,t) \mu_i(x,t)$. Applying the law of total variance, the total variance in the measured transcriptional activity $\sigma^2$ is given by
\begin{equation}
\sigma^2(x,t) = \sigma_{\rm emb}^2(x,t) + \sum_{i=1}^{N_e} w_i(x,t) \sigma_i^2(x,t),
\end{equation}
where the last term corresponds to intra-embryo variability, i.e. variability across nuclei (plus a small contribution coming from measurement error, see Section \ref{sec:meas_error}). To determine the contribution of the embryo-to-embryo variability on the measured total variance, we compute the ratio $\sigma_{\rm emb}^2(x,t)/\sigma^2(x,t) \leq 1$ for all gap genes in NC13 and NC14. As it has been previously measured in fixed embryos \cite{Zoller:2018gj}, the embryo-to-embryo variability remains small compared to the total variance throughout developmental time, overall averaging to less than 10\% of the total variance (see fig. S1F for specific time points).

\paragraph{Pooling nuclei across embryos.} Having shown that embryo-to-embryo variability represents only a small fraction ($\sim10$\%) of the total variability, we can treat (to a good approximation) individual embryos as independent samples drawn from the same underlying distribution. Thus, to characterize the fluctuations in transcriptional activity for a given gene, we pool together all the nuclei (at the same time and position along the AP axis) from all measured embryos (aligned in space and time beforehand, see Section \ref{sec:alignment}) generating a large sample of nuclei ($\sim200$ nuclei). Specifically, once the embryos are spatiotemporally aligned, we define common spatial bins along the AP axis (bin sizes of 2.5\% and 1.5\% egg length in NC13 and NC14, respectively) allowing us to pool nuclei at same position and at each time point. Pooling provides a large gain in sample size (by a factor $N_e=$10--20 embryos), facilitating the estimation of the mean profiles and higher moments of the alleles' transcriptional activity (Section \ref{sec:calibration}), as well as the estimation of the bursting parameters (Section \ref{sec:analyis_modeling}). Moreover, embryo pooling is justified as 1) embryo-to-embryo variability is almost negligible, and 2) we focus explicitly
 on variability across nuclei, which mostly reflects transcription dynamics and the underlying bursts (provided the measurement noise is small). Hence we assume that for a given AP bin nuclei sample the same distribution function in each embryo. 

\subsection{Calibration of absolute units in live imaging data}
\label{sec:calibration}

We calibrate our live measurements to absolute units, which are defined as cytoplasmic units (C.U.) and correspond to equivalent single mRNA molecule counts at the site of transcription. The calibration is achieved by matching mean activity profiles in NC13 from live and previously calibrated fixed smFISH measurements \cite{Zoller:2018gj}. Here we explain this procedure in detail.

\paragraph{Expected live activity.} First, we use calibrated smFISH measurements of nascent transcriptional activity (in C.U.) for all trunk gap genes in late NC13 ($10-15\,$min into NC13) to compute the expected live activity. From these calibrated fixed measurements, we compute mean activity profiles for each gene by pooling the nuclei from multiple embryos ($n=10-20$) within AP bins of $2.5\%$ egg length ($\sim\!2$ nuclei wide region along the AP axis) (see Zoller et al. \cite{Zoller:2018gj}, Fig. 1C). The resulting smFISH mean activity profiles $\bar{A}_F(x)$ predict the expected levels of live mean activity $\bar{A}_L(x)$ (in C.U.) as a function of position $x$ along the AP axis. We compute the expected $\bar{A}_L$ by adjusting $\bar{A}_F$ for differences in gene copy number ($n_F=2$ alleles imaged in fixed vs $n_L=1$ allele in live), gene length (endogenous length in fixed versus an additional $\sim\!1.5\,$kb length due to the insertion of the MS2 cassette in live measurements), and the specificity of targeted locations by FISH probes and stem-loops. We thus get 
\begin{equation*}
\bar{A}_L=\frac{n_L \tilde{L}_L}{n_F \tilde{L}_F} \bar{A}_F,
\end{equation*}
where $\tilde{L}_F$ and $\tilde{L}_L$ are effective gene lengths that account for the effect of probe locations on the signal.

\paragraph{Computing effective lengths.} Second, we compute the effective gene length $\tilde{L}$ for all constructs. Knowing the exact location of the binding sequences for the fluorescent smFISH probes (fixed) and of the MS2 sequences (live) along the gene, we calculate the relative contribution of a single nascent transcript (compared to a fully tagged transcript corresponding to one C.U.) to the signal $s(l)$ as a function of its length $l$:
\begin{equation}
s(l) = \frac{1}{m}\sum_{i=1}^m H(l-l_i),
\label{equ:transcript_signal}
\end{equation}
where $H$ is the (Heaviside) unit step function, $l_i$ the end position of the $i^\text{th}$ probe binding sequences and $m$ the total number of probes. Integrating $s(l)$ over the gene length $L_g$ (from TSS to polyA site) leads to the effective length:
\begin{equation}
\tilde{L}=\int_0^{L_g} s(l) dl = \frac{1}{m}\sum_{i=1}^m (L_g-l_i) = L_g-\avg{l_i}.
\label{equ:effective_length}
\end{equation}
The resulting physical gene length $L_g$ and the effective gene length $\tilde{L}$ are given in Table \ref{tab:gene_length}. The physical length $L_g$ is slightly longer than the annotated gene length, as it accounts for a small retention time ($\sim\!30\,$s) of nascent transcripts at the transcription site, due to termination and possibly Pol II running further than the annotated 3'UTR. We have previously estimated the equivalent quantity of this retention time as an extra transcribed length in dual color smFISH measurements \cite{Zoller:2018gj}.

\begin{table}
\begin{center}
 \begin{tabular}{| c || c | c | c |} 
 \hline
 Gene & Annotated length [bp] & Physical length $L_g$ [bp] & Effective length $\tilde{L}$ [bp] \\
 \hline\hline
 \emph{hb}-intron & 4464 & 4464 & 3851 \\ 
 \hline
 \emph{hb}-3utr & 3516 & 3516 & 2903 \\ 
 \hline
 \emph{Kr}-intron & 3765 & 4565 & 3952 \\ 
 \hline
 \emph{Kr}-3utr & 1915 & 2715 & 2102 \\ 
 \hline
 \emph{kni} & 3049 & 3849 & 3236 \\ 
 \hline
 \emph{gt} & 2940 & 3740 & 3127 \\ 
 \hline
\end{tabular}
\caption{\label{tab:gene_length} Gene length of the measured gap genes. The annotated gene length was obtained from UCSC Genome Browser (BDGP Release 6 + ISO1 MT/dm6) Assembly and  it includes the length of the inserted MS2 cassette. The physical gene length is obtained by further adding 800bp to all genes except \emph{hb}, and it accounts for the retention of elongated nascent transcripts at the sites. This extra length was estimated from dual color smFISH measurements \cite{Zoller:2018gj}. The effective gene length is calculated from the physical gene length according to Equ. \ref{equ:effective_length}.}
\end{center}
\end{table}

\paragraph{Estimating conversion factor for profile matching.} Third, we compute the non-calibrated live mean activity profiles $\bar{S}(x)$ for all constructs by averaging the single-allele activity time series over all nuclei in a given $2.5\%$ AP bin over a $10-15\,$min time window in NC13, effectively pooling together nuclei from $n=10-20$ embryos. The conversion factor $\nu$ (allowing us to express $S$ in absolute units) is estimated by defining a pseudo-likelihood function for a given gene $g$:
\begin{equation*}
\mathcal{L}_g(\bar{S}|\nu,\Delta x) = \prod_{i=1}^{N_x} \mathcal{N}(\nu \bar{S}(x_i) | \mu=\bar{A}_L(x_i- \Delta x), \sigma^2=\sigma_A^2(x_i- \Delta x)+\nu^2 \sigma_S^2(x_i)),
\end{equation*}
where $\mathcal{N}$ is the normal probability density function (i.e. a Gaussian), $N_x$ the number of positions along the profiles, and $\Delta x$ a possible small shift in space registration (typically $<1\%$ egg length) between fixed and live experiments. $\sigma_A$ and $\sigma_S$ are standard errors of the mean based on the smFISH prediction and live averaging, respectively. Typically, $\nu$ accounts for the different imaging parameters (such as quantum yield and PMT amplification, etc.) between the fixed (confocal microscopy) and live (two-photon microscopy) experiments. The inference of the factor $\nu$ was performed globally over all genes at once, i.e., by maximizing
\begin{equation*}
\mathcal{L}(\{\bar{S}_g\}|\nu,\{\Delta x_g\})=\prod_g \mathcal{L}_g(\bar{S}_g|\nu,\Delta x_g),
\end{equation*}
leading to a single $\nu$ that was used to define our global calibration unit $1/\nu$. We also inferred $\nu$ per individual gene to assess our calibration error (fig. S1B).

\paragraph{Comparison of higher cumulants.} To further validate the quantitative nature and the proper calibration of our measurements, we compute the mean-to-higher cumulant relationships and compare them to previously extracted relationships where we used smFISH  \cite{Zoller:2018gj}. The first four cumulants of the measured activity $A$ are defined as
\begin{equation}
\begin{aligned}
\kappa_1 &\equiv \mu = \avg{A} \\
\kappa_2 &\equiv \sigma^2 = \avg{A^2}- \avg{A}^2 \\
\kappa_3 &= \avg{A^3} -3 \avg{A^2} \avg{A} +2 \avg{A}^3 \\
\kappa_4 &= \avg{A^4} -4 \avg{A^3} \avg{A} -3 \avg{A^2}^2 +12 \avg{A^2} \avg{A}^2 -6 \avg{A}^4.
\end{aligned}
\label{equ:a_cumulants}
\end{equation}
The cumulants are extensive, meaning that the contributions from independent random variables are additive, which is convenient as the measured transcriptional activities result from the sum of $n_g$ independent gene copies. Assuming replicated sister chromatids, $n_g=2$ for our live data (only one allele per nucleus has MS2 stem-loops) and $n_g=4$ for smFISH data (both alleles in each nucleus are tagged with probes). Using the definition above, we compute the higher cumulants of the calibrated activity from both, smFISH and live measurements. The same pooling strategy as for the means is applied (see previous paragraphs), using $2.5\%$ AP bins over a $10-15\,$min time window in NC13.

We have previously demonstrated that the higher cumulants can be normalized, assuming Poisson background, such that we account for differences in gene length, probe locations, and gene copy number \cite{Zoller:2018gj}. We compute the normalized cumulants $\kappa_k'$ for a single gene copy as follows:
\begin{equation}
\kappa_k' =\frac{1}{n_g C_k} \left( \frac{L'}{L_g} \right)^k \kappa_k,
\label{equ:norm_cumu}
\end{equation}
where $n_g$ is the gene copy number, $L_g$ the original gene length, $L'$ the normalized gene length. The $k$-th order coefficients $C_k$ are defined as
\begin{equation*}
C_k = \frac{1}{L_g} \int_0^{L_g} s(l)^k dl =\sum_{i=1}^m \left( \frac{i}{m} \right)^k \frac{l_{i+1}-l_i}{L_g},
\end{equation*}
where $s(l)$ is the relative contribution of a single nascent transcript (Eq. \ref{equ:transcript_signal}). Note that for $k=1$, we have $C_1 L_g= \tilde{L}$, which is the effective gene length.

We compare the normalized cumulants computed from live and smFISH data using a normalized gene length of $L'=3.3$ kb (which corresponds to the average gap gene length in the absence of stem-loops, also used in Zoller et al. \cite{Zoller:2018gj}). Once normalized, the first cumulant $\kappa_1'$ corresponds to the mean number of nascent transcripts $\avg{g}$ on a single gene copy of length $L'$. We plot the normalized higher cumulants $\kappa_2'$, $\kappa_3'$, and $\kappa_4'$ as a function of that $\avg{g}$ (fig. S1C). We further normalized all the cumulants by $g_0^k$. The value $g_0$ corresponds to the largest intercept between a $k$-th order polynomial fit to mean-cumulant relationships (continuous line for live and dotted line for smFISH data) and the straight line of the Poisson background (dashed line) set by $\kappa_k'=\avg{g}$. The number $g_0$ provides the maximal possible value of $\avg{g}$ based on the assumption that transcriptional variability $\kappa_2'$ cannot be lower than Poisson. Thus, $g_0$ can be interpreted as the mean number of nascent transcripts (on a 3.3 kb long gene) at maximal activity. We get $g_0 = 13.6$ from live and $g_0 = 15.2$ with smFISH measurements, a difference of 12\% based on higher cumulants, not too far from the 5\% error obtained from calibrating the means (fig. S1B). Overall, the higher cumulants versus mean relationships obtained from live (fig. S1C left column) and from smFISH (fig. S1C right column) measurements match almost perfectly (black solid versus dotted line), strongly confirming the quantitative nature and the proper calibration of both our live and fixed assay.

\subsection{Measurement error}
\label{sec:meas_error}

We evaluate imaging error through two distinct approaches: a direct measurement conducted through a specially designed experiment, and an auto-correlation analysis of the recorded transcription time series. The utilization of these two independent methods not only provides a robust internal control but also validates the reliability and effectiveness of our chosen approach. The convergence of results obtained from both approaches strengthens the credibility of our findings.

\paragraph{Imaging error via direct dual color measurement.} In order to experimentally assess imaging or measurement error, we performed the same experiment twice and simultaneously with two different colors. A discrepancy in both experiments from a perfect match can be assigned to measurement error. We inserted a cassette of interlaced (alternating) MS2 and PP7 stem-loops in the first intron of \emph{Kr}. The MS2 stem-loops were tagged with MCP-GFP (green) and the PP7 stem-loops with PCP-mCherry (red). By plotting the green channel against the red channel (fig. S1D, left), we characterize the spread of the data along the expected line of slope one (when correctly calibrated and without noise both channels are expected to perfectly correlate). We build a simple effective model to describe measurement noise in each channel:
\begin{equation*}
P(A|G)=\mathcal{N}(A | G,\sigma^2(G)),
\end{equation*}
where, for a given channel, the measured transcriptional activity $A$ (in C.U.) is normally distributed with mean $G$ and variance $\sigma^2(G)$. $G$ is the corresponding activity (in C.U.) in the absence of noise, and the variance $\sigma^2$ explicitly depends on $G$ to account for the heteroscedasticity (i.e., heterogeneity of variance) in the data as the observed spread in the data increases with transcriptional activity.

To estimate the variance $\sigma^2$, we fit a straight line $y=\alpha x+\gamma$ assuming error on both $x\equiv A_{\rm green}$ and $y\equiv A_{\rm red}$. We expand the variance as a function of the scalar projection along the line $v$:
\begin{align*}
\sigma^2(v) &=\sigma_b^2  + \beta_1  v + \beta_2  v^2, \\
v &= \frac{x+\alpha(y-\gamma)}{\sqrt{1+\alpha^2}}.
\end{align*}
Assuming the same error for \emph{green} (x) and \emph{red} (y), we maximize the following likelihood to estimate the set of parameters $\theta=\{ \alpha,\gamma,\sigma_b,\beta_1,\beta_2 \}$:
\begin{equation*}
\mathcal{L}(\{x_i,y_i\}|\theta) = \prod_{i=1}^{N_d} \frac{1}{\sqrt{2\pi \sigma^2(v_i)}}\exp{\left(-\frac{(y_i-\alpha x_i-\gamma)^2}{2(1+\alpha^2)\sigma^2(v_i)}  \right)},
\end{equation*}
where $N_d$ is the total number of pairs of red-green data points. 

Since only the green channel is calibrated in absolute units, we normalize the red channel such that the slope $\alpha=1$. Using the Akaike information criterion \cite{deLeeuw:1992}, we find that the best model was parameterized by only $(\sigma_b,\beta_1)$ with $\alpha=1$, $\gamma=0$ and $\beta_2=0$. The best fitting parameters are $\sigma_b = 1.22$ and $\beta_1=0.11$. The imaging error derived from the noise measurement model follows:
\begin{equation}
\sigma(G)=\sqrt{\sigma_b^2 + \tilde{\beta}_1 G},
\label{equ:meas_noise}
\end{equation}
with $\tilde{\beta}_1=\sqrt{2} \beta_1=0.16$. The resulting imaging noise is shown in fig. S1D and S1E. The functional form of the imaging error is exactly what one would expect from laser scanning microscopy, that is a background term $\sigma_b^2$ and a Poisson shot noise term $\tilde{\beta}_1 G$, the latter being proportional to the activity $G$.

\paragraph{Imaging error from correlation-based approach.} Due to the elongation of individual nascent transcripts adding a persistent signal to the measured activity, we expect a strong temporal correlation in the measured single allele activity $a(t)$. Indeed, based on gene length, i.e. Table \ref{tab:gene_length}), and elongation rate, i.e. $K_{\rm elo}=1.8$kb/min, (see Section \ref{sec:validation_deconv} for validation of deconvolution and estimation of elongation rate), imaged transcripts should persist at the transcription site for at least $\tau=2\,$min on average. Thus, at a short timescale near our sampling time $\Delta t=10\,$s ($\Delta t \ll \tau$), most biological variability should be correlated in time. Consequently, the remaining uncorrelated variability $\sigma_u^2$ is likely related to the imaging error as the latter is also expected to be temporally uncorrelated.

We estimate the uncorrelated variability $\sigma_u^2$ from the transcriptional activity time series $A(t)$ within each AP bin for each gene individually. To this end, we compute the time-dependent mean activity $\mu(t)=\avg{A(t)}$, variance $\sigma^2(t)=\avg{(A(t)-\mu(t))^2}$ and covariance $\text{Cov}(t,t+\Delta t)=\avg{(A(t)-\mu(t))(A(t+\Delta t)-\mu(t+\Delta t))}$, where $\Delta t$ is the sampling time (and the averaging is performed over all alleles in each AP bin pooled from multiple embryos). For a $\Delta t$ sufficiently small compared to the correlation time $\tau$ imposed by elongation, $\sigma_u^2(t)=\sigma^2(t)-\text{Cov}(t,t+\Delta t)$ should be a good estimate of the uncorrelated variability in the data.

When we plot $\sigma_u^2(t)$ as a function $\mu(t)$ for all genes and AP bins across NC13 and NC14 (fig. S1D left), we observe a strong linear relationship between $\mu(t)$ and $\sigma_u^2(t)$. That is, $\sigma_u^2(t) = \sigma_b^2+\beta \mu(t)$ with $\sigma_b=0.75$ and $\beta=0.19$, which is reminiscent of the functional form that we find using the interlaced approach above (Eq. \ref{equ:meas_noise}). Indeed, we find again that the variance is given by a background $\sigma_b^2$ and a Poisson shot noise term $\beta \mu$ that is proportional to the mean activity. The interlaced- and the correlation-based quantifications of the imaging error are numerically very close over the entire range of measured activities with a sensitivity close to a single mRNA (fig. S1E).

\section{Data analysis and modeling}
\label{sec:analyis_modeling}

\subsection{Deconvolution of the single-allele transcription rate}
\label{sec:single_transrate}

\paragraph{Modeling initiation events.} \label{sec:model_ini} We designed a simple model to reconstruct the individual initiation events of productive transcription by Pol II from the calibrated single allele transcriptional activity $A(t)$ (in C.U.). Our model for measurement noise $\sigma(G)$ (Eq. \ref{equ:meas_noise}) links the activity $A(t)$ to the signal in the absence of measurement noise $G(t)$ by:
\begin{equation}
A(t)=G(t)+\sigma(G(t))\eta(t),
\label{equ:adding_noise}
\end{equation}
where $\eta(t)$ is an uncorrelated Gaussian white noise with zero mean and unit standard deviation. The signal $G(t)$ is expressed as a convolution between a kernel $\kappa(t)$ (modeling the fluorescent signal resulting from the Pol II elongation process through the stem-loop cassette) and a function $I(t)$ that models Pol II initiation events:
\begin{equation}  
G(t)=\kappa(t)*I(t).
\label{equ:conv}
\end{equation}
The initiation function I(t) can be expressed as
\begin{equation}
I(t)=\sum_{i=1}\delta (t-t_i),
\label{equ:ini_fonction}
\end{equation}
where $\delta$ are unit pulses ($\delta$-Dirac) representing to individual initiation events at time $t_i$. The kernel function $\kappa(t)$ is built according to the signal contribution of individual transcripts (see Eq. \ref{equ:transcript_signal}):
\begin{equation}
\kappa(t) = \frac{1- H \left( t-\frac{L_g}{K_{\rm elo}} \right)}{m} \sum_{i=1}^m H \left( t -\frac{l_i}{K_{\rm elo}} \right),
\label{equ:elong_kernel}
\end{equation}
where $H$ is the unit (Heaviside) step function, $L_g$ the physical length of the gene, $l_i$ the end position of the $i^{\text th}$ MS2 stem-loop sequence and $m$ the total number of loops (in our construct $m=24$). The key assumptions behind this kernel are i) constant elongation rate $K_{\rm elo}$, ii) deterministic elongation (limited fluctuations, no pausing in the gene body), iii) fast termination, and iv) no co-transcriptional splicing (with short gene lengths, facilitating fast transcription). Importantly, all the parameters in Eq. \ref{equ:elong_kernel} are determined by the DNA sequence, except the elongation rate $K_{\rm elo}$ that needs to be inferred from measurements.

\paragraph{Single allele Bayesian deconvolution of transcription initiation events.} \label{sec:deconv} We performed Bayesian deconvolution to reconstruct the possible configuration of initiation events $I$ from individual single allele activity time series $A=(A_1,A_2,\ldots,A_{N_t})$, where $A_i\equiv A(t_i )$, $t_{i+1}-t_i=\Delta t\approx10$s $\forall i$ and $N_t$ the number of time points. We first wrote the likelihood of the measured activity time series $A$ given the true signal time series $G$ (discretely sampled at the same time points as $A$):
\begin{equation}
P(A|G)=\prod_{i=1}^{N_t}\mathcal{N}(A_i | G_i,\sigma^2(G_i)).
\label{equ:obs_prob}
\end{equation}
Individual time points, $A_i$, are assumed to be normally distributed with mean $G_i$ and variance $\sigma^2(G_i)$ given by our measurement noise model (see Eq. \ref{equ:meas_noise}). 

According to our model above (Eq. \ref{equ:conv}), the signal $G$ results from the convolution of $\kappa*I$. To compute this convolution, we first discretize the time variable $t$ in the kernel $\kappa(t)$ and the initiation function $I(t)$ (see Eq. \ref{equ:ini_fonction} and \ref{equ:elong_kernel}) using a smaller sampling time $\Delta t'=f/(2K_{\rm elo})=1\,$s, where $f=60\,$bp is a conservative estimate on the Pol II footprint on the DNA, $K_{\rm elo}=1.8\,$kb/min is the measured elongation rate (see Section \ref{sec:validation_deconv}; Validation of deconvolution and estimation of elongation rate) and the factor 2 accounts for having two unresolved active sister chromatids. $\Delta t'=1\,$s is the minimal physically possible time interval between two consecutive Pol II loading events (oversampling beyond $\Delta t'$ is superfluous). When discretized, the initiation configuration $I$ becomes a binary vector $I=(I_1,I_2,\ldots,I_{N_I})$ with $I_i\equiv I(t_i')\in \{0,1\}$, $t_{i+1}'-t_i'=\Delta t'$ $\forall i$ and $N_I= \lceil N_t \Delta t/\Delta t'\rceil-1$. Thus $G$ can be computed as a discrete convolution:
\begin{equation*}
G_i=\sum_{j=1}^{\min{(k,i)}} I_{i-j+1} \kappa_j,
\end{equation*}
where $k=\lceil L_g/(K_{\rm elo} \Delta t')\rceil+1$ is the number of time points needed to discretize the kernel $\kappa$ and $L_g$ the physical length of the gene (see Table \ref{tab:gene_length}). The resulting time series $G$ is then downsampled to evaluate $P(A|G)$ (Eq. \ref{equ:obs_prob}) at the same time points as the measured activity time series $A$. We thus have effectively designed a simple procedure to evaluate $P(A|I)$, i.e., the likelihood of $A$ given an initiation configuration $I$:
\begin{equation}
P(A|I) = \sum_G P(A|G) P(G|I),
\label{equ:deconv_likelihood}
\end{equation}
where $P(G|I)=1$ if $G=\kappa*I$ and $P(G|I)=0$ otherwise.

Applying Bayes' theorem gives us the posterior distribution $P(I|A)$, i.e., the distribution of possible configurations of initiation events $I$ given a measured single allele activity time series $A$:
\begin{equation}
P(I|A) = \frac{P(A|I)P(I)}{\sum_{I}P(A|I)P(I)},
\label{equ:bayes_theorem}
\end{equation}
where $P(I)$ is the prior distribution on the configuration of initiation events (see Section \ref{sec:prior}; Setting the prior). The posterior $P(I|A)$ is our Bayesian representation of the deconvolved initiation configurations $I$ and naturally provides the uncertainty on these configurations as a probability distribution. Importantly, $P(I|A)$ accounts for measurement noise and incorporates our assumptions about the elongation process underlying the fluorescent signal. Overall, the Bayesian approach to our deconvolution problem has a few advantages over maximum likelihood estimation (MLE). First, in the presence of measurement noise, there could be a risk with MLE that the unique resulting configuration I (the one that maximizes $P(I|A)$) overfits the noise. In contrast, Bayesian estimation readily provides all the possible configurations compatible with measurement noise. Second, when the transcriptional activity is high and many Pol II/transcripts contribute to the measured activity $A$, the underlying initiation configurations $I$ may be degenerate (meaning that multiple $I$ once convolved with the kernel $\kappa$ would lead to the same or very close $A$. In that context, choosing a single configuration seems arbitrary, as MLE would do, whereas Bayesian estimation naturally captures the symmetries underlying the initiation configurations.

\paragraph{Setting the prior.} \label{sec:prior} A prior $P(I)$ over the initiation configurations needs to be defined in order to calculate the posterior distribution $P(I|A)$ in Eq. \ref{equ:bayes_theorem}. As $I_i\in\{0,1\}$, expressing the prior $P(I)$ as a product of multiple independent Bernoulli distributions represents a simple and natural solution:
\begin{equation}
P(I)=\prod_{i=1}^{N_I}p_i^{I_i}(1-p_i)^{1-I_i},
\label{equ:prior_bernouli}
\end{equation}
with the parameters $p_i\in[0,1]$ for each time point. For a given gene and AP bin, we set these parameters to $p_i=R(t_i )\Delta t'$, where $R(t)$ is the time-dependent mean transcription rate in the AP bin. A good proxy for $R(t)$ is given by $R(t)\approx \mu(t+\tau_{\rm elo}/2)/\tau_{\rm elo}$, where $\mu(t)$ is the mean activity in the AP bin and $\tau_{\rm elo}$ the effective elongation time of the gene given by $\tilde{L}/K_{\rm elo}$ (see Eq. \ref{equ:mean_proxyr} and Table \ref{tab:gene_length}). Thus, our prior mainly encapsulates our knowledge about the mean transcription rate and not much about the underlying single allele initiation configurations. Interestingly, a constant prior $P(I)=\text{cst.}$ implies that $p_i=0.5$ $\forall i$, which is equivalent to a constant mean transcription rate set at $R_{\rm max}/2$ with $R_{\rm max}=1/\Delta t'=60\,$mRNA/min the maximal physical rate given the Pol II footprint and the measured elongation rate. However, in practice, the prior only has a small impact on inferred initiation configurations as the data is strong, i.e., given our imaging sensitivity and a rather short elongation time the likelihood $P(A|I)$ strongly constrains the possible configurations $I$ allowed by the prior.

\paragraph{MCMC sampling of initiation configurations.} \label{sec:MCMC} We designed a Monte-Carlo Markov Chain (MCMC) algorithm \cite{Andrieu:2003wb}, to directly sample from the posterior distribution $P(I|A)$ in Eq. \ref{equ:bayes_theorem}. This method relies on building a Markov Chain $P(I'|I)$, whose stationary distribution $\pi(I)$ is the target distribution, i.e., $\pi(I)\equiv P(I|A)$. Such a Markov chain is obtained by imposing detailed balance $P(I'|I) \pi(I)=P(I|I') \pi(I')$, which ensures that the chain has the correct stationary distribution. Once the chain has asymptotically reached after a few iterations the stationary distribution, samples drawn from $P(I'|I)$ at each subsequent iteration are equivalent to samples directly drawn from $P(I'|A)$ (where $I'$ is the new sample given the previous one $I$).

In practice, the sampling procedure at each iteration is performed in two steps. First, a new sample $I'$ is drawn from a proposal distribution $Q(I'|I)$. Second, the proposed sample $I'$ is accepted ($I'\rightarrow I$) or rejected ($I\rightarrow I$), with acceptance probability $\alpha(I',I)$:
\begin{equation}
\alpha(I,I') = \min{\left( 1, \frac{P(A|I') P(I') Q(I,I')}{P(A|I)P(I)Q(I',I)} \right)},
\end{equation}
where $P(A|I)$ is the likelihood (Eq. \ref{equ:deconv_likelihood}) and $P(I)$ is the prior (Eq. \ref{equ:prior_bernouli}). The proposal distribution $Q(I'|I)$ must be carefully chosen to ensure a good acceptance probability (a good acceptance is typically between $10$--$70\%$) \cite{Andrieu:2008kh,Rosenthal:2010}. Indeed, if the acceptance is too high ($I'$ is too close to $I$) or too low ($I'$ is too far from $I$), the chain will generate highly correlated samples (poor mixing), and the sampling will be inefficient (slow convergence). In principle, a good proposal is a distribution that closely mimics the target distribution. Since the target distribution is a priori unknown, we use an adaptative method to learn $Q(I'|I)$ on the fly such that it approaches the target distribution \cite{Andrieu:2008kh,Rosenthal:2010}. In that case, the proposal distribution can typically be chosen as independent from the previous sample $Q(I'|I)=Q(I')$, and our prior (Eq. \ref{equ:prior_bernouli}) represents a decent candidate proposal distribution:
\begin{equation}
Q(I)=\prod_{i=1}^{N_I}p_i^{I_i}(1-p_i)^{1-I_i}
\label{equ:proposal}
\end{equation}
Here, the goal is to learn the parameters $q_i\in[0,1]$ while sampling the configurations $I$. As initial values for the $q_i$, we chose $q_i=\tilde{A}(t_i' ) \Delta t'$, where $\tilde{A}(t)=A(t+\tau_{\rm elo}/2)/\tau_{\rm elo}$ with $A(t)$ the activity time series and $\tau_{\rm elo}$ the elongation time. After each iteration of the MCMC sampler (after the acceptance or rejection of the proposed $I'$), we then update the $q_i$ according to the following adaptative scheme:
\begin{equation*}
q_i'=q_i(1-l)+l I_i,
\end{equation*}
where $q_i'$ is the new $q_i$, $I$ the current configuration, and the learning rate $l=1/\sqrt{3j}$ with $j$ the iteration number. The function $l,$ which vanishes when $j \rightarrow \infty$, is chosen such that the Markov chain converges despite adaptation. Importantly, we enforce $q_i>0.05$ $\forall i$ at each iteration to ensure that all time points of configuration $I$ are sampled.

In practice, even with adaptation, attempting to update the whole configuration $I$ at once according to the proposal (Eq. \ref{equ:proposal}) can lead to sub-optimal acceptance. Indeed, the $I_i$'s are often quite correlated, whereas our proposal assumes independent $I_i$. A simple workaround is to update the $I_i$'s by block and increase the number of iterations: only the $I_i$ from a subset $i\in \{b,\ldots,b+w-1\}\subset\{1,\ldots,N_I\}$ are drawn from the proposal (Eq. \ref{equ:proposal}) at each iteration (while the remaining $I_i$ are left unchanged), with $b$ a random integer and $w$ the size of the block. The size $w$ is then tuned to achieve the desired acceptance ratio. We found that block size $w$ covering approximately 6 min led to a good compromise between a decent acceptance ratio and sampling speed. To sample $I$, we performed $N_s N_I / w$ iterations with $N_s = 3500$ and we discarded the first $N_b N_I / w$ iteration, where $N_b=500$.

\paragraph{Estimating single allele transcription rate.} \label{sec:singler_estimate} Once the posterior distribution $P(I|A)$ of possible initiation configuration has been sampled by MCMC, we compute the single-cell transcription rate $r$. For each configuration $I$ (Eq. \ref{equ:ini_fonction} and \ref{equ:deconv_likelihood}), we can compute a single $r(t)$ as follows:
\begin{equation}
r(t)= \frac{1}{v \Delta t} \int_{t-v \Delta t}^{t} I(t') dt' = \frac{1}{v \Delta t} g(t),
\label{equ:singler_estimate}
\end{equation}
where $g(t)$ is the number of initiation events during the time interval $t\in[t-v\Delta t,t]$ given by
\begin{equation*}
g(t)=\sum_{ i\in\{i:t-v \Delta t \leq t_i' \leq t \} } I_i.
\end{equation*}
In the equations above, $\Delta t\approx10\,$s is our sampling time, and $v$ is a positive integer defining the time window over which $r$ is estimated. Except when explicitly specified otherwise, we estimated $r$ using $v=1$ which we defined as the instantaneous single allele transcription rate. 

Using our MCMC samples, $P(r|A)$ is easily estimated by
\begin{equation}
P(r|A)=\sum_I P(r|I) P(I|A),
\label{equ:singler_distrib}
\end{equation}
where $P(r|I)=1$ if Eq. \ref{equ:singler_estimate} is satisfied, and $P(r|I)=0$ otherwise. The distribution $P(r|A)$ enables the computation of the instantaneous mean and standard deviation of $r(t)$ for individual single-allele activity time series $A$ (fig. S2B). Of note, $P(r|A)$ is actually a discrete probability distribution since we have a finite amount of MCMC samples and $r$ is discrete (as is $g$).

\paragraph{Validation of deconvolution and estimation of elongation rate.} \label{sec:validation_deconv} To validate the hypothesis behind our kernel-based deconvolution (see Section \ref{sec:model_ini}; Modeling initiation events), we tested our approach on dual color data, where the MS2 and PP7 stem-loop cassettes were inserted in the first intron and 3'UTR of \emph{hb}, respectively. Similarly, our approach was tested on a data set for an additional gene, \emph{Kr}, where the stem-loop insertions were permuted (PP7 in intron and MS2 in 3'UTR). In both cases MS2 stem-loops were tagged with MCP-GFP (green) and PP7 stem-loop  via PCP-mCherry (red).

After calibrating the green and red channels in absolute units (see Section \ref{sec:calibration}), we expressed the likelihood of obtaining the green $A_{\rm g}$ and the red activity $A_{\rm r}$ together, which is simply $P(A_{\rm g} | G_{\rm g}) P(A_{\rm r}|G_{\rm r})$, where $G_{\rm g}$ and $G_{\rm r}$ are the true signals in each channel. Both $P(A_{\rm g}|G_{\rm g})$ and $P(A_{\rm r}|G_{\rm r})$ are given by Eq. \ref{equ:obs_prob}, with $\sigma^2(G)$ that is specific to each channel (see Section \ref{sec:meas_error}). In principle, the true signals $G_{\rm g}$ and $G_{\rm r}$ should be highly correlated since they should result from the same initiation configuration $I$. $G_{\rm g}=\kappa_{\rm g}*I$ and $G_{\rm r}=\kappa_{\rm r}*I$ contain two kernels $\kappa_g$ and $\kappa_r$ that are determined from Eq. \ref{equ:elong_kernel}. Whereas these kernels mainly differ in the location of the stem-loop insertions, they share one free parameter, i.e., the elongation rate $K_{\rm elo}$ that is a priori unknown. Importantly, the overall delay between $G_{\rm g}$ and $G_{\rm r}$ should be given by the relative position $| l_{\rm g}-l_{\rm r} |$ of the MS2 and PP7 stem-loop cassettes divided by $K_{\rm elo}$. It implies that $K_{\rm elo}$ can be extracted from these dual-color data. With these ingredients, the likelihood function for the pair $A_{\rm g}$ and $A_{\rm r}$ is as follows:
\begin{equation}
P(A_{\rm g}, A_{\rm r} | I, K_{\rm elo}) = \sum_{G_{\rm g}, G_{\rm r}} P(A_{\rm g} | G_{\rm g}) P(A_{\rm r} | G_{\rm r}) P(G_{\rm g}, G_{\rm r} | I, K_{\rm elo}),
\label{equ:dualc_likelihood}
\end{equation}
where $P(G_{\rm g}, G_{\rm r} | I, K_{\rm elo})$ is trivially determined from $G_{\rm g}=\kappa_{\rm g}(K_{\rm elo})*I$ and $G_{\rm r}=\kappa_{\rm r}(K_{\rm elo})*I$ (see Eq. \ref{equ:deconv_likelihood}).

Next, from our dual-color data, we inferred $I$ and $K_{\rm elo}$ for each nucleus using the likelihood in Eq. \ref{equ:dualc_likelihood}. Specifically, we sampled from the posterior distribution $P(I,K_{\rm elo} |A_{\rm g},A_{\rm r})$ according to our MCMC sampler (see Section \ref{sec:MCMC}; MCMC sampling of initiation configurations). We set the prior of $K_{\rm elo}$ to log-uniform $P(K_{\rm elo})\sim1/K_{\rm elo}$. For both \emph{hb} and \emph{Kr} data, we assessed the quality of the deconvolution by computing the normalized residuals for each channel ${\rm c}\in \{{\rm g}, {\rm r}\}$ and each nucleus:
\begin{equation*}
z_{\rm c}=\frac{A_{\rm c} - \bar{G}_{\rm c} }{\sigma_{\rm c}(\bar{G}_{\rm c})},
\end{equation*}
where $\sigma_{\rm c}(G)$ is the measurement error of channel $\rm c$, and $\bar{G}_{\rm c}$ the mean reconstructed signal computed as the marginal mean of
\begin{equation*}
P(G_{\rm g}, G_{\rm r} | A_{\rm g}, A_{\rm r}) = \sum_I \int P(G_{\rm g}, G_{\rm r} | I, K_{\rm elo}) P(I, K_{\rm elo} | A_{\rm g}, A_{\rm r}) dK_{\rm elo}.
\end{equation*}
Since the residuals $z_{\rm c}$ are given by a time series, we thus reported the mean $\mu(z_{\rm c})$ and the standard deviation of $\sigma(z_{\rm c})$ in the two channels for each nucleus (fig. S2E). We then compared the $\mu(z_{\rm c})$ and $\sigma(z_{\rm c})$ to a perfect model for which the $z_{\rm c}$ have the same length than our data and each $z_{\rm c}(t)$ is independent and normally distributed with $\mu=0$ and $\sigma=1$. We computed the 95\% confidence ellipse of $\mu(z_{\rm c})$ and $\sigma(z_{\rm c})$ for both our data (solid black line in fig. S2E) and the perfect model (dotted black line in fig. S2E). Our reconstructions are close to the perfect model, meaning that the two channels can be reconstructed from the same initiation configurations within confidence intervals, overall justifying our kernel assumptions. Indeed, the good agreement between the predicted and the measured dual color signal implies that the elongation process must be close to deterministic (with approximately constant elongation rate) and co-transcriptional splicing must be limited for these short genes. Otherwise, we would have observed a striking mismatch between the model prediction and the measurements in the 5'-to-3' amplitude ratio and the 5'-to-3' delay.

Lastly, we estimate the elongation rate $K_{\rm elo}$ for each individual nucleus numerically as the mean of the marginal posterior distribution $P(K_{\rm elo}|A_{\rm g},A_{\rm r})$. Within each embryo and for both $\emph{hb}$ and \emph{Kr}, the average $K_{\rm elo}$ across nuclei at different AP bins remains stable along the AP axis (fig. S2F). We estimated the global mean $K_{\rm elo}$ by averaging all nuclei for each embryo, the resulting averages are very similar across genes and embryos (fig. S2G). The overall mean elongation rate is $\bar{K}_{\rm elo}=1.8\pm0.1\,$kb/min.

\subsection{Gap gene mean expression pattern}

\paragraph{Transcription pattern.} From our deconvolution-based estimates of the single-allele transcription rate $r(t)$, we reconstruct the mean expression pattern as a further test for the validity of our approach. For each gene, we align the imaged embryos in space and time, allowing us to pool nuclei from different embryos (see Section \ref{sec:pooling}; Embryo-to-embryo variability and pooling). Thus, for a given gene and nuclear cycle, we obtain a set $\mathcal{A}(x,t)$ of activity time series $A\in \mathcal{A}(x,t)$ coming from all the nuclei in the same AP bin at time $t$ (of width 2.5\% and 1.5\% egg length in NC13 and NC14, respectively) centered at position $x$. The pooling procedure results in the distribution of single-allele transcription rate $P(r|x,t)$ defined as the following mixture:
\begin{equation}
P(r|x,t) = \sum_{A\in \mathcal{A}(x,t)} P(r(t)|A)P(A),
\label{equ:r_mixture}
\end{equation}
where $P(r|A)$ is the distribution of deconvolved $r$ given a single-allele time series $A$ (defined in Eq. \ref{equ:singler_distrib}) and $P(A)=1/N_A(x,t)$ with $N_A(x,t)=\# \mathcal{A}(x,t)\sim200$ the total number of nuclei in the AP bin at position $x$ and time $t$. From $P(r|x,t)$, we compute the time-dependent mean transcription rate $R(x,t)$ at position $x$:
\begin{equation}
R(x,t)=\avg{r(x,t)}=\sum_r r P(r|x,t).
\label{equ:R_mean}
\end{equation}
Using $R(x,t)$, we reconstruct the mean profiles for all genes as a function of time $t$, which are displayed for three different time points in fig. S3A (color dots).

To validate our estimation of the mean transcription rate $R(x,t)$, we compare it to an alternative approximation that is independent of the deconvolution procedure:
\begin{equation}
R(x,t)\approx \frac{\mu(x,t+\tau_{\rm elo}/2)}{\tau_{\rm elo}},
\label{equ:mean_proxyr}
\end{equation}
where $\mu(x,t)$ is the time-dependent mean activity in the AP bin centered at position $x$ and $\tau_{\rm elo}$ the effective elongation time of the gene given by $\tilde{L}/K_{\rm elo}$ (see Table \ref{tab:gene_length}). We see in fig. S3A that both approaches (color and black dashed line for deconvolution approach and approximation via Eq. \ref{equ:mean_proxyr}, respectively) strongly agree and can thus be used interchangeably to estimate the mean transcription rate $R(x,t)$. 

\paragraph{Transcription and protein pattern comparison.} We further tested whether our mean gap gene patterns measured directly from the transcriptional output, i.e., MS2-signal (fig. S4A, left column) are consistent with previously measured protein patterns (fig. S4A, right column) \cite{Dubuis2013}. To this end, we extrapolate protein accumulation from the characterized transcriptional mean output using a minimal model (fig. S4A, middle column). 

We model the accumulation of protein as a convolution between the mean transcription rate $R(x,t)$ and a kernel $\kappa(x,t)$ that accounts for diffusion, degradation, and a delay in protein production:
\begin{equation}
\kappa(x,t)=\left\{
\begin{aligned}
\exp{ \left( -\frac{t-\delta}{\tau_{m} } \right) } &\exp {\left(-\frac{x^2}{4D(t-\delta)} \right) } &\text{if}\quad t-\delta \geq 0\\
&0 &\text{otherwise}\quad
\end{aligned}
\right.
\label{equ_kernel_prot}
\end{equation}
Thus $\kappa(x,t)$ above is parameterized by three free parameters: a diffusion constant $D$, a mean lifetime $\tau_m$, and a delay $\delta$. We compute the predicted protein pattern $\tilde{P}(x,t)$ as bidimensional convolution (along $x$ and $t$, hence $**$ symbol) $\tilde{P}(x,t)=R(x,t)$$**$$\kappa(x,t)$, where $R(x,t)$ is estimated using Eq. \ref{equ:mean_proxyr}. During NC14, we multiply $R(x,t)$ by a factor of two to account for the doubling of nuclei during mitosis 13. In addition, the resulting $\tilde{P}(x,t)$ is normalized such that its maximum value is one to facilitate direct comparison with previously similarly normalized protein data \cite{Dubuis2013}.

To estimate the three kernel parameters $\{D,\tau_m,\delta \}$ (Eq. \ref{equ_kernel_prot}), we minimize the squared difference between the predicted protein pattern $\tilde{P}(x,t)$ and the measured protein pattern $P(x,t)$. We performed the optimization for all gap genes individually and together (fig. S4B). The resulting parameters are in line with previous estimates and the modeled accumulation is consistent with previously measured protein patterns (fig. S4A). \textbf{Given our minimal model, the agreement between the modeled and measured protein patterns suggests that protein levels are mainly determined by transcription.}

\subsection{Transcription fluctuation analysis}

\paragraph{Distributions of single-allele transcription rates.} To characterize the transcriptional fluctuations around the mean transcription rate $R(x,t)$, we aimed to reconstruct the conditional distribution $P(r|R)$ of single-allele transcription rate $r$ for each gene. In principle, $P(r|R)$ should exhibit key signatures of the transcriptional regime at play. Importantly, to magnify these signatures, here we specifically estimate the single-allele transcription rate $r$ over a $\tau=1$--min time window ($v=6$, in Eq. \ref{equ:singler_estimate}). Indeed, with limited time-averaging, i.e., $r$ estimated over $\Delta t=10$s, the distribution $P(r|R)$ should approach the Poisson distribution at all $R$.

For each gene and nuclear cycle, our deconvolution method enables estimating the distribution of single-allele transcription rate $P(r|x,t)$ by pooling nuclei at position $x$ and time point $t$ (see Eq. \ref{equ:r_mixture}). The mean transcription rate is then given by $R(x,t)=\avg{r(x,t)}=\sum_r r P(r|x,t)$ (Eq. \ref{equ:R_mean}). 

We reconstruct the conditional distribution $P(r|R)$ by pooling together all nuclei from the distributions $P(r|x,t)$ at all position $x$ and time $t$ that satisfy a specific range of means $|R(x,t)-R_0|\leq \epsilon$ (where $\epsilon$ is a positive parameter). By defining the set $\mathcal{V}(R_0)=\{(x,t):|R(x,t)-R_0|\leq \epsilon\}$, we express the conditional as a mixture of the single-allele transcription rate distributions $P(r|x,t)$:
\begin{align}
P(r|R_0) &= \sum_{(x,t)\in\mathcal{V}(R_0)} P(r|x,t) P(x,t) \label{equ:reconstruction} \\
P(x,t) &= \frac{N_A(x,t)}{\sum_{(x,t)\in\mathcal{V}(R_0)} N_A(x,t)}, \nonumber
\end{align}
where $N_A(x,t)$ is the number of nuclei in the bin at position $x$ and time $t$. Importantly, a reconstruction of $P(r|R_0)$ through pooling is only valid under the condition that the distributions $P(r|x,t)$ for which $(x,t)\in \mathcal{V}(R_0)$ are mostly similar. 

To verify the validity of our reconstruction of $P(r|R)$ for each gene, we binned the range of observed $R$ values in 15 equal bins from 0 to 16 mRNA/min. Each respective $R_0$ is thus defined as the centers of these $R$-bins, which sets $\epsilon=0.53$ (the half width of the $R$ bins). We then computed the overall $P(r|R)$ by applying Eq. \ref{equ:reconstruction} for each $R_0$. Given $P(r|R_0)$, we determined the 95\% confidence envelope on its cumulative distribution. We assessed what fraction of the individual $P(r|x,t)$ (with $(x,t)\in \mathcal{V}(R_0)$) are within the global 95\% envelope. It turns out that when pooling alleles irrespective of developmental time, we observe a significant fraction of $(x,t)$-bins (position and time point) in NC14 for which $P(r|x,t)$ significantly deviates from the global distribution (fig. S3B, bottom right). When performing the same operation on three separate periods -- NC13 ($t \geq 6.5$ min after mitosis) plus early NC14 ($7.5\leq t<20.5$ min), mid-NC14 ($20.5\leq t<34.5$ min) and late NC14 ($34.5\leq t < 48$ min) -- the agreement between the individual $P(r|x,t)$ and the overall distribution $P(r|R)$ improves significantly (fig. S3B). This suggests that transcription might be undergoing slow kinetic changes over developmental time, warranting  a partitioning of NC14 into broad periods that are roughly stationary.

Having checked the pooling validity underlying the reconstruction of $P(r|R)$, we then analyze the single-allele transcription rates underlying the period encompassing NC13 and early NC14 ($7.5\leq t<20.5$ min). Within this period and for each of the 15 $R_0$ values defined as the center of the $R$-bin, we reconstruct the conditional $P(r|R)$ distribution according to Eq. \ref{equ:reconstruction}. We show the resulting distributions for each gene in fig. S3C. Strikingly, the conditional $P(r|R)$ at a given $R=R_0$ value looks very similar across genes and periods (NC13 vs early NC14). Moreover, these distributions differ from the typical distribution one would expect in a constitutive (i.e., non-bursting) regime.

\paragraph{Constitutive regime.} In a constitutive regime, initiation events are assumed to happen as a simple Poisson process at a constant rate of $R$. Thus over some time $\tau$, we would expect to observe $g$ initiation events with probability given by the Poisson distribution:
\begin{equation*}
P_P(g | \lambda=R \tau) = \frac{\lambda^{g}\exp(-\lambda)}{g!}.
\end{equation*}
Given our definition of the single-allele transcription rate (see Eq. \ref{equ:singler_estimate}), we have $r=g/\tau$ implying that in a constitutive regime $P(r|R)=P_P(g=r\tau|\lambda=R \tau)$. Such a description of the constitutive regime implicitly assumes that Pol II has no footprint, meaning that initiation could happen in very quick succession. However, this is unrealistic as the Pol II footprint $f=60$bp (conservative estimate) should set a refractory period $\Delta t'= f/(2K_{\rm elo})=1$s below which successive initiations cannot occur (see Section \ref{sec:deconv}; Single allele Bayesian deconvolution of transcription initiation). As a simple correction to account for such a refractory period, we discretize the period $\tau> \Delta t'$ using the interval $\Delta t'$ as we did for our deconvolution procedure. In that case, the probability to observe $g$ initiation events over some time $\tau$ would be given by the Binomial distribution:
\begin{equation}
P_B(g| n_t= \lceil \tau/\Delta t' \rceil, p=R\Delta t') = \binom{n_t}{g} p^{g} (1-p)^{n_t-g}.
\label{equ:constit_bino}
\end{equation}
Thus, considering the Pol II footprint, the distribution of single-allele transcription rate in a constitutive regime should be well approximated by $P(r|R)=P_B(g=r\tau | n_t= \lceil \tau/\Delta t' \rceil, p=R\Delta t')$. Of note, when $\tau \gg \Delta t' $ ($\Delta t' \rightarrow 0$), we recover the Poisson distribution and $P(r|R)=P_P(g=r\tau|\lambda=R \tau)$. This means that the binomial correction to the Poisson regime only matters when $\Delta t' \lesssim \tau$ and it becomes negligible when $\Delta t' \ll \tau$.

To highlight the difference between the empirically determined conditional distributions $P(r|R)$ and the constitutive ``Poisson'' regime, we plotted $P_B(g=r\tau | n_t= \lceil \tau/\Delta t' \rceil, p=R\Delta t')$ in fig. S3C (dashed line), where we used $\tau=1\,$min. We also computed the $2^\text{nd}$ to $4^\text{th}$ cumulants from the empirically determined conditional distributions $P(r|R)$ and the constitutive ``Poisson'' regime as a function of $R$. The cumulants are defined as follows:
\begin{align*}
\sigma^2(R) &= \avg{r^2}- \avg{r}^2 \nonumber \\
\kappa_3(R) &= \avg{r^3} -3 \avg{r^2} \avg{r} +2 \avg{r}^3 \\
\kappa_4(R) &= \avg{r^4} -4 \avg{r^3} \avg{r} -3 \avg{r^2}^2 +12 \avg{r^2} \avg{r}^2 -6 \avg{r}^4 \nonumber
\label{equ:r_cumulants}
\end{align*}
where $\avg{r^k}=\sum_ r^k P(r|R)$. Assuming a constitutive regime (Eq. \ref{equ:constit_bino}), we get the following cumulants:
\begin{align*}
\sigma^2(R) &= n_tp(1-p)/\tau^2 = R (1- R\Delta t')/\tau \\
\kappa_3(R) &= n_tp(1-p)(1-2p) /\tau^3 = R (1- R\Delta t')(1-2R\Delta t')/\tau^2  \\
\kappa_4(R) &= n_tp(1-p)(1-6p(1-p)) /\tau^4 = R (1- R\Delta t')(1-6R\Delta t'(1-R\Delta t'))/\tau^3
\end{align*}
We plot the cumulants from the empirically determined conditional distributions $P(r|R)$ (color) and the constitutive ``Poisson'' regime (black dashed line) as a function of $R$ in fig. S3D. Overall, the empirical cumulants deviate significantly from the ``Poisson'' regime, except at the extreme ends of the $R$ spectrum where we observe a convergence towards the constitutive regime. This can be interpreted as evidence for a bursting regime spanning all the way from $P_{\rm ON}=0$ to $P_{\rm ON}=1$ \cite{Zoller:2018gj}. Errors on $R$ and the cumulants were computed from bootstrapping.

\paragraph{Transcription rate auto-correlation.} \label{sec:ac_estimation} To characterize the transcriptional fluctuations over time, we compute the auto-correlation of the instantaneous single-allele transcription rate $r(t)$ ($v=1$, in Eq. \ref{equ:singler_estimate}). For a given gene, nuclear cycle, and for each AP-bin (centered at position $x$), pooling gives a set $\mathcal{A}(x)$ of activity time series $A\in \mathcal{A}(x)$ for all nuclei in that particular AP bin. Our Bayesian deconvolution approach enables sampling from the posterior distribution $P(r|A)$. It generates MCMC samples of the instantaneous single-allele transcription rate given an activity time series $A$ (see Eq. \ref{equ:singler_distrib}). We denote such a sample $r_s(t)|A$, where $s$ labels a specific MCMC sample drawn from $P(r|A)$. Thus, for a given AP-bin (centered at position $x$), we compute the normalized auto-correlation function $\rho_s(\tau)$ as follows:
\begin{equation}
\rho_s(\tau) = \frac{\avg{(r_s(t)|A-\mu_s(t))(r_s(t+\tau)|A-\mu_s(t+\tau)}}{\sigma_s(t)\sigma_s(t+\tau)},
\label{equ:ac_estimate}
\end{equation}
where $\tau$ is the lag (which is a multiple of $\Delta t$) and the averaging is performed over all the nuclei in the AP-bin, i.e., $\avg{F(r_s(t)|A)}=\frac{1}{N_A(x)}\sum_{A\in \mathcal{A}(x)} F(r_s(t)|A)$ with $N_A(x)=\# \mathcal{A}(x,t)\sim200$ nuclei. Hence, $\mu_s(t)=\avg{r_s(t)|A}$ is the time-dependent mean and $\sigma_s(t)=\sqrt{\avg{(r_s(t)|A-\mu(t))^2}}$ the time-dependent standard deviation in the AP-bin. The final estimate of the normalized auto-correlation $\rho(\tau)$ and the corresponding error $\sigma_{\rho}(\tau)$ are obtained by averaging over MCMC samples:
\begin{equation}
\begin{aligned}
\rho(\tau)&=\frac{1}{N_s}\sum_{s=1}^{N_s} \rho_s(\tau) \\
\sigma_{\rho}^2(\tau)&=\frac{1}{N_s}\sum_{s=1}^{N_s} \left(\rho_s(\tau) -\rho(\tau)\right)^2,
\end{aligned}
\end{equation}
where $N_s = 10^3$ is the total number of samples used per allele.

We show auto-correlation functions $\rho(\tau)$ for \emph{hb} in early NC14 ($7.5\leq t<20.5$ min after mitosis) in Fig. 1E (color code stands for the AP-bin). An auto-correlation is characterized by two key quantities: the magnitude of correlated fluctuations $\Sigma_{\rm AC}:=\rho(\Delta t)$ (where $\Delta t=10\,$s is the sampling time) and the correlation time $\tau_{\rm AC}$ that captures the temporal scale of the fluctuations. We estimated $\Sigma_{\rm AC}$ and $\tau_{\rm AC}$ for all gap genes at all AP positions in NC13 and early NC14 (fig. S5C and Fig. 1F). The correlation time $\tau_{\rm AC}$ was estimated by fitting $\rho(\tau)$ with the following exponential function for $\Delta t \leq \tau <10\,$min:
\begin{equation}
\tilde{\rho}(\tau)=\Sigma_{\rm AC}(1-\beta) \exp{\left(-\frac{\tau-\Delta t}{\tau_{\rm AC}}\right)} +\beta,
\label{equ:ac_fit}
\end{equation}
where $\Sigma_{\rm AC}$ is estimated from $\rho(\Delta t)$ and $\{ \tau_{\rm AC},\beta \}$ are obtained by maxmimum likelihood. The parameter $\beta$ accommodates for a possible noise floor in $\rho(\tau)$, which typically results from residual embryo-to-embryo variability; on average $\beta = 0.04$. Errors on $\Sigma_{\rm AC}$ and $\tau_{\rm AC}$ are computed from bootstrapping the fit to $\rho_s(\tau)$ (Eq. \ref{equ:ac_estimate}) over MCMC samples.

\paragraph{Two-state model auto-correlation.} \label{sec:2state} To help us interpret the auto-correlation function, we investigated the prediction of the 2-state model of transcriptional bursting \cite{Zoller:2018gj,Peccoud1995}. In the 2-state model, the gene toggles stochastically between an ON ($n=1$) and OFF ($n=0$) states with rates $k_{\rm ON}$ and $k_{\rm OFF}$, respectively. In the ON state, the gene promoter can initiate transcription with initiation rate $k$ leading to the synthesis of $g$ transcripts. Within that model, we further define the switching correlation time $T=1/(k_{\rm ON}+k_{\rm OFF})$ and steady-state mean occupancy $\eta=k_{\rm ON}/(k_{\rm ON}+k_{\rm OFF})$. In what follows, we focus on the stationary case ($k_{\rm ON}$, $k_{\rm OFF}$, and $k$ don't change over time), which is easier to interpret. From the master equation \cite{Lestas:2008hy,Walczak:2012}, we derive the equations that describe the time evolution of the moments:
\begin{equation}
\left\{
\begin{aligned}
\frac{d}{dt} \avg{n(t)} &= \frac{1}{T} (\eta-\avg{n(t)}) \\
\frac{d}{dt} \avg{g(t)} &= k \avg{n(t)} \\
\frac{d}{dt} \avg{g(1,t)} &= k \avg{n(t)} + \frac{1}{T}(\eta \avg{g(t)} - \avg{g(1,t)}) \\
\frac{d}{dt} \avg{g^2(t)} &= k (\avg{n(t)} + 2\avg{g(1,t)}) 
\end{aligned}
\right.
\label{equ:time_dep_moment}
\end{equation}
$\avg{n(t)}$ is the mean occupancy of the ON-state (or equivalently the ON-probability $P_{\rm ON}$), $\avg{g(t)}$ is the mean number of initiation events (or transcripts), $\avg{g(1,t)}=\avg{g(t)}-\avg{g(0,t)}$ is the mean number of initiation events conditioned on the gene being ON, and $\avg{g^2(t)}$ is the second moment of $g$. Solving the equations above (Eq. \ref{equ:time_dep_moment}) with initial conditions $\avg{n(t=0)}=n_0$, $\avg{g(t=0)}=\avg{g(1,t=0)}=\avg{g^2(t=0)}=0$, we determine the mean $\mu(t)\equiv\avg{g(t)}$ and the variance $\sigma^2(t) = \avg{g^2(t)}-\avg{g(t)}$ of the number of transcription initiation events $g$ over the period $[0,t]$. In the special case $t=\tau_{\rm elo}$ (the elongation time), $\mu(\tau_{\rm elo})$ and $\sigma^2(\tau_{\rm elo})$ correspond to the mean and the variance of the number of nascent transcripts on the gene (i.e., the transcriptional activity), under the assumption that elongation is a deterministic process at constant rate $K_{\rm elo}$.

Assuming the gene is initially at steady-state ($n_0=\eta$), we solve Eq. \ref{equ:time_dep_moment} and get:
\begin{equation}
\begin{aligned}
\avg{g(t)} &= k\eta t \\
\avg{g(1,t)} &= k\eta^2 t + k\eta (1-\eta) T(1-\exp{(-t/T)}) \\
\avg{g(0,t)} &= k\eta(1-\eta) t - k\eta(1-\eta) T(1-\exp{(-t/T)})  \\
\avg{g^2(t)} &= k\eta t + k^2\eta^2 t^2 + 2 k^2 \eta(1-\eta) (T^2 \exp{(-t/T)} + t T -T^2).
\end{aligned}
\label{equ:sol_moment}
\end{equation}
To compute the auto-correlation, we further need the transient of the first moments obtained by solving Eq. \ref{equ:time_dep_moment} with generic initial condition $\avg{n(t=0)}=n_0\in\{0,1\}$:
\begin{align}
\avg{n(t|n_0)} &= \eta (1-\exp{(-t/T)}) + n_0 \exp{(-t/T)} \nonumber \\
\avg{g(t|n_0)} &= k\eta t + k(n_0-\eta)T(1-\exp{(-t/T)}). \label{equ:sol_transient}
\end{align}

The transcriptional activity at time $t$ is given by $g(t;t-\tau_{\rm elo})$, which is the total number of nascent transcripts (on the gene) initiated over the period $[t,t-\tau_{\rm elo}]$. In the stationary case, the transcriptional activity is simply given by $\avg{g(t;t-\tau_{\rm elo})}=\avg{g(\tau_{\rm elo})}$. Using this definition, we then define the normalized auto-correlation of the transcriptional activity as
\begin{align}
\rho_A(\tau) &= \frac{\avg{(g(t;t-\tau_{\rm elo})-\mu(\tau_{\rm elo}))(g(t+\tau;t+\tau-\tau_{\rm elo})-\mu(\tau_{\rm elo}))}}{\sigma^2(\tau_{\rm elo})} \nonumber \\
&= \frac{\avg{g(t+\tau;t+\tau-\tau_{\rm elo})g(t;t-\tau_{\rm elo})}-\mu^2(\tau_{\rm elo})}{\sigma^2(\tau_{\rm elo})}. \label{equ:rhoA}
\end{align}
And similarly for the normalized auto-correlation of the transcription rate $r(t)=g(t;t-\Delta t)/\Delta t$ (where $\Delta t=10$s is the sampling time), we have:
\begin{align}
\rho_R(\tau) &= \frac{\avg{(g(t;t-\Delta t)-\mu(\Delta t))(g(t+\tau;t+\tau-\Delta t)-\mu(\Delta t))}/(\Delta t)^2 }{\sigma^2(\Delta t)/(\Delta t)^2} \nonumber \\
&= \frac{\avg{g(t+\tau;t+\tau-\Delta t) g(t;t-\Delta t)}-\mu^2(\Delta t)} {\sigma^2(\Delta t)}. \label{equ:rhoR}
\end{align}
As it turns out, these auto-correlation functions are almost identical and only differ on the integration period $\tilde{\tau}$, which is set to $\tilde{\tau}=\tau_{\rm elo}$ to compute $\rho_A(\tau)$ and to $\tilde{\tau}=\Delta t$ to compute $\rho_R(\tau)$. In addition, the mean $\mu(t)\equiv\avg{g(t)}$ and the variance $\sigma^2(t) = \avg{g^2(t)}-\avg{g(t)}$ are fully determined by the solutions in Eq. \ref{equ:sol_moment}. However, we still need to calculate the term $\avg{g(t+\tau;t+\tau-\tilde{\tau}) g(t;t-\tilde{\tau})}$. This term is easily calculated by splitting the calculation into two separate cases:
\begin{enumerate}
\item When $g(t+\tau;t+\tau-\tilde{\tau})$ and $g(t;t-\tilde{\tau})$ share the same underlying trajectory, i.e., when $0\leq\tau<\tilde{\tau}$.
\item When $g(t+\tau;t+\tau-\tilde{\tau})$ and $g(t;t-\tilde{\tau})$ are disjoint, i.e., when $\tau\geq\tilde{\tau}$.
\end{enumerate} We can then take advantage of the fact that $g(t_1;t_3)=g(t_1;t_2) + g(t_2;t_3)$, provided we preserve the underlying connection of the gene state $n$ at $t_2$.

\underline{Case 1: $0\leq\tau<\tilde{\tau}$.} Introducing $t_1:=t+\tau$, $t_2:=t$, $t_3:=t-\tilde{\tau}+\tau$ and $t_4:=t-\tilde{\tau}$ such that $t_1\geq t_2 \geq t_3 \geq t_4$, we get:
\begin{align}
\avg{g(t+\tau;t-\tilde{\tau}+\tau)g(t;t-\tilde{\tau})} &= \avg{(g(t_1;t_2) + g(t_2;t_3)) (g(t_2;t_3) + g(t_3;t_4))} \nonumber \\
&= \avg{g(t_1-t_2)}_{n|n} \avg{g(t_2-t_3)} + \avg{g(t_1-t_2)}_{n|n'}  \avg{g(t_3-t_4)} \label{equ:corr_te} \\
&+ \avg{g(t_2-t_3)}_{n|n}  \avg{g(t_3-t_4)} + \avg{g^2(t_2-t_3)}, \nonumber 
\end{align}
where the symbol $_{n|n}$ means that the products are ``connected'' through the temporal correlation in $n$. These products are computed as follows:
\begin{align*}
\avg{g(t_1-t_2)}_{n|n} \avg{g(t_2-t_3)} &= \sum_{n=0}^1 \avg{g(\tau|n)} \avg{g(n,\tilde{\tau}-\tau)} \\
\avg{g(t_1-t_2)}_{n|n'} \avg{g(t_3-t_4)} &= \sum_{n,n'=0}^1 \avg{g(\tau|n)} P(n|n';\tilde{\tau}-\tau) \avg{g(n',\tau)} \\
\avg{g(t_2-t_3)}_{n|n}  \avg{g(t_3-t_4)} &= \sum_{n=0}^1 \avg{g(\tilde{\tau}-\tau|n)} \avg{g(n,\tau)},
\end{align*}
where both $\avg{g(t|n)}$ and $\avg{g(n,t)}$ have been calculated above (Eq. \ref{equ:sol_moment} and \ref{equ:sol_transient}), and $P(n|n';t)$ is the propagator (i.e., the transition probabilities) of the telegraph process:
\begin{equation}
P(n|n';t) = (n\eta+(1-n)(1-\eta))(1-\exp{(-t/T)})+\delta_{nn'} \exp{(-t/T)},
\label{equ:tele_propa}
\end{equation}
where $\delta_{nn'}$ is the Kronecker delta. We thus have determined all the terms in Eq. \ref{equ:corr_te} and we can finally obtain:
\begin{align}
\avg{g(t+\tau;t-\tilde{\tau}+\tau)g(t;t-\tilde{\tau})} &= k\eta(\tilde{\tau}-\tau) + k^2\eta^2\tilde{\tau}^2 + 2 k^2 \eta (1-\eta)T(\tilde{\tau}-\tau) \label{equ:gg_taulesstilde}\\
&+ k^2 \eta(1-\eta)T^2 \left( \exp{(-(\tilde{\tau}+\tau)/T)} + \exp{(-(\tilde{\tau}-\tau)/T)} -2\exp{(-\tau/T)} \right). \nonumber
\end{align}
When $\tau=0$, the expression above reduces to $\avg{g^2(\tilde{\tau})}$, consistent with the fact that all the terms in Eq. \ref{equ:corr_te} vanish except $\avg{g^2(t_2-t_3)}=\avg{g^2(\tilde{\tau})}$. Since, $\avg{g^2(\tilde{\tau})}-\avg{g(\tilde{\tau})}^2=\sigma^2(\tilde{\tau})$, we thus properly recover $\rho(0)=1$ as expected.

\underline{Case 2: $\tau\geq\tilde{\tau}$.} Introducing $t_1:=t+\tau$, $t_2:=t+\tau-\tilde{\tau}$, $t_3:=t$ and $t_4:=t-\tilde{\tau}$ such that $t_1\geq t_2 \geq t_3 \geq t_4$, we get:
\begin{align*}
\avg{g(t+\tau;t-\tilde{\tau}+\tau)g(t;t-\tilde{\tau})}&= \avg{(g(t_1;t_2) g(t_3;t_4)} \\
&= \avg{g(t_1-t_2)}_{n|n'}  \avg{g(t_3-t_4)},
\end{align*}
where the connected product is given by
\begin{equation*}
\avg{g(t_1-t_2)}_{n|n'} \avg{g(t_3-t_4)} = \sum_{n,n'=0}^1 \avg{g(\tilde{\tau}|n)} P(n|n';\tau-\tilde{\tau}) \avg{g(n',\tilde{\tau})}.
\end{equation*}
Again, using the expression for $\avg{g(t|n)}$ and $\avg{g(n,t)}$ (Eq. \ref{equ:sol_moment} and \ref{equ:sol_transient}), and for $P(n|n';t)$ (Eq. \ref{equ:tele_propa}), we finally get:
\begin{align}
\avg{g(t+\tau;t-\tilde{\tau}+\tau)g(t;t-\tilde{\tau})} &= k^2\eta^2\tilde{\tau}^2 + k^2\eta(1-\eta)T^2 (1-\exp{(-\tilde{\tau}/T)})^2 \exp{(-\tau/T)}
\label{equ:gg_taugreatertilde}
\end{align}

\underline{Conclusion.} We have calculated the normalized auto-correlation functions $\rho_A(\tau)$ (Eq. \ref{equ:rhoA}, $\tilde{\tau}=\tau_{\rm elo}$) and $\rho_R(\tau)$ (Eq. \ref{equ:rhoR}, $\tilde{\tau}=\Delta t$) using the two-state model. Indeed, using Eq. \ref{equ:sol_moment}, \ref{equ:gg_taulesstilde} and \ref{equ:gg_taugreatertilde}, we found
\begin{equation}
\rho(\tau)=\left\{
\begin{aligned}
&\frac{\tilde{\tau}-\tau +  k(1-\eta)\left( 2 (\tilde{\tau}-\tau)T + T^2 \left( \exp{(-\frac{\tilde{\tau}+\tau}{T})} + \exp{(-\frac{\tilde{\tau}-\tau}{T})} -2\exp{(-\frac{\tau}{T})} \right) \right)} {\tilde{\tau} + 2 k (1-\eta) \left(T^2 \exp{(-\frac{\tilde{\tau}}{T})} + \tilde{\tau} T -T^2 \right)} &\text{if}\quad \tau&<\tilde{\tau}\\
&\quad\quad\quad\quad\quad\quad\quad\quad  \frac{ k(1-\eta)T^2 \left(1-\exp{(-\frac{\tilde{\tau}}{T})} \right)^2 \exp{(-\frac{\tau}{T})}} {\tilde{\tau} + 2 k (1-\eta) \left(T^2 \exp{(-\frac{\tilde{\tau}}{T})} + \tilde{\tau} T -T^2 \right)} &\text{if}\quad \tau&\geq\tilde{\tau}
\end{aligned}
\right.
\label{equ:autocorr_2s}
\end{equation}
Using the expression above, we plot $\rho_A(\tau)$ and $\rho_R(\tau)$ for different values of $\eta$ in fig. S5A (left and right respectively). For $\rho_A(\tau)$, we clearly see two regimes: one that is dominated by the elongation process $\tau<\tau_{\rm elo}$ and one that is only dictated by the switching process $\tau\geq\tau_{\rm elo}$. As $\rho_R(\tau)$ is unencumbered by the elongation process, only the exponential decay due to promoter switching remains. Importantly, these auto-correlation functions are identical for two independent gene copies (2 sister chromatids), since the correlated contributions and the variances add (in the independent case).

From Eq. \ref{equ:autocorr_2s}, the auto-correlation of the single-allele transcription rate $\rho_R(\tau)$ (with $\tilde{\tau}=\Delta t$) can be further simplified. Indeed, since for real data $\tau$ becomes a multiple of the sampling time $\Delta t$, we simply get
\begin{equation}
\rho_R(\tau) = \left\{
\begin{aligned}
&\quad\quad\quad\quad1 &\text{if}\quad \tau &=0\\
&\Sigma_{\rm AC} \exp{\left(-\frac{\tau-\Delta t}{T}\right)} &\text{if}\quad \tau&\geq\Delta t
\end{aligned}
\right.
\label{equ:autocorr_R}
\end{equation}
Here the amplitude of the correlated variability $\Sigma_{\rm AC}$ is given by
\begin{equation}
\Sigma_{\rm AC} = \frac{k\Delta t (1-\eta) \phi_2(\Delta t/T)}{1+k\Delta t (1-\eta) \phi_1(\Delta t/T)},
\label{equ:sigma_AC}
\end{equation}
with filtering functions $\phi_1(x)\in[0,1]$ and $\phi_2(x)\in[0,1]$ given by
\begin{align*}
\phi_1(x) &= 2\frac{\exp{(-x)}+x-1}{x^2} \\
\phi_2(x) &= \frac{\left(1-\exp{(-x)}\right)^2}{x^2}.
\end{align*}
From Eq. \ref{equ:autocorr_R}, we clearly see that in the 2-state model, the time scale of the exponential decay is solely dictated by the switching correlation time $T$. This suggests that the measured correlation time $\tau_{\rm AC}$ results from bursts and is in fact related to $T$ (Fig. 1E and 1F). Importantly, $\Sigma_{\rm AC}$ displays two opposite behaviors depending on whether the mean transcription rate $R=k\eta$ is varied through $k$ or $\eta$ (Eq. \ref{equ:sigma_AC} and fig. S5B). Indeed, an increase in $k$ leads to an increase in $\Sigma_{\rm AC}$, while an increase in $\eta$ leads to a decrease $\Sigma_{\rm AC}$. Interestingly, the scaling of $\Sigma_{\rm AC}$ with $R$ observed in data is very much in line with the latter scenario (fig. S5C), i.e. changes in $R$ driven by the ON-probability $\eta$. Lastly, we also see that $\Sigma_{\rm AC}$ vanishes when $\Delta t \rightarrow 0$ whereas the exponential decay is lost when $\Delta t \gtrsim T$. This implies there is a range of suitable $\Delta t$ that provide sufficient integration $\Delta t \gg 0$ but still remain sufficiently small $\Delta t <T$ to observe the correlation in the signal. Given, the range of measured $\Sigma_{\rm AC}\sim0.2$--$0.6$ and $\tau_{\rm AC}\sim1$--$2$min (see Fig. 1F and fig. S5C), it appears that our sampling time $\Delta t \approx 10\,$s is adequate. 

\paragraph{Validating auto-correlation on simulated data.} To assess our ability to estimate the auto-correlation (AC) function properly, we performed single allele deconvolution on simulated data mimicking experimental conditions. Specifically, we compared the estimated AC parameters ($\Sigma_{\rm AC}$ and $\tau_{\rm AC}$) obtained through deconvolution to the known input parameters used to generate the simulated data. For each combination of input parameters, we generated 200 synthetic alleles and 50 min long activity time series using the Gillespie algorithm \cite{Gillespie:2007bx}, including measurement noise consistent with real data and assuming two independent gene copies (i.e., sister chromatids) described by the 2-state model (See Section \ref{sec:validation_bursting} ``Validation using synthetic data'' for further details). Overall, we probed 56 combinations of (single gene copy) input parameters with $k\equiv K^{(1)}=8$ mRNA/min (consistent with real data), $\eta \equiv P_{\rm ON}^{(1)} \in \{ 0.03,0.12,0.23,0.36,0.52,0.78,0.90 \}$ and $T \equiv T_{\rm C}^{(1)} \in \{ 0.5,1,2,3,4,5,7,10 \}$ min. 

For each combination of input parameters, we calculated the input $\Sigma_{\rm AC}$ according to Eq. \ref{equ:sigma_AC}, while the input $\tau_{\rm AC}$ is simply given by the input $T$. Using the simulated data, we estimated $\Sigma_{\rm AC}$ and $\tau_{\rm AC}$ from the instantaneous single-allele transcription rate, as described in Section \ref{sec:ac_estimation} ``Transcription rate auto-correlation''. In fig. S5D, we show that for both $\Sigma_{\rm AC}$ and $\tau_{\rm AC}$, the input and the estimated AC parameters strongly correlate over a large range of $P_{\rm ON}^{(1)}$ and $T_{\rm C}^{(1)}$ values, demonstrating our ability to faithfully determine $\Sigma_{\rm AC}$ and $\tau_{\rm AC}$ from deconvolved instantaneous single-allele transcription rates. Thus, the striking collapse of $\Sigma_{\rm AC}$ and $\tau_{\rm AC}$ observed in real data (Fig. 1F and fig. S5C), across genes and activity, should not be markedly affected by our approach and appears to be a solid result.

Lastly, to assess the potential impact of errors in the elongation rate on the deconvolution process and the auto-correlation estimation, we compared the estimated AC parameters using the correct deconvolution kernel (used to generate the simulated data and computed here with $K_{\rm elo}=2\,$mRNA/min) and incorrect kernels (computed with either with $K_{\rm elo}=1.5$ or $2.5\,$mRNA/min, i.e., a $\pm 25\%$ error). We focused on the analysis of the $T_{\rm C}^{(1)}=2\,$min simulated data set (that is closest to real data) and show the estimated $\tau_{\rm AC}$ and $\Sigma_{\rm AC}$ using the three kernels for all values of input $P_{\rm ON}^{(1)}$ (in fig. S5E). The analysis determines that while the effect of a 25\% deviation on $K_{\rm elo}$ has only a very limited impact on $\Sigma_{\rm AC}$ (resulting error on $\Sigma_{\rm AC}\sim$4\%), the estimation of $\tau_{\rm AC}$ is almost entirely unaffected by a 25\% overestimation of $K_{\rm elo}$ (error on $\tau_{\rm AC}\sim$7\%). However, an underestimation of $K_{\rm elo}$ leads to a significant underestimation of $\tau_{\rm AC}$, albeit with errors on $\tau_{\rm AC}$ not exceeding 25\% on average. Overall, this demonstrates the robustness of our approach to changes in the elongation rate.

\subsection{Burst calling and transcriptional parameter estimation}
\label{sec:calling_burst}

\paragraph{Calling bursts from deconvolved time series.} To call bursts from individual single-allele transcriptional activity time series $A$, we opt for a simple approach with minimal mechanistic assumptions. We simply aim to cluster the deconvolved configurations of initiation events $I$ underlying $A$ over time (see Eq. \ref{equ:bayes_theorem} and Fig. 1D). We first compute the single-allele transcription rate $r(t)$ from $I$ (see Eq. \ref{equ:singler_estimate}). We then perform a (centered) temporal moving average of $r(t)$ using the following kernel:
\begin{equation}
\kappa(t) = \frac{1}{Z(w)}\exp{\left(-\left(\frac{2t}{w}\right)^4\right)},
\label{equ:clustering_kernel}
\end{equation} 
where $w$ is the width of the moving window and $Z(w)=\sum_{i=1}^{N_t} \kappa(t_i)$ is the normalization. We compute the moving average through discrete convolution resulting in the following rate $r_w(t) = r(t)*\kappa(t)$. Subsequently, a threshold $r_b=g_b/w$ is applied to the resulting rate estimate $r_w$, where $g_b$ corresponds to the desired minimum number of initiations events for a burst. This threshold allows us to determine the cluster of initiation events (bursts) defining the allele's ON-state $n(t)\in\{0,1\}$:
\begin{equation}
n(t) = \left\{
\begin{aligned}
1&\quad\text{if}& r_w(t) &\geq r_b\\
0&\quad\text{if}& r_w(t) &< r_b
\end{aligned}
\right.
\label{equ:n_threshold}
\end{equation}
It follows that the clustering approach described above and the ensuing definition of the ON state $n(t)$ depend on two free parameters: the width of the averaging window $w$ and the minimal number of initiation events $g_b$. As a rule of thumb, $w$ needs to be large enough to average the Poisson initiation fluctuations and prevent the calling of spurious bursts, but small enough to preserve the temporal correlations attributed to ON-OFF switching in the data. In practice, we set $w=5\Delta t\approx50\,$s, which is slightly below the measured correlation time $\tau_{\rm AC}\sim1$--$2\,$min (Fig. 2F). Regarding the threshold, we set $g_b=2$ corresponding to a detection of at least 2 initiation events within the moving window, which is motivated by our detection sensitivity of around 1--2 mRNAs (see fig. S1E. and Section \ref{sec:meas_error}). Notably, the shape of the kernel (Eq. \ref{equ:clustering_kernel}) and the chosen parameter values ($w=5\Delta t$ and $g_b=2$) set a lower bound on the minimal ON interval duration that can be reliably detected, which is $3\Delta t\approx30\,$s. This value corresponds to the duration of the ON interval calculated from $n(t)$ with input $I(t)=g_b \delta(t)$. To verify the impact of our choice of clustering parameters on the estimation of bursting parameters go to Section \ref{sec:validation_bursting} Validation using synthetic data.

Importantly, our estimation of the ON state $n$ through clustering can be applied to each individual initiation configuration $I$ sampled from the posterior distribution $P(I|A)$ (Eq. \ref{equ:bayes_theorem}). It implies that we can easily reconstruct $P(n|A)$ for each allele, as we did for the instantaneous distribution of the single-allele transcription rate $P(r|A)$ (see Section \ref{sec:singler_estimate}; Estimating single allele transcription rate). Using our MCMC samples we get
\begin{equation}
P(n|A) = \sum_I P(n|I) P(I|A),
\end{equation}
where $P(n|I)=1$ when Eq. \ref{equ:n_threshold} is satisfied and $P(n|I)=0$ otherwise. The distribution $P(n|A)$ allows us to assess the uncertainty on burst calling for each time series of transcriptional activity $A$.

\paragraph{Estimating transcriptional parameters.} \label{sec:estimating_param} To characterize the bursting dynamics of each gap gene in space and time, we define a set of bursting parameters and estimate them empirically from the single-allele transcription rates $r(t)$ and the ON-state $n(t)$ (see fig. S6A-B). For a given single-allele transcriptional activity time series $A$, we sampled individual initiation configurations $I_s$ from the posterior distribution $P(I|A)$ using MCMC (see Section \ref{sec:single_transrate}). The index $s\in \{1,2,\ldots,N_s\}$ denotes one specific MCMC sample drawn from the posterior, where $N_s=10^3$ is the total number of samples used per allele. For each $I_s$, we further computed $r_s(t)|A$ (Eq. \ref{equ:singler_estimate}) and $n_s(t)|A$ (Eq. \ref{equ:n_threshold}), where the index $s$ and the notation $|A$ specify that these come from the sample $I_s$ given the time series $A$. We then defined the mean transcription rate $R$ and the ON-probability $P_{\rm ON}$ over all nuclei in a specific spatiotemporal $(x,t)$-bin, that is over all the $A\in \mathcal{A}(x,t)$. For a single MCMC sample $s$ per $A$ we get
\begin{align}
R_s(x,t) &= \frac{1}{N_A(x,t)} \sum_{A\in \mathcal{A}(x,t)} r_s(t)|A \\
P_{{\rm ON},s}(x,t) &= \frac{1}{N_A(x,t)} \sum_{A\in \mathcal{A}(x,t)} n_s(t)|A,
\end{align}
where $N_A(x,t) = \#\mathcal{A}(x,t)$. We then compute a single point-estimate per bursting parameter $\Theta$ for each position $x$ and time $t$ by averaging over our MCMC samples as follows
\begin{align}
\Theta(x,t) :=& \avg{\Theta_s(x,t)} = \frac{1}{N_s} \sum_{s=1}^{N_s} \Theta_s(x,t) \label{equ:mcmc_avg} \\
\sigma_{\Theta}(x,t) =& \sqrt{\avg{(\Theta_s(x,t)-\Theta(x,t))^2}},
\end{align}
where $\sigma_{\Theta}$ is our estimation of the error on $\Theta$. Thus, the mean transcription rate and the ON-probability are given by $R(x,t)=\avg{R_s(x,t)}$ and $P_{\rm ON}(x,t)=\avg{P_{{\rm ON},s}(x,t)}$ (see Fig. 2A and fig. S6C). Importantly, this construction of $R(x,t)$ is equivalent to our previous calculation using the mixture of posterior distributions (see Eq. \ref{equ:R_mean}).

The mean initiation rate $K$ is defined over all nuclei in a spatiotemporal $(x,t)$-bin by averaging the single-allele transcription rate $r(t)$ conditioned on the allele being ON ($n(t)=1$). It is expressed as
\begin{equation}
\begin{aligned}
K_{s}(x,t) &= \frac{1}{Z_K} \sum_{A\in \mathcal{A}(x,t)} r_s(t)|A \cdot n_s(t)|A \\
Z_K &= \sum_{A\in \mathcal{A}(x,t)} n_s(t)|A,
\end{aligned}
\label{equ:ini_rate}
\end{equation}
where $Z_K$ counts the number of ON alleles at time $t$. Averaging over MCMC samples (Eq. \ref{equ:mcmc_avg}) leads to the the mean initiation rate $K(x,t)=\avg{K_s(x,t)}$ (see fig. S6D). By construction $R=K\cdot P_{\rm ON}$, which we verify empirically (see fig. S6F). As additional control, we define a ``leaking'' rate $K_L$ by averaging the single-allele transcription rate $r(t)$ conditioned on the allele being OFF ($n(t)=0$):
\begin{equation}
\begin{aligned}
K_{L,s}(x,t) &= \frac{1}{Z_L} \sum_{A\in \mathcal{A}(x,t)} r_s(t)|A \cdot (1-n_s(t)|A) \\
Z_L &= \sum_{A\in \mathcal{A}(x,t)} 1-n_s(t)|A.
\end{aligned}
\end{equation}
We verified that $K_L(x,t)=\avg{K_{L,s}(x,t)}$ is indeed negligible compared to $K(x,t)$, supporting our ability to identify well-demarcated bursts (fig. S6G).

To calculate the mean durations of the ON and OFF periods ($T_{\rm ON}$ and $T_{\rm OFF}$) for each spatiotemporal $(x,t)$-bin, we first determine the boundaries of the ON ($n(t)=1$) and OFF ($n(t)=0$) periods for individual alleles. Using $n(t)$, we built a function $b(t)$ such that $b(t)=1$ whenever time point $t$ corresponds to an ON-OFF boundary: 
\begin{equation}
b(t) = \left\{
\begin{aligned}
1&\quad\text{if}& |n(t+\Delta t)-n(t)|&>0\\
0&\quad\text{if}& |n(t+\Delta t)-n(t)|&=0
\end{aligned}
\right.
\end{equation}
where $\Delta t$ is the sampling time in our measurements. $b(t)$ defines ON-OFF boundaries that are consistent with our definition of the instantaneous single-allele transcription rate $r(t)$ (see Eq. \ref{equ:singler_estimate}), i.e., we associate the state $n(t)$ and the rate $r(t)$ to the same time interval $(t-\Delta t,t]$. For convenience, we also enforce $b(t_1)=1$ and $b(t_{N_t})=1$ for the first and last time point $t_1$ and $t_{N_t}$ of the time series. 

From $b(t$) we construct the set of boundary time points, i.e, $\mathcal{T}=\{t:b(t)=1\}= \{t_{b_1},\dots,t_{b_k}\}$ such that $b_j<b_{j+1}$ $\forall j \in \{1,2,\ldots,k-1\}$ where $k\geq2$ (in each case $t_{b_1}=t_1$ and $t_{b_k}=t_{N_t}$). To measure the durations of individual ON and OFF periods, we introduce a function $\tau(t)$ such that
\begin{equation}
\tau(t) = \left\{
\begin{aligned}
t_{b_{2}}-t_{b_{1}} &\quad\text{if}\quad t = t_1\\
t_{b_{j+1}}-t_{b_{j}} &\quad\text{else if}\quad t\in(t_{b_{j}},t_{b_{j+1}}]
\end{aligned}
\right.
\end{equation}
We further introduce a function $\pi(t)=\tau(t)/\Delta t$ that counts the number of $\Delta t$ intervals underlying $\tau(t)$. We calculate $T_{\rm ON}$ from the activity time series $A$ as
\begin{equation}
\begin{aligned}
T_{{\rm ON},s}(x,t) &= \frac{1}{Z_{\rm ON}} \sum_{A\in \mathcal{A}(x,t)} \tau_s(t)|A \cdot \frac{n_s(t)|A}{\pi_s(t)|A} \\
Z_{\rm ON} &= \sum_{A\in \mathcal{A}(x,t)} \frac{n_s(t)|A}{\pi_s(t)|A}, 
\end{aligned}
\label{equ:mean_ton}
\end{equation}
where $Z_{\rm ON}$ counts the number of ON alleles at time $t$ weighted by the inverse of the length of the underlying ON periods ($s$ denotes an individual MCMC sample as before). The rationale behind the $\frac{n_s(t)|A}{\pi_s(t)|A}/Z_{\rm ON}$ weights is as follows: as we aim to estimate mean durations across nuclei at time $t$ (fig. S8A), we need to account for the length of each period $\tau(t)$ and correct for their overall contribution to $\pi(t)$ time points (e.g. longer periods contribute to more time points, thus we need weights that balance over-counting). 

Finally, by averaging over MCMC samples (Eq. \ref{equ:mcmc_avg}), we obtain the mean ON duration $T_{\rm ON}(x,t)=\avg{T_{{\rm ON},s}(x,t)}$ (see Fig. 2C). Likewise, for $T_{\rm OFF}$, we have
\begin{equation}
\begin{aligned}
T_{{\rm OFF},s}(x,t) &= \frac{1}{Z_{\rm OFF}} \sum_{A\in \mathcal{A}(x,t)} \tau_s(t)|A \cdot \frac{1-n_s(t)|A}{\pi_s(t)|A} \\
Z_{\rm OFF} &= \sum_{A\in \mathcal{A}(x,t)} \frac{1-n_s(t)|A}{\pi_s(t)|A}.
\end{aligned}
\label{equ:mean_tof}
\end{equation}
Again, averaging over MCMC samples (Eq. \ref{equ:mcmc_avg}) leads to the mean OFF duration $T_{\rm OFF}(x,t)=\avg{T_{{\rm OFF},s}(x,t)}$ (see Fig. 2C).

When estimating the initiation rate $K$ (Eq. \ref{equ:ini_rate}), and the mean durations of the ON and OFF periods $T_{\rm ON}$ (Eq. \ref{equ:mean_ton}) and $T_{\rm OFF}$ (Eq. \ref{equ:mean_tof}), we sometimes encounter missing values, e.g. absence of ON or OFF periods over all nuclei at a given time point $t$. This typically occurs when $P_{\rm ON}(t)\rightarrow0$ or $P_{\rm ON}(t)\rightarrow1$. Therefore, we filter each bursting parameter estimate $\Theta_s(x,t)$ using a Gaussian kernel over a short timescale of 1 min. This ``smoothing" interpolates the eventual missing values and prevents gaps in our time-dependent parameter estimation. 

Using $K$, $T_{\rm ON}$ and $T_{\rm OFF}$, we can further define the following bursting parameters:
\begin{align}
T_C :=& \frac{T_{\rm ON} T_{\rm OFF}}{T_{\rm ON}+T_{\rm OFF}} \label{equ:def_tc} \\
B :=& K \cdot T_{\rm ON} \\
F :=& \frac{1}{T_{\rm ON}+T_{\rm OFF}},
\end{align}
where $T_C$ is the switching correlation time, $B$ the burst size and $F$ the burst frequency. These parameters were directly computed from our estimates of $K$, $T_{\rm ON}$, and $T_{\rm OFF}$ as defined above. Importantly, we tested extensively our ability to estimate static and time-dependent bursting parameters correctly using simulated data (see Section \ref{sec:validation_bursting}; Validation using synthetic data).

\paragraph{Sister chromatids and single copy parameters} \label{sec:sgc_param} In the blastoderm phase of the Drosophila embryo, DNA replication occurs during the first few minutes of each nuclear cycle (2--3$\,$min after mitosis). Therefore, as our microscopy does not allow for separating individual sister chromatids, imaged spots decorated with fluorescently labeled nascent transcripts typically correspond to two optically unresolved sites of transcription. Assuming the measured activity results from two \textit{identical} and \textit{independent} alleles, we transform our effective bursting parameters into single-gene copy (SGC) parameters. In that case, the mean SGC transcription rate $R^{(1)}$ is simply given by $R^{(1)}=R/2$, which is half the effective one. In addition, the SGC ON-probability $P_{\rm ON}^{(1)}$ must satisfy
\begin{equation*}
\underbrace{(1-P_{\rm ON}^{(1)})^2}_{P(0 \text{ copy active})} + \underbrace{2P_{\rm ON}^{(1)}(1-P_{\rm ON}^{(1)}) +{P_{\rm ON}^{(1)}}^2}_{P(1\geq \text{ copy active})} = 1,
\end{equation*}
where the probability to observe at least one active gene copy corresponds to the effective $P_{\rm ON}$. Thus, $1-(1-P_{\rm ON}^{(1)})^2 = P_{\rm ON}$ and we obtain the SGC ON-probability given by
\begin{equation}
P_{\rm ON}^{(1)}=1-(1-P_{\rm ON} )^{1/2}.
\label{equ:sgc_pon}
\end{equation}
Regarding the SGC mean initiation rate $K^{(1)}$, it must satisfy $K^{(1)}=R^{(1)}/P_{\rm ON}^{(1)}= K P_{\rm ON}/(2P_{\rm ON}^{(1)})$. Thus, using Eq. \ref{equ:sgc_pon} we get
\begin{equation}
K^{(1)}=\frac{K}{2} \left(1+(1-P_{\rm ON})^{1/2} \right).
\label{equ:sgc_k}
\end{equation}

For a single gene copy, the switching correlation time $T_{\rm C}^{(1)}$ is exactly given by $T_{\rm C}^{(1)}=1/(k_{\rm ON}+k_{\rm OFF})$, where $k_{\rm ON}$ and $k_{\rm OFF}$ are the ON and OFF rates of the 2-state model (see Section \ref{sec:2state}). For a system made of two identical and independent gene copies we can write the following kinetic equation:
\begin{equation}
\lambda_0 \overset{2 k_{\rm ON}}{\underset{k_{\rm OFF}} \rightleftharpoons} \lambda_1 \overset{k_{\rm ON}}{\underset{2 k_{\rm OFF}} \rightleftharpoons} \lambda_2,
\label{equ:3state}
\end{equation}
where $\lambda_0$, $\lambda_1$, and $\lambda_2$ are states corresponding to zero, one, and two active gene copies, respectively. Again, the effective OFF-state is equivalent to $\lambda_0$, while the effective ON-state is either $\lambda_1$ or $\lambda_2$. Further assuming steady-state, from Eq. \ref{equ:3state} we calculate the mean residence time in the OFF and ON states (using phase-type distributions), i.e., $T_{\rm OFF}$ and $T_{\rm ON}$ as a function of $k_{\rm ON}$ and $k_{\rm OFF}$:
\begin{equation}
\begin{aligned}
T_{\rm OFF}&=\frac{1}{2 k_{\rm ON}} \\
T_{\rm ON}&=\frac{k_{\rm ON} + 2k_{\rm OFF}}{2 k_{\rm OFF}^2}.
\end{aligned}
\label{equ:times}
\end{equation}
From our definition of the effective switching correlation time $T_{\rm C}$ (Eq. \ref{equ:def_tc}), and the steady state relationships $k_{\rm ON}=P_{\rm ON}^{(1)}/T_{\rm C}^{(1)}$ and $k_{\rm OFF}=(1-P_{\rm ON}^{(1)})/T_{\rm C}^{(1)}$, we get
\begin{equation*}
T_{\rm C} := \frac{T_{\rm ON} T_{\rm OFF}}{T_{\rm ON}+T_{\rm OFF}} = \frac{k_{\rm ON}+2k_{\rm OFF}}{2(k_{\rm ON}+k_{\rm OFF})^2} = \frac{1}{2}T_{\rm C}^{(1)}(2-P_{\rm ON}^{(1)}).
\end{equation*}
Using Eq. \ref{equ:sgc_pon}, we express the SGC switching correlation time as a function of $T_{\rm C}$ and $P_{\rm ON}$:
\begin{equation}
T_{\rm C}^{(1)}= \frac{2 T_{\rm C}}{1+(1-P_{\rm ON})^{1/2}}
\label{equ:sgc_tc}
\end{equation}
Together Eq. \ref{equ:sgc_pon}, \ref{equ:sgc_k} and \ref{equ:sgc_tc} provide the key relationships to transform our effective bursting parameters to SGC parameters (see fig. S10E-H, S7B-G and S12B). Note that the relationships between $T_{\rm OFF}$, $T_{\rm ON}$ and $T_{\rm C}^{(1)}$ (Eq. \ref{equ:times} and \ref{equ:sgc_tc}) are in principle valid only near steady state. Nevertheless, these relationships provide good approximations in the non-stationary case given that the temporal changes of the underlying kinetics over developmental time are slow (see Section \ref{sec:validation_bursting}; Validation using synthetic data).

\paragraph{Predicting time-dependent bursting parameters.} \label{sec:predicting_param} To test our deconvolution and burst calling approaches on simulated data (modeled via the master equation), we need to predict the time dependence of the effective bursting parameters (ground truth) given the underlying input kinetic parameters. Indeed, most bursting parameters (i.e., $P_{\rm ON}(t)$, $T_{\rm ON}(t)$, $T_{\rm OFF}(t)$ and $T_{\rm C}(t)$; but not $K(t)$) result from non-trivial time integration of the underlying gene switching rates $k_{\rm OFF}(t)$ and $k_{\rm ON}(t)$. In addition, the time scales $T_{\rm ON}(t)$, $T_{\rm OFF}(t)$ and $T_{\rm C}(t)$ when estimated from data are typically subjected to biases due to discreteness (e.g., missing events) and the finite nature (e.g., censoring) of the measured time intervals. To account for these effects, we compute the expected $P_{\rm ON}(t)$, $T_{\rm ON}(t)$, $T_{\rm OFF}(t)$ and $T_{\rm C}(t)$ given a 2-state model of transcription (for either a single allele or for each of the two independent alleles of the replicated sister chromatids) and the kinetic rates $k_{\rm OFF}(t)$ and $k_{\rm ON}(t)$ \cite{Zoller:2018gj,Peccoud1995}.

Starting with the time evolution of the gene state $n$ (with $n\in \{0,1\}$ for a single gene copy and $n\in \{0,1,2\}$ in the two gene copies system), the non-stationary master equation of the system \cite{Lestas:2008hy,Walczak:2012} is given by
\begin{equation}
\frac{d}{dt} \mathbf{P}(t) = \hat{M}(t) \mathbf{P}(t),
\label{equ:master_t}
\end{equation}
where $\mathbf{P}(t)$ is the time-dependent vector whose components are the probabilities to find the system in any of the states $n$ at time $t$. Thus, $\sum_n P_n(t) =1$ with the sum taken over all components of $\mathbf{P}(t)$. $\hat{M}(t)$ is the time-dependent state transition matrix containing the propensity functions of the different reactions. For a single gene copy, $\hat{M}(t)$ is given by
\begin{equation}
\hat{M}_1(t)= \begin{pmatrix}
  -k_{\rm ON}(t) & k_{\rm OFF}(t)\\
  k_{\rm ON}(t) & -k_{\rm OFF}(t)\\
 \end{pmatrix}.
 \label{equ:switch_operator}
\end{equation}
For two independent and identical gene copies, $\hat{M}(t)$ is given by
\begin{equation}
\hat{M}_2(t)= \begin{pmatrix}
  -2 k_{\rm ON}(t) & k_{\rm OFF}(t) & 0\\
  2 k_{\rm ON}(t) & -(k_{\rm OFF}(t)+k_{\rm ON}(t)) & 2 k_{\rm OFF}(t) \\
  0 & k_{\rm ON}(t) & -2 k_{\rm OFF}(t) \\
 \end{pmatrix}.
 \label{equ:switch_operator_2Al}
\end{equation}

The solution of the non-stationary master equation for the system (Eq. \ref{equ:master_t}) is given by:
\begin{equation}
\mathbf{P}(t) = \hat{U}(t,t_1) \mathbf{P}(t_1),
\label{equ:transition_prob}
\end{equation}
where $\mathbf{P}(t_1)$ is the initial condition vector, and the propagator $\hat{U}(t,t_1)\equiv{\rm OE}[\hat{M}](t,t_1)$ is the time-ordered exponential of $\hat{M}(t)$. Intuitively, $\hat{U}(t,t_1)$ is a matrix describing the time evolution of the system whose entries correspond to the transition probabilities, i.e., $\hat{U}_{nn'}(t,t_1) = P(n;t|n';t_1)$, such that $P_n(t)=\sum_{n'}\hat{U}_{nn'}(t,t_1)P_{n'}(t_1)$. There are multiple ways to define the time-ordered exponential ${\rm OE}[\hat{M}](t,t_1)$; here we define it such that our computation becomes an infinite product of matrix exponential:
\begin{equation}
{\rm OE}[\hat{M}](t,t_1) = \lim_{N\rightarrow \infty} \left( \exp{\left(\int_{t_{N-1}}^{t_N} \hat{M}(t')dt'\right)} \ldots \exp{\left(\int_{t_{2}}^{t_{3}} \hat{M}(t')dt'\right)} \exp{\left(\int_{t_{1}}^{t_{2}} \hat{M}(t')dt'\right)} \right),
\end{equation}
where $t_i=t_1+(i-1)\Delta t$ for $i\in\{1,\ldots,N\}$ and $\Delta t = (t-t_1)/(N-1)$. Provided the temporal changes in $k_{\rm OFF}(t)$ and $k_{\rm ON}(t)$ are small over individual $\Delta t$, the product above provides a good approximation to the time-ordered exponential when $N$ is finite. In practice, we set $\Delta t$ to our sampling time ($\sim\!10\,$s) and we approximated the integral using the trapezoidal rule such that $\int_{t_{i-1}}^{t_i} \hat{M}(t)dt\approx \frac{1}{2}(\hat{M}(t_i)+\hat{M}(t_{i-1}))\Delta t$ leading to
\begin{equation}
{\rm OE}[\hat{M}](t_i,t_1)\approx \exp{\left( \frac{1}{2}(\hat{M}(t_i)+\hat{M}(t_{i-1}))\Delta t \right)} \ldots \exp{\left( \frac{1}{2}(\hat{M}(t_2)+\hat{M}(t_1))\Delta t \right)}.
\label{equ:time_ord}
\end{equation}
This approximation is quite convenient for discretized time intervals as is often the case for real data. Moreover, the matrix exponential is straightforward to compute and rather well-behaved (i.e., it preserves the normalization of the columns to one) \cite{Sidje:1998uq}.

We compute $P_{\rm ON}(t)$ using the transition probabilities determined by the propagator (Eq. \ref{equ:transition_prob} and \ref{equ:time_ord}), namely $P_{\rm ON}(t)=\sum_{n>0}P_n(t)$ with $P_n(t)=\sum_{n'}\hat{U}_{nn'}(t,t_1)P_{n'}(t_1)$. Incidentally, for a single gene copy in the stationary case ($\hat{M}$ given by Eq. \ref{equ:switch_operator} with $k_{\rm ON}$ and $k_{\rm OFF}$ constant in time), the computation of $P_{\rm ON}(t)$ above becomes equivalent to solving the moment equation for $\avg{n(t)}$, introduced previously (see Eq. \ref{equ:time_dep_moment}).

To determine the expected $T_{\rm ON}(t)$ and $T_{\rm OFF}(t)$, we compute the discrete residence time distribution of the ON and OFF states in the non-stationary case. We define the sets of residence states $\mathcal{V}_r$ and absorbing states $\mathcal{V}_a$ such that $n \in \mathcal{V}_r \cup \mathcal{V}_a$ with $\mathcal{V}_r \cap \mathcal{V}_a = \text{\O}$. For the ON-time we have $\mathcal{V}_r=\{1\}$ and $\mathcal{V}_a=\{0\}$ (1 gene copy) or $\mathcal{V}_r=\{1,2\}$ and $\mathcal{V}_a=\{0\}$ (2 gene copies). For the OFF-time we have $\mathcal{V}_r=\{0\}$ and $\mathcal{V}_a=\{1\}$ (1 gene copy) or $\mathcal{V}_r=\{0\}$ and $\mathcal{V}_a=\{1,2\}$ (2 gene copies). We define the probability $P_r(\tau=t_j-t_i|t_i)$ that the system has just settled at time $t_i$ in any of the residence states ($n(t_i)\in\mathcal{V}_r$) and is found for the first time after a duration $\tau$ in any absorbing state ($n(t_j)\in\mathcal{V}_a$). Given that time is discrete and finite, with $t_{i+1}-t_{i}=\Delta t$ $\forall i\in \{1,\ldots,N_t-1\}$ and $N_t$ the total number of time points, the residence time (or first-passage time) distribution $P_r(\tau=t_j-t_i|t_i)$ can be written as
\begin{equation}
P_r(\tau=t_j-t_i|t_i) \equiv W_{ji},
\end{equation}
where $W$ is a triangular matrix (with $W_{ji}=0$ if $j<i$) whose columns correspond to truncated discrete phase-type distributions (truncated due to a finite number of time points $N_t$). The coefficients $W_{ji}$ are defined as follows
\begin{equation}
W_{ji}=\left\{
\begin{aligned}
\sum_{n \in \mathcal{V}_a} Q_n(t_j,t_i)&&\text{if}\quad i\leq j < N_t \\
1-\sum_{k=i}^{N_t-1} W_{ki}&&\text{if}\quad j=N_t 
\end{aligned}\right.
\end{equation}
where the value for $j=N_t$ ensures that the $W_{ji}$ form normalized distributions along $j$, i.e. $\sum_{j=i}^{N_t} W_{ji}=1$. In practice, these distributions were computed by defining $Q_n(t_j,t_i)$ recursively for $1\leq i < j \leq N_t$ using the propagator (Eq. \ref{equ:transition_prob} and \ref{equ:time_ord}):
\begin{equation}
Q_n(t_j,t_i)=\sum_{n'\in \mathcal{V}_r} \hat{U}_{nn'}(t_j,t_{j-1})Q_{n'}(t_{j-1},t_i)
\end{equation}

The initial condition of our recursion $Q_{n}(t_i,t_i)$ (when $i=j$) corresponds to the probability that the system has just settled in any of the residence states at time $t_i$. For $1<i\leq N_t$, this probability is given by
\begin{equation}
Q_n(t_i,t_i) = \left\{
\begin{aligned}
\frac{1}{Z_i} \sum_{n' \in \mathcal{V}_a}  \hat{U}_{nn'}(t_i&,t_{i-1})P_{n'}(t_{i-1}) &\text{if}\quad n\in \mathcal{V}_r \\
0& &\text{otherwise}
\end{aligned}\right.
\label{equ:ini_qi}
\end{equation}
where $Z_i=\sum_{n \in \mathcal{V}_r} \sum_{n' \in \mathcal{V}_a}  \hat{U}_{nn'}(t_i,t_{i-1})P_{n'}(t_{i-1})$ with $P_{n}(t)$ the occupancy of the states given by Eq. \ref{equ:transition_prob}. For $i=1$, $Q_n(t_1,t_1)$ is actually ill-defined and we chose
\begin{equation}
Q_n(t_1,t_1) = \left\{
\begin{aligned}
\frac{1}{Z_1} P_{n}&(t_{1}) &\text{if}\quad n\in \mathcal{V}_r \\
0& &\text{otherwise}
\end{aligned}\right.
\label{equ:ini_q1}
\end{equation}
where $Z_1=\sum_{n \in \mathcal{V}_r} P_{n}(t_{1})$. Consistent with our definition of $P_r(\tau|t_i)$, the probability that the system remains in a given state for a null duration is zero. Indeed, by construction $P_r(\tau=0|t_i) \equiv W_{ii}=0$ since $W_{ii}=\sum_{n \in \mathcal{V}_a} Q_n(t_t,t_i)$ where $Q_n(t_t,t_i)=0$ $\forall n\in \mathcal{V}_a$.

To compute the expected $T_{\rm ON}(t)$ and $T_{\rm OFF}(t)$ according to our definition (see Eq. \ref{equ:mean_ton} and \ref{equ:mean_tof}), we renormalize the computed residence time distributions. Specifically, we want to compute the average of $\tau=t_j-t_i$ for all possible $t_i$ given $t_j$, meaning we must properly normalize the $W_{ij}$ along $i$ (they only form distributions along $j$). To do so, we first need to weight each $t_i$ according to the probability that the system settled in any residence state at that particular time, which is given by the previously computed $Z_i$ (see Eq. \ref{equ:ini_qi} and \ref{equ:ini_q1}). In addition, we need to take into account that any $\tau$ contributes to $j-i$ time points along $j$. Thus, with that in mind, we defined the following renormalized coefficient $\tilde{W}_{ji}$ for $1\leq i < j \leq N_t$:
\begin{equation}
\tilde{W}_{ji} = \frac{W_{ji}}{j-i} \frac{Z_i}{\tilde{Z}_j},
\end{equation}
with $\tilde{Z}_j = \sum_{i<j} \frac{W_{ji}}{j-i} \frac{Z_i}{\tilde{Z}_j}$, such that the $\tilde{W}_{ji}$ properly form distributions along $i$. Note that in practice, the $Z_i$'s in the equation above cancel out with those in Eq. \ref{equ:ini_qi} and \ref{equ:ini_q1}. The expected $T_{\rm ON}(t)$ and $T_{\rm OFF}(t)$ are then computed using the $\tilde{W}_{ji}$ as
\begin{equation}
T(t_j) = \sum_{i<j} \tilde{W}_{ji} (t_j-t_i),
\label{equ:T_theo}
\end{equation}
where the appropriate residence state ($n\in \mathcal{V}_r$) and absorbing state ($n\in \mathcal{V}_a$) have to be used to determine the $\tilde{W}_{ji}$. Thus, Eq. \ref{equ:T_theo} allows us to computationally predict the expected $T_{\rm ON}(t)$ and $T_{\rm OFF}(t)$, while accounting for discrete and finite number of time points time.

\paragraph{Validation using synthetic data.}
\label{sec:validation_bursting}

To validate our Bayesian deconvolution and burst calling approaches, we tested our method on synthetic data mimicking experimental conditions. In essence, we used a 2-state model to generate single-gene copy initiation events $I$ given a set of input parameters $\Theta = \{k,k_{\rm ON},k_{\rm OFF} \}$, where $k$ is the initiation rate and $k_{\rm ON}$ and $k_{\rm OFF}$ are the gene switching rates. The resulting initiation events are convolved with the elongation kernel $\kappa$ of the gene \emph{hb} (see Eq. \ref{equ:elong_kernel}) to generate transcriptional activity time series $G$, i.e., $G= \kappa * I$. Assuming two independent sister chromatids as in real data, we sum over two independently generated transcriptional activity time series $G$ per nucleus. Finally, we add the characterized measurement noise on the resulting activity according to Eq. \ref{equ:meas_noise} and \ref{equ:adding_noise}, thus generating realistic single allele activity time series $A$ per simulated nucleus.

\underline{Stationary bursting parameters.} We first investigate the performance of our method in recovering stationary bursting parameters (that do not vary in time). We generate an extensive simulated data set using the following input parameter ranges: $k\equiv K^{(1)}=8\,$mRNA/min (consistent with real data), $k_{\rm ON} = \eta/T$ and $k_{\rm OFF} = (1-\eta)/T$ with $\eta \equiv P_{\rm ON}^{(1)} \in \{ 0.03,0.12,0.23,0.36,0.52,0.78,0.90 \}$ and $T \equiv T_{\rm C}^{(1)} \in \{ 0.5,1,2,3,4,5,7,10 \}$ min. For each of the 56 combinations of input parameters $\{K^{(1)},T_C^{(1)},P_{\rm ON}^{(1)}\}$, we generate 200 alleles and $50\,$min long activity time series $A$ at $10\,$s intervals (as in real data). For each allele, the underlying initiation events of the 2-state model are sampled using the Gillespie algorithm \cite{Gillespie:2007bx}. 

We next perform single-allele deconvolution and burst calling (see Sections \ref{sec:single_transrate} and \ref{sec:calling_burst}) to estimate the effective mean transcription parameters ($R$, $K$, $P_{\rm ON}$, $T_{\rm ON}$, $T_{\rm OFF}$ and $T_{\rm C}$) for each simulated set of alleles. As an example, we show both the predicted parameters based on input (fig. S9A-B top, also see \ref{sec:predicting_param}) and the resulting estimated parameters (fig. S9A-B bottom, also see \ref{sec:estimating_param}) for $T=2\,$min as a function of time and $P_{\rm ON}$ (color coded). Overall, the estimated parameters (bottom) mimic the predicted ones (top) well, albeit with fluctuations around the mean, which is expected from a finite-sized sample (200 alleles). 

We observe that the parameters $T_{\rm ON}$ and $T_{\rm OFF}$ are affected by different biases (fig. S9B): 1) bias due to censoring that lead to underestimation and the bending observed near the beginning and the end of the time interval, 2) bias due to missing short ON/OFF periods compared to the sampling time leading to overestimation for small values of $T_{\rm ON}$ and $T_{\rm OFF}$, 3) bias due to the finite length of the recording leading to underestimation for values of $T_{\rm ON}$ and $T_{\rm OFF}$ that are close or larger than the recording duration. Nevertheless, the estimates for $T_{\rm ON}$ and $T_{\rm OFF}$ remain good for the range of probed values. Interestingly, biases on $T_{\rm C}$ are rather limited, whereas fluctuations seem larger than observed for $T_{\rm ON}$ and $T_{\rm OFF}$.

To fully characterize the errors on our estimates, we compare the estimated effective bursting parameters to the predicted ones for all simulated data sets as a function of the burst calling parameters $w$ and $r_b$ (see Eqn. \ref{equ:n_threshold}).
Starting with our default burst calling parameters, i.e. $w=1u$ and $r_b=2/u$, where $u=5\Delta t=5/6\,$min, we compute the median (over time) and 68\% confidence intervals of the estimated bursting parameters for each data set (i.e., a combination of input parameters). Comparing the median bursting parameters with the predicted ones for all data sets (see fig. S9C) demonstrates excellent agreement between the two quantities for a large range of input $P_{\rm ON}^{(1)}\in[0.02,0.90]$ and input $T_{\rm C}^{(1)}\in[0.5,10]\,$min. We quantified the median relative error between the estimated and predicted parameters, both for each individual parameter and globally, as a function of the input $T_{\rm C}^{(1)}$ (see fig. S9D). Although a fixed moving window of size $w=1u$ was used to call bursts, our approach still recovers bursting parameters properly (global median relative error $<$ 31\%) over a rather large range of correlation times ($T_{\rm C}\sim 0.5$--$8\,$min) that encompasses the values estimated from real data ($T_{\rm C}\sim 1$--$2\,$min). Thus, our choice of burst calling parameters did not prevent us from detecting possibly smaller or larger values of switching correlation times than observed in real data (see Fig. 3). Lastly, we investigate the impact of different values of the burst calling parameters $w$ and $r_b$ on the global median relative error (see fig. S9D). This analysis shows that our default choice of burst calling parameters leads to the lowest global relative error over the probed burst calling parameters.

\underline{Time-dependent bursting parameters.} We assess the performance of our method for estimating time-dependent bursting parameters in a realistic setting by generating a data set that mimics the spatiotemporal transcriptional output of \emph{hb} in NC14 within $x/L\sim$0.20--0.46. The corresponding time-dependent input bursting parameters for 18 virtual AP bins (color coded) are displayed in fig. S10E-F top. For each virtual position, we generate 200 alleles and 50 min long activity time series $A$ at $10\,$s intervals (as in real data). Instead of sampling the initiation times series $I$ given the 2-state model using a time-dependent Gillespie algorithm, we use the propagator (Eq. \ref{equ:transition_prob}) to sample both the allele state $n$ and the resulting initiation events $g$ at fixed and short time interval such that $I_j=g(t_{j})$. 

To compute the full propagator of the system, we expand the previously defined state transition matrix (Eq. \ref{equ:switch_operator_2Al}) to also include the initiation process. The new state transition matrix $\hat{M}'(t)$ can be written using tensor products:
\begin{equation}
\hat{M}'(t) = \hat{I}_g \otimes \hat{M}_2(t) + \hat{K}_g(t) \otimes \hat{R}_2,
\label{equ:lagrangian_operator}
\end{equation}
where $\hat{I}_g$ is the identity matrix of size $N_g$, $\hat{M}_2(t)$ is the transition matrix of the allele state for two independent sister chromatids (see Eq. \ref{equ:switch_operator_2Al}), $\hat{K}_g(t)$ is a square matrix of size $N_g$ that describes the initiation of transcripts and is given by
\begin{equation*}
\hat{K}_g(t) = \begin{pmatrix}
  -k(t) & 0 & \hdots & \hdots & 0 \\
  k(t) & -k(t) & \ddots & \ddots  & \vdots \\
  0 & k(t) & \ddots & \ddots & \vdots \\
  \vdots & \ddots & \ddots & -k(t) & 0 \\
 0 & \hdots & 0  & k(t) & -k(t)
 \end{pmatrix},
\end{equation*}
where $k \equiv K^{(1)}$ corresponds to the single gene copy (SGC) initiation rate, and lastly the matrix $\hat{R}_2$ indicates (based on the allele state) how many gene copies participate in the initiation process:
\begin{equation*}
\hat{R}_2 = \begin{pmatrix}
  0 & 0 & 0\\
  0 & 1 & 0 \\
  0 & 0 & 2\\
 \end{pmatrix}.
 \label{equ:r_operator_2Al}
\end{equation*}
Given the expanded state transition matrix (Eq. \ref{equ:lagrangian_operator}), we compute the complete time-dependent propagator of the system using the trapezoidal rule:
\begin{equation}
\hat{U}(t_{j+1},t_{j}) \approx \exp{\left( \frac{1}{2} \left( \hat{M}' (t_{j+1}) + \hat{M}'(t_{j}) \right)\Delta t' \right)}.
\label{equ:full_propa}
\end{equation}
For accurate sampling, we set $t_{j+1}-t_{j}=\Delta t'$ to the minimal discretized time interval $\Delta t'= f/(2K_{\rm elo})=1$s given by the Pol II footprint of $f=60\,$bp and the measured elongation rate $K_{\rm elo}=1.8\,$kb/min. In addition, we set $N_g=6$, which is a safe cutoff given the probed input parameters and the small $\Delta t'$.

Once the propagator has been computed (Eq. \ref{equ:full_propa}), sampling the allele state and the initiation events becomes straightforward. Given a specific allele state $n(t_j)\in \{0,1,2\}$ at time $t_j$, we define the expanded state $\lambda=n(t_j) N_g+1$ and the corresponding zero vector $\mathbf{\Lambda}(t_j)$ of size $3N_g$, whose $\lambda$-th entry has value 1. The vector $\mathbf{\Lambda}(t_j)$ and the propagator enables computing the probability $P_{\lambda'}(t_{j+1})$ to find the system in any state $\lambda'=n(t_{j+1})N_g + g(t_{j+1})+1\in \{1,2,\ldots,3N_g\}$ at time $t_{j+1}=t_i+\Delta t'$, where $g(t_{j+1})\in \{0,1,\ldots,N_g-1 \}$ is the number of initiation events during the time interval $\Delta t'$. We compute the vector $\mathbf{P}(t_{j+1})$ whose entries are the probability $P_{\lambda'}(t_{j+1})$ as
\begin{equation}
\mathbf{P}(t_{j+1}) = \hat{U}(t_{j+1},t_{j}) \mathbf{\Lambda}(t_j).
\end{equation}
Using $\mathbf{P}(t_{j+1})$, we sample the state $\lambda'$ at time $t_{j+1}$ by drawing a single random number $\lambda' \in \{1,2,\ldots,3N_g\}$ according to $P_{\lambda'}(t_{j+1})$. The corresponding new allele state is then given by $n(t_{j+1})=\lfloor (\lambda'-1)/N_g \rfloor$ and the number of initiation events during $\Delta t'$ by $g(t_{j+1})=(\lambda'-1) \mod N_g$. Setting the new state vector $\mathbf{\Lambda}(t_{j+1})$ as the zero vector whose $\lambda''$-th entry is set to 1, with $\lambda''=n(t_{j+1}) N_g+1$, defines an iterative procedure to generate times series according to the 2-state model with time-dependent input parameters. Specifically, this approach enables sampling single initiation time series $I$ such that $I_j=g(t_{j})$, i.e., the number of initiation events during the time interval $(t_{j_1},t_j]$. Given the short interval $\Delta t'=1$s and the probed value of the initiation rate $k\leq8$ mRNA/min, the sampled number of initiation events $g(t_{j})$ is usually comprised between 0 and 1, and almost never exceeds 2.

Capitalizing on the sampling procedure described above, we generate 200 initiation time series $I$ and the corresponding activity time series $A$ at $10\,$s intervals for each virtual position. We first compared the estimated time-dependent bursting parameters with the predicted ones (fig. S10A-D). The temporal changes in effective bursting parameters are recovered well (fig. S10A-B). Indeed, both the predicted and estimated parameters are highly correlated (fig. S10C-D). We further estimate the single gene copy (SGC) input parameters from the estimated effective ones using the relationships derived in Section \ref{sec:sgc_param} ``Sister chromatids and single copy parameters''. Even though the SGC input parameters are deeply buried within the data, our deconvolution and burst calling approaches lead to a well-behaved time-dependent estimation of the latter (fig. S10E-H). Albeit with some distortions, we recover the input parameters adequately, confirming our ability to estimate bursting parameters in a real-case scenario.

\subsection{Exploring the generality of bursting rules.}

We test the validity of our derived bursting relationships beyond the \emph{Drosophila} gap genes. To this end, we extract bursting parameters from four other studies performed in different laboratories on different genes and systems. 

\paragraph{\emph{Drosophila} data.} While our study mainly focusses on the \emph{Drosophila} gap genes, two other studies performed in early \emph{Drosophila} embryos (i.e., one on dorsoventral genes \cite{Hoppe2020} and another one on synthetic core promoter modifications \cite{Pimmett2021}) generated a similar set of bursting parameters that allow for a test of our established rules. For both studies, we assume that the estimated parameters are effective ones (for two indistinguishable sister chromatids) and not single gene copy parameters. 

From the first study (Figure 5D in \cite{Hoppe2020}), we recover the ON-probability (occupancy) $P_{\rm ON}$, the initiation rate (loading rate) $K$, and the burst frequency $F$ for \emph{ush}-wt and st2-\emph{dpp} for the ten positions along the dorsoventral axis considered in that study. We calculate the mean ON-time $T_{\rm ON}=P_{\rm ON}/F$, the mean OFF-time $T_{\rm OFF}=(1-P_{\rm ON})/F$, and the switching correlation time and plot the resulting parameters in yellow in Fig. 5A.

From the second study \cite{Pimmett2021}, we directly use the estimated parameters ($T_{\rm OFF}$, $T_{\rm ON}$, $K$ and $P_{\rm ON}$) for the 2-state model provided in Supplementary Table 2. These correspond to the following seven constructs: snaE$>$\emph{sna}$>$24xMS2-y, snaE$>$\emph{snaTATAlight}$>$24xMS2-y, snaE$>$\emph{snaTATAmut}$>$24xMS2-y,\\snaE$>$\emph{snaTATAlight+INR}$>$24xMS2-y, snaE$>$\emph{kr-INR1}$>$24xMS2-y, snaE$>$\emph{kr-INR2}$>$24xMS2-y and\\ snaE$>$\emph{Ilp4-INR}$>$24xMS2-y. The resulting seven data points are plotted in pink in Fig. 5A. We did not use the estimated parameters for the 3-state model due to some arbitrariness in the potential definitions of an equivalent 2-state $T_{\rm OFF}$ in that particular model.

\paragraph{Yeast data.} We test estimated parameters in a recent study using yeast genes \cite{Brouwer2023}. We recover the mean ON-time (burst duration) $T_{\rm ON}$ and the mean OFF-time (time between bursts) $T_{\rm OFF}$ for \emph{GAL10} under multiple conditions from their Extended Data Fig. 6. We calculated the ON-probability assuming steady state as $P_{\rm ON}=T_{\rm ON}/(T_{\rm ON}+T_{\rm OFF})$. We determine the initiation rate $K$ from the uncalibrated Burst intensity $\tilde{K}$. To this end, we estimate the residence time of nascent mRNA on \emph{GAL10} from the estimated genomic gene length (2200 bp + 862 bp for the PP7 cassette) and the estimated elongation rate of 3.9 kb/min \cite{Donovan2019}. As there might be some retention of nascent transcripts at the transcription site, we round up the gene length to 3.9kb, leading to a residence time for nascent mRNAs of approximately 1 min. We use smFISH analysis results (in triplicate) of the -RSC (DMSO) experiment \cite{Brouwer2023} to calibrate the burst intensity by calculating the mean number of nascent transcripts on the gene. With a conservative definition of transcription sites having at least 5 nascent transcripts, we obtain on average $13.2\pm0.2$ nascent transcripts. Since the residence time of nascent mRNA on this yeast gene is short $~$1 min compared to $T_{\rm ON}\sim$1.7 min, we assume as a first approximation that the mean number of nascent transcripts corresponds only to the ON period. We thus estimate the initiation rate as $K=13.2\tilde{K}/\avg{\tilde{K}}$, which is a lower bound on the true initiation rate. We plot the resulting parameters for the 42 different conditions as orange circles in Fig. 5A. While these yeast parameters are extracted from a single gene copy system, the difference between effective and single gene copy is not very pronounced below $P_{\rm ON}<5$ (see Section \ref{sec:sgc_param}).

\paragraph{Human data.} We re-analyzed single-cell transcriptional time traces of eleven endogenous human genes obtained through live imaging of a MS2 reporter \cite{Rodriguez2019,Wan2021}. Two key reasons motivated such an analysis: 1) the definition of the ON period in the original studies is different from ours, conflating both the mRNA dwell-time (resulting from elongation, cleavage and termination) with the promoter ON-time; 2) the fitted generalized transcription model relies on different assumptions (3 promoter states, maximum number of nascent mRNA between 2 and 5, non-trivial dwell-time distribution due to early intron ejection and initiation prevented by early PolII engagement), which make the mapping of their parameters onto ours difficult. Thus, we chose to apply our deconvolution method on the original data and reconstruct the initiation events, allowing us to estimate bursting parameters whose definition are identical to ours.

The single-cell time traces from the original studies were obtained using a 24xMS2 stem-loops cassette inserted either in the 3'UTR (\emph{TFF1}, \cite{Rodriguez2019}) or within introns (10 other genes, \cite{Wan2021}). The fluorescent signal resulted from MCP-GFP imaged using a confocal microscope (Zeiss LSM780), with a temporal resolution of $\Delta t=100$ s. The original dataset comprised 35 to 123 cells per gene, whose time traces last on average 11.5 hours. As these human genes are lowly expressed, estimating the fluorescent background in the data was straightforward. Indeed, only a few bursts are noticeable in each single-cell time trace and the majority of the time points correspond to background fluctuations. We thus computed the histogram of intensities for each cell, resulting in a mixture of a Gaussian (background fluctuation) and a heavy-tailed distribution (nascent mRNA counts plus noise). We kernel smoothed the resulting distributions to robustly estimate the mode and the 2.5$^\text{th}$ percentile. For each cell, we defined the mean of the background $\mu_b$ as the mode of the mixture and the standard deviation of the background $\sigma_b$ as the half difference between the mode and 2.5$^\text{th}$ percentile. We checked the consistency of our assessment of the background distribution across all cells for a given gene, noticing only a very limited number of outlier time traces corresponding to highly active allele. For these traces, we replaced $\mu_b$ and $\sigma_b$ by $\text{med}(\mu_b)$ and $\text{med}(\sigma_b)$, the median over all cells of the corresponding gene. Lastly, we subtracted the cell specific background $\mu_b$ to the intensities of all traces, such that in absence of transcription the signal is centered around the true zero.

Next, we calibrated the fluorescent signal in absolute units, i.e., nascent mRNA counts. We used the gene \emph{TFF1} as nascent mRNA counts were characterized by smFISH (see Fig. S2C-D in \cite{Rodriguez2019}). Importantly, these smFISH distributions do not include non transcribing alleles (as with dual color smFISH, the detection of transcription sites necessitates intron signal). Thus, to match the histogram of live fluorescent intensities with the smFISH counts for calibration, one needs to account for the unknown fraction of non-transcribing alleles. We fitted the live histogram of intensities (maximum likelihood) with a mixture distribution composed of Gaussian background noise $\mathcal{N}(s|\mu, \sigma_b)$ and smFISH mRNA count distribution $P(g)$, which is given by
\begin{equation*}
P(s|\rho,\alpha) = \frac{1}{Z} \left( \rho \mathcal{N}(s|0, \sigma_b) + (1-\rho) \sum_{g=1}^\infty \mathcal{N}(s|\alpha g, \sigma_b) P(g) \right),
\end{equation*}
where $s$ is the live intensity, $\rho\in[0,1]$ the fraction of time nascent mRNA are present at the locus, and $\alpha\in[0,+\infty)$ the calibration factor (i.e., the live intensity of a single nascent mRNA). $Z$ is the normalization constant such that $\int P(s)ds=1$ and $P(g)$ is normalized such that $\sum_{g=1}^{\infty} P(g)=1$ (thus excluding $P(g=0)$ corresponding to the unknown fraction of non-transcribing alleles). By maximizing the likelihood above for the \emph{TFF1} data, we obtained $\rho=0.81$ and $\alpha=38$. We managed to calibrate the other genes in absence of available smFISH data by extrapolation. We first noticed that $\text{med}(\sigma_b)$ (median over all cells for a given gene) scales linearly with $\text{med}(\mu_b)$ across genes and over a 2-fold range ($R^2=0.96$), which points mainly to a gain difference between datasets. As the different genes were imaged using the same MCP-GFP in the same condition, the background fluctuations should in principle be the same. We thus rescaled the background subtracted signal of the other 10 genes using $\text{med}(\sigma_b)/(\alpha \text{med}(\sigma_b^{\text{\emph{TFF1}}}))$, giving us absolute units for all the dataset.

In order to deconvolve the single-cell time traces, we first need to characterize the gene-specific mRNA dwell-time and assess measurements noise. For each gene, we computed the auto-correlation (AC) function using all the time traces. From the AC (at lag $\tau=0$), we estimated the fraction of uncorrelated variability $\sigma_{\rm img}^2/\sigma^2$, where $\sigma^2$ is the total variance (Table \ref{tab:larson_data}, see Fig. S1D for comparison with our data). Measurements noise represents 34$\%$ of total variability on average (5$\%$ in ours). Given these numbers, we extrapolated a simple measurement noise model in absolute units, i.e., $\sigma(g)=\sqrt{\sigma_b^2 + \beta g}$, with $\sigma_b=0.5$ and $\beta=0.38$. This noise model includes Poisson shot noise as in ours (see Equ. \ref{equ:meas_noise}) and complies with the single mRNA sensitivity claimed in the original studies (though no dual color control were available to us to verify it). 

Regarding the correlated component of the AC (for lag $\tau>0$), it is well approximated by a single exponential that decays with the lag $\tau$ over a timescale $T_{\rm d}$. We thus fitted these AC functions with an exponential estimating $T_{\rm d}$ for each gene (Table \ref{tab:larson_data}). Notably, these time scales are much larger than what we observed in flies (6-fold larger on average). For the gap genes, we know that the mRNA dwell-time is rather small ($\sim$2 min), due to the rapid elongation of short mRNAs (see Section \ref{sec:validation_deconv} and Table \ref{tab:gene_length}) and their small retention time at the locus \cite{Zoller:2018gj}. On the other hand, it has been demonstrated in the original studies \cite{Rodriguez2019,Wan2021} that the mRNA-dwell time of these human genes is rather long (typically around 10 min, even for short mRNAs), due to a combination of transcription elongation, splicing and termination. In accordance with this observation, we assumed that the measured $T_{\rm d}$ for each gene mainly correspond to mRNA dwell-times, thus setting an upper bound on possible switching correlation time (see Sec. \ref{sec:ac_estimation} and Equ. \ref{equ:autocorr_2s}).

Having calibrated the data and characterized both the measurement noise and the mRNA dwell-time, we proceeded to re-analyzed all the transcriptional time traces according to our deconvolution approach (see Section \ref{sec:single_transrate}). We assumed a squared elongation kernel whose duration was set by $T_{\rm d}$ (Table \ref{tab:larson_data}), thus neglecting the discrete effect of the individual MS2 loops. Indeed, the whole MS2 sequence length is $\sim1.4$ kb long, and thus transcribed in $0.7$ min ($K_{\rm elo}=2$kb/min in humans \cite{Wan2021}, which is negligible compared to the total dwell-time. To sample the initiation events, we used $\Delta t'=10$ s (instead of $\Delta t'=1$ for our data), which was small enough given the temporal resolution ($\Delta t=100$ s) and the maximal amount of nascent mRNAs ($\sim 10$). Once the initiation events have been reconstructed, we performed our burst calling routine (see Section \ref{sec:calling_burst}). For the burst calling, we set the averaging window to $w=3\Delta t=300$ s and $g_b=1$. The window $w$ is roughly half the average dwell-time ($\sim 10$ min), which is the limit of what we can reliably detect based on simulated data (Fig. S9). The threshold $g_b$ corresponds to one mRNA, consistent with their claimed single mRNA sensitivity \cite{Rodriguez2019,Wan2021}. We then estimated the bursting parameters as described in Section \ref{sec:estimating_param}. For individual time points, the bursting parameter relative errors is on average 6-fold larger than for our data (27\% vs 4\% relative error on average). However, since these human genes are at steady state (their bursting parameters are stable in time), we only report the time averaged parameters. These parameters are given in Table \ref{tab:larson_param1} and \ref{tab:larson_param2}, and shown in Fig. 5. 

\begin{table}
\begin{center}
    \begin{tabular}{ |c|c|c| } 
    \hline
    Gene & $\sigma_{\rm img}^2/\sigma^2$ & $T_{\rm d}$ [min] \\ 
    \hline
    \emph{TFF1} & $0.21\pm0.02$ & $10.5\pm0.8$ \\ 
    \hline
    \emph{CANX} & $0.53\pm0.04$ & $7.4\pm1.5$ \\ 
    \hline
    \emph{DNAJC5} & $0.29\pm0.02$ & $10.4\pm0.9$ \\ 
    \hline
    \emph{ERRFI1} & $0.49\pm0.06$ & $15.7\pm2.6$ \\ 
    \hline
    \emph{KPNB1} & $0.26\pm0.03$ & $11.8\pm1.4$ \\
    \hline
    \emph{MYH9} & $0.24\pm0.03$ & $6.2\pm0.8$ \\ 
    \hline
    \emph{RAB7A} & $0.25\pm0.02$ & $12.4\pm1.3$ \\ 
    \hline
    \emph{RHOA} & $0.32\pm0.03$ & $9.0\pm1.2$ \\ 
    \hline
    \emph{RPAP3} & $0.62\pm0.04$ & $7.9\pm1.7$ \\ 
    \hline
    \emph{SEC16A} & $0.39\pm0.04$ & $15.5\pm2.1$ \\ 
    \hline
    \emph{SLC2A1} & $0.15\pm0.03$ & $14.9\pm2.0$ \\ 
    \hline
    \end{tabular}
    \caption{\label{tab:larson_data} Estimated fractional imaging noise and dwell-time for the 11 human genes from \cite{Rodriguez2019,Wan2021}.}
\end{center}
\end{table}

\begin{table}
\begin{center}
    \begin{tabular}{ |c|c|c|c|c|c|c|c|c| } 
    \hline
    Gene & $R$ [1/min] & $K$ [1/min]  & $P_{\rm ON}$ & $T_{\rm OFF}$ [min] & $T_{\rm ON}$ [min] & $T_{\rm C}$ [min] \\ 
    \hline
    \emph{TFF1} & $0.027\pm0.001$ & $0.54\pm0.02$ & $0.050\pm0.016$ & $71.3\pm11.7$ & $4.4\pm0.7$ & $4.1\pm0.6$ \\ 
    \hline
    \emph{CANX} & $0.037\pm0.017$ & $0.55\pm0.03$ & $0.065\pm0.035$ & $37.7\pm11.2$ & $2.7\pm0.6$ & $2.5\pm0.5$ \\ 
    \hline
    \emph{DNAJC5} & $0.084\pm0.016$ & $0.52\pm0.01$ & $0.163\pm0.029$ & $17.0\pm1.9$ & $3.5\pm0.4$ & $2.9\pm0.3$ \\ 
    \hline
    \emph{ERRFI1} & $0.008\pm0.004$ & $0.57\pm0.03$ & $0.009\pm0.007$ & $213.6\pm44.7$ & $2.8\pm1.2$ & $2.8\pm1.2$ \\ 
    \hline
    \emph{KPNB1} & $0.101\pm0.041$ & $0.55\pm0.04$ & $0.182\pm0.067$ & $19.9\pm4.4$ & $4.7\pm0.8$ & $3.8\pm0.5$ \\
    \hline
    \emph{MYH9} & $0.081\pm0.033$ & $0.51\pm0.02$ & $0.157\pm0.056$ & $24.9\pm4.6$ & $5.5\pm1.2$ & $4.5\pm0.8$ \\ 
    \hline
    \emph{RAB7A} & $0.052\pm0.017$ & $0.53\pm0.02$ & $0.097\pm0.032$ & $25.3\pm5.7$ & $3.0\pm0.5$ & $2.6\pm0.4$ \\ 
    \hline
    \emph{RHOA} & $0.117\pm0.032$ & $0.51\pm0.01$ & $0.231\pm0.060$ & $12.1\pm2.4$ & $3.8\pm0.6$ & $2.8\pm0.3$ \\ 
    \hline
    \emph{RPAP3} & $0.011\pm0.005$ & $0.57\pm0.03$ & $0.018\pm0.012$ & $117.2\pm31.3$ & $2.4\pm0.6$ & $2.3\pm0.5$ \\ 
    \hline
    \emph{SEC16A} & $0.036\pm0.015$ & $0.55\pm0.02$ & $0.065\pm0.026$ & $37.1\pm9.7$ & $2.9\pm0.5$ & $2.6\pm0.4$ \\ 
    \hline
    \emph{SLC2A1} & $0.067\pm0.029$ & $0.53\pm0.03$ & $0.122\pm0.049$ & $32.9\pm8.0$ & $5.5\pm1.6$ & $4.5\pm1.2$ \\ 
    \hline
    \end{tabular}
    \caption{\label{tab:larson_param1} Estimated bursting parameters using our deconvolution approach for the 11 human genes from \cite{Rodriguez2019,Wan2021}.}
\end{center}
\end{table}

\begin{table}
\begin{center}
    \begin{tabular}{ |c|c|c| } 
    \hline
    Gene & $F$ [1/min] & $B$ \\ 
    \hline
    \emph{TFF1} & $0.014\pm0.002$ & $2.33\pm0.39$ \\ 
    \hline
    \emph{CANX} & $0.026\pm0.007$ & $1.50\pm0.28$ \\ 
    \hline
    \emph{DNAJC5} & $0.049\pm0.005$ & $1.79\pm0.19$ \\ 
    \hline
    \emph{ERRFI1} & $0.005\pm0.001$ & $1.59\pm0.58$ \\ 
    \hline
    \emph{KPNB1} & $0.042\pm0.007$ & $2.62\pm0.53$ \\
    \hline
    \emph{MYH9} & $0.034\pm0.005$ & $2.83\pm0.66$ \\ 
    \hline
    \emph{RAB7A} & $0.037\pm0.007$ & $1.58\pm0.25$ \\ 
    \hline
    \emph{RHOA} & $0.064\pm0.009$ & $1.91\pm0.27$ \\ 
    \hline
    \emph{RPAP3} & $0.009\pm0.002$ & $1.33\pm0.26$ \\ 
    \hline
    \emph{SEC16A} & $0.026\pm0.007$ & $1.58\pm0.30$ \\ 
    \hline
    \emph{SLC2A1} & $0.027\pm0.005$ & $2.94\pm0.87$ \\ 
    \hline
    \end{tabular}
    \caption{\label{tab:larson_param2} Estimated burst frequency and size using our deconvolution approach for the 11 human genes from \cite{Rodriguez2019,Wan2021}.}
\end{center}
\end{table}

\paragraph{Mouse scRNA-seq data} We reanalyzed single-cell RNA-seq data from mouse fibroblasts and embryonic stem cells \cite{Larsson2019}. We followed a very similar approach as in the original study to estimate bursting parameters from UMI (unique molecular identifier) counts. Namely, we fitted the steady state distribution of UMI counts for each gene through maximum likelihood, using a Beta-Poisson distribution (the steady state distribution of the 2-state model of transcription). We also used mRNA lifetime estimates \cite{Herzog2017} to get physical time units. Unlike the original study, we explicitly imposed a prior on the switching correlation time $T_{\rm C}$, effectively regularizing the parameter inference. Indeed, for the vast majority of genes, the empirical steady state distribution is not sufficient to determine all the three independent parameter of the 2-state model ($K$, $P_{\rm ON}$, $T_{\rm C}$). Typically, using mature mRNA counts, one cannot resolve a $T_{\rm C}$ value that is smaller than the mRNA lifetime $T_{\rm M}$, which is of the order of a few hours ($\text{med}(T_{\rm M})=5.3$ h). To alleviate this identifiability issue, we imposed a prior on $T_{\rm C}$ such that the effective log-likelihood $\log{\tilde{\mathcal{L}}}$ is given by
\begin{equation}
\log{\tilde{\mathcal{L}}(D|K,P_{\rm ON},T_{\rm C})} = \log{\mathcal{L}(D|K,P_{\rm ON},T_{\rm C})} -\frac{1}{2}\left(\frac{\log_{10}{T_{\rm C}}-\mu}{\sigma}\right)^2,
\label{equ:sandberg_likelihood}
\end{equation}
where $\log{\mathcal{L}(D|\Theta)}$ is the Beta-Poisson distribution (see \cite{Larsson2019}), and the prior parameters are set to $\mu=0$ and $\sigma=0.5$. The prior corresponds to a Gaussian in log-space centered around $T_{\rm C}=1$ min, whose 95\% coverage spans two order of magnitude ($T_{\rm C}\in[0.1,10]$ min). Given the range of observed $T_{\rm C}$ values (Fig. 5A), this prior is not strongly informative and simply enforces a plausible range compatible with observations.

For each gene, we estimated the bursting parameters $K$, $P_{\rm ON}$ and $T_{\rm C}$ by maximizing the effective likelihood (Equ. \ref{equ:sandberg_likelihood}) on UMI counts. We used the following range for the optimization: $P_{\rm ON}\in[10^{-5},1]$, $T_{\rm C}\in[10^{-2},10^{3}]$ min and $K\in[10^{-2.5},10^{2.5}]$ min$^{-1}$. To assess inference errors $\sigma_{\Theta}$, we bootstrapped the estimated parameters by resampling the UMI distributions. We only reported genes whose parameters satisfy $2\sigma_{\Theta}<0.5$ in log-space and $B=K T_{\rm C}/(1-P_{\rm ON})>1$. This effectively removes 32\% of the genes, which is comparable to the effect of the thresholding applied in the original study. We show the mean transcription rate $R=K P_{\rm ON}$, the burst frequency $F=P_{\rm ON}(1-P_{\rm ON})/T_{\rm C}$ and the burst size $B$ in Fig. 5B. Importantly, as in the orginal study \cite{Larsson2019}, we did not attempt to correct the UMI counts for low mRNA recovery rate (typically $\sim$10--30\% recovery rate in recent scRNA-seq). Thus, parameters such as $R$, $P_{\rm ON}$ and $F$ are very likely underestimated by a factor 3 to 10. Lastly, we assessed the impact of our prior on the maximum log-likelihood values. The median penalty imposed by our prior on the log-likelihood is 0.06, and for 80\% of the genes this penalty is less than one. It turns out that the resulting best-fit distributions are barely different from the ones obtained without priors, suggesting that indeed the mRNA distributions contain very little information about $T_{\rm C}$ when $T_{\rm C}\ll T_{\rm M}$ (the mRNA lifetime).
 
\end{document}